%% file: main.tex
\documentclass[twocolumn]{aastex701}
\usepackage{placeins} 

\usepackage{tcolorbox}
\usepackage{soul}

\input{vectors}
\input{units}
\input{symbols}
\input{nuclides}

\input{formatting}

\input{derivatives}

\input{concepts}

\input{code}


\input{abbrev}

\submitjournal{ApJ}

\begin{document}

\title{Impact of 3D macro physics and nuclear physics on the \pnucn{} in \oxygen[]-\carbon[] shell mergers}

\author[0000-0002-1283-6636,gname=Joshua,sname=Issa]{Joshua Issa}
\affiliation{Astronomy Research Centre, Department of Physics \& Astronomy, University of Victoria, Victoria, BC, V8W 2Y2, Canada}
\affiliation{NuGrid Collaboration, \url{http://nugridstars.org}}
\email[show]{joshuaissa@uvic.ca}  

\correspondingauthor{Joshua Issa}

\author[0000-0001-8087-9278,gname=Falk,sname=Herwig]{Falk Herwig}
\affiliation{Astronomy Research Centre, Department of Physics \& Astronomy, University of Victoria, Victoria, BC, V8W 2Y2, Canada}
\affiliation{NuGrid Collaboration, \url{http://nugridstars.org}}
\email{fherwig@uvic.ca}

\author[0000-0001-6120-3264]{Pavel A. Denissenkov}
\affiliation{Astronomy Research Centre, Department of Physics \& Astronomy, University of Victoria, Victoria, BC, V8W 2Y2, Canada}
\affiliation{NuGrid Collaboration, \url{http://nugridstars.org}}
\email{pavelden@uvic.ca}

\author[0000-0002-9048-6010]{Marco Pignatari}
\affiliation{HUN REN Konkoly Observatory, CSFK, H-1121, Budapest, Konkoly Thege M. \'ut 15–17, Hungary}
\affiliation{CSFK, MTA Centre of Excellence, Budapest, Konkoly Thege Miklós út 15-17., H-1121, Hungary}
\affiliation{University of Bayreuth, BGI, Universitätsstraße 30, 95447 Bayreuth, Germany}
\affiliation{NuGrid Collaboration, \url{http://nugridstars.org}}
\email{mpignatari@gmail.com}

\received{July 22, 2025}
\revised{October 8, 2025}
\accepted{October 8, 2025}

\begin{abstract}
\oxygen[]-\carbon[] shell mergers in massive stars are a site for producing the \pnucn{} by the \gprn{}, but 1D stellar models rely on mixing length theory, which does not match the radial velocity profiles of 3D hydrodynamic simulations.
We investigate how 3D macro physics informed mixing impacts the nucleosynthesis of \pnucn{}.
We post-process the \oxygen[] shell of the $\mzams=\unit{15}{\Msun}$, $Z=0.02$ model from the NuGrid stellar data set.
Applying a downturn to velocities at the boundary and increasing velocities across the shell as obtained in previous results, we find non-linear, non-monotonic increase in \pnuc{} production with a spread of \unit{0.96}{\dex}, and find that isotopic ratios can change. 
Reducing \carbon[]-shell ingestion rates as found in 3D simulations suppresses production, with spreads of $1.22${--}\unit{1.84}{\dex} across MLT and downturn scenarios. 
Applying dips to the diffusion profile to mimic quenching events also suppresses production, with a \unit{0.51}{\dex} spread. 
We analyze the impact of varying all photo-disintegration rates of unstable \nt-deficient isotopes from \selenium[]{--}\polonium[] by a factor of $10$ up and down.
The nuclear physics variations for the MLT and downturn cases have a spread of $0.56${--}\unit{0.78}{\dex}.
We also provide which reaction rates are correlated with the \pnucn{}, and find few correlations shared between mixing scenarios.
Our results demonstrate that uncertainties in mixing arising from uncertain 3D macro physics are as significant as nuclear physics and are crucial for understanding \pnuc{} production during \oxygen[]-\carbon[] shell mergers quantitatively. 
\end{abstract}


\section{Introduction} \label{sec:intro}
The origin of the solar pattern of the 35 stable \pnucn{} is a long-standing problem for understanding nucleosynthesis.
\cite{burbidgeSynthesisElementsStars1957} first identified these isotopes as a distinct group, produced primarily not by the $s$ or \rprn, but instead by a \pprn{} through $(\pt,\gamma)$, $(\gamma,\pt)$, and $(\gamma,\nt)$ reactions occurring during Type II supernovae and possibly Type I supernovae.
Based on the first generations of stellar computational models, \cite{arnouldPossibilitySynthesisProtonrich1976} found that the \pprn{} could be driven by all photo-disintegration reactions $(\gamma,\nt)$, $(\gamma,\alpha)$ and $(\gamma,\pt)$ during the most advanced evolutionary stages of massive stars. 
\cite{woosleyPprocessesSupernovae1978} found these photo-disintegrations could create the distribution of \ppr{} nuclei during the passage of the supernova shock over the internal progenitor structure, which they called the \gprn{}.
Following works better defined \gprn{} production in Type II supernovae and more in general in core-collapse supernovae \cite[CCSNe e.g.,][]{prantzosPprocessSN1987A1990, rayetPprocessTypeII1995, travaglioRoleCorecollapseSupernovae2018,choplinPprocessExplodingRotating2022, robertiGprocessNucleosynthesisCorecollapse2023, robertiGprocessNucleosynthesisCorecollapse2024b}.

A variety of additional processes and astrophysical sites have been discussed, and no single mechanism produces all the \pnucn{}.
\cite{woosleyAlphaProcessRProcess1992} found that \selenium[74]{--}\molybdenum[92] could be produced during the $\alpha$-rich freezeout of a supernova.
\cite{frohlichNeutrinoInducedNucleosynthesisA>642006} found high neutrino fluxes during a supernova can create \selenium[74]{--}\cadmium[108] by $(\nt,\pt)$, $(\nt, \gamma)$, and $(\pt, \gamma)$ reactions in a $\nu\pt$ process.
\cite{schatzRpprocessNucleosynthesisExtreme1998} suggested that a hydrogen-rich accretion disk around a neutron star could undergo a series of rapid proton captures in a $r p$ process to produce \selenium[74]{--}\ruthenium[98].
\cite{xiongProduction$p$Nuclei2024} proposed that neutrino induced reactions of \rpr{} material in a $\nu r$ process could produce \krypton[78]{--}\lanthanum[138] in the winds of a proto-neutron star. 
\cite{gorielyHedetonationSubChandrasekharCO2002} proposed that a proton-poor and neutron boosted region could undergo proton-captures could produce all \pnucn{} by a $pn$ process during He-detonation of a \carbon[]-\oxygen[] white dwarf's ejected envelope.
\cite{rauscherNucleosynthesisMassiveStars2002} found that the \gprn{} can produce the \pnucn{} in massive stars during core-collapse supernovae, and also beforehand if the \oxygen[] shell merges with the \carbon[] shell, but that the \gprn{} underproduces \molybdenum[92,94] and \ruthenium[96,98].
\cite{ritterConvectivereactiveNucleosynthesisSc2018} and \cite{robertiGprocessNucleosynthesisCorecollapse2023} confirmed these results and studied the \pprn{} triggered by \oxygen[]-\carbon[] shell merger.

The \gprn{} describes the flow of $(\gamma,\nt)$, $(\gamma,\pt)$, and $(\gamma,\alpha)$ reactions on the stable isotopic seeds that are already present in the shell at temperatures of $1.5 \leq T \leq \unit{\natlog{3}{9}}{K}$ \citep{rauscherConstrainingAstrophysicalOrigin2013}.
However, the \gprn{} in massive stars underproduces not only the \molybdenum[] and \ruthenium[], but all \pnucn{} with $A=90{-}130$ \citep{arnouldPprocessStellarNucleosynthesis2003, woosleyNucleosynthesisRemnantsMassive2007}.
\cite{travaglioTypeIaSupernovae2011} showed that the \gprn{} could produce all \pnucn{} during a Type Ia supernova from the \spr{} material synthesized during stellar evolution without the underproduction of \molybdenum[92,94] and \ruthenium[96,98].
Furthermore, \cite{travaglioTestingRoleSNe2015} found that modifying the distribution of \spr{} material significantly influenced the production of the \pnucn{}, especially the heaviest ones, and that the lightest three were strongly dependent on the metallicity.
\cite{battinoHeavyElementsNucleosynthesis2020} additionally found that the H-flashes of rapidly accreting white dwarfs which undergo the \iprn{} could modify the seed distribution to produce \pnucn{} with $96 < A < 196$ by the \gprn{} during the subsequent SNIa.

Since the first isotopic classification made by \cite{burbidgeSynthesisElementsStars1957}, it has been found that not all \pnucn{} in the solar pattern are produced by a single process.
\cite{bisterzoSprocessLowmetallicityStars2011} state that \gadolinium[152], \erbium[164], and \tantalum[180] have significant contributions of $70.5\%$, $75.5\%$, and $74.5\%$ from the \sprn{}.
\cite{dillmannPProcessSimulationsModified2008} found that \indium[113] and \tin[115] are made by $\beta$-decays after the \rprn{} through isomeric states.
\cite{gorielyPuzzleSynthesisRare2001} argue that $(\gamma,\nt)$ was too weak to produce \lanthanum[138], and instead that it is made by $\nu_e$-capture on \barium[138] during the CCSN, and \cite{arnouldPprocessStellarNucleosynthesis2003} similarly say that \tantalum[180m] could also have $\nu$-induced contributions.
\cite{sieverdingNProcessLightImproved2018} also found that the $\nu$-process is important for the nucleosynthesis of \indium[113], \lanthanum[138], and \tantalum[180m].

The \oxygen[] shell where \pnucn{} are produced during a merger is a convective-reactive environment where mixing and nuclear burning timescales are equal \citep{ritterConvectivereactiveNucleosynthesisSc2018,yadavLargescaleMixingViolent2020a}.  
If there is significant energy released the flow can be modified \citep{dimotakisTurbulentMixing2005}, such as \hydrogen[]-ingestion into \helium[]-burning shell \citep{herwigCONVECTIVEREACTIVEPROTON2011,herwigGLOBALNONSPHERICALOSCILLATIONS2014} or \oxygen[]-\carbon[] shell mergers causing violent mixing \citep{andrassy3DHydrodynamicSimulations2020,yadavLargescaleMixingViolent2020a}.  
The Damköhler number \citep{dimotakisTurbulentMixing2005} quantifies the ratio of these timescales and is defined as:  
\begin{equation}  
    D_\alpha \equiv \frac{\tau_{\mathrm{mix}}}{\tau_{\mathrm{react}}}  
\end{equation}  
where $\tau_{\mathrm{mix}}$ is the mixing timescale and $\tau_{\mathrm{react}}$ is the nuclear reaction timescale.  
Convective regions where the mixing timescale is much faster than the nuclear burning timescale have $D_\alpha \ll 1$, and species are well-mixed across the region.  
Convective-reactive regions where timescales are equal have $D_\alpha \sim 1$, and species can either react at a location or advect to another location and react with the material there, and as a consequence are not well-mixed.  
The mixing and reaction timescales can be given by:  
\begin{equation}  
    \tau_{\mathrm{mix}} = \frac{\ell^2}{D_{\mathrm{MLT}}}; \quad  \tau_{\mathrm{react}} = \frac{1}{\rho N_A \langle \sigma v \rangle Y_j}  
\end{equation}  
where $\ell$ is the mixing length, $D_{\mathrm{MLT}}$ is the mixing diffusion coefficient, $\rho$ is the local density, $N_A$ is Avogadro's number, $\langle \sigma v \rangle$ is the thermally averaged reaction rate, and $Y_j$ is the molar abundance of the interacting species.  
The diffusion coefficient is $D_{\mathrm{MLT}} = \frac{1}{3} v_{\mathrm{MLT}}\cdot\ell$, where $v_{\mathrm{MLT}}$ is the convective velocity and the mixing length is $\ell= \alpha \cdot H_p$ where $H_p$ is the pressure scale height and $\alpha$ is a free parameter \citep[][as reviewed in \citealp{arnettMIXINGLENGTHTHEORYSTEP2015}]{vitenseWasserstoffkonvektionszoneSonneMit1953, bohm-vitenseUberWasserstoffkonvektionszoneSternen1958, kippenhahnStellarStructureEvolution2013}.

Existing massive star models that calculate \pnuc{} nucleosynthesis are 1D and rely on mixing length theory (MLT) to describe convection. 
However, multi-dimensional hydrodynamic simulations of convective \oxygen[]-shell burning predict higher convective velocities than MLT and show a gradual downturn in the mixing efficiency profile at shell boundaries \citep{meakinTurbulentConvectionStellar2007,jonesIdealizedHydrodynamicSimulations2017}. 
3D simulations reveal features absent in 1D, such as asymmetric nuclear burning \citep{bazanConvectionNucleosynthesisCore1994, yadavLargescaleMixingViolent2020a}, large-scale non-radial density asymmetries, and potentially lower \carbon[]-shell ingestion rates during \oxygen[]-\neon[]-\carbon[] shell mergers \citep{andrassy3DHydrodynamicSimulations2020,yadavLargescaleMixingViolent2020a}. 
In 1D models like those from \cite{ritterNuGridStellarData2018}, \oxygen[]-\carbon[] shell mergers occur because the upper boundary of the \oxygen[] shell entrains \carbon[12] and \neon[20] as it burns, flattening the entropy gradient. 
In contrast, 3D simulations by \cite{rizzutiShellMergersLate2024a} find the lower boundary of the \carbon[] shell extending downward and engulfing the \oxygen[] shell. 
Similarly, \cite{yadavLargescaleMixingViolent2020a} show that entropy generation in nuclear burning hotspots within the \neon[] shell leads to downdrafts that raise the entropy in the \oxygen[] shell. 
Dynamic behaviour shortly before the core collapse for these supernova progenitors late in their evolution are not captured in 1D \citep{arnettRealisticProgenitorsCorecollapse2011, mullerStatusMultiDimensionalCoreCollapse2016, yadavLargescaleMixingViolent2020a}. 
While 3D hydrodynamic simulations may not be solving all relevant equations, such as a robust nuclear network, 1D models fundamentally fail to represent the non-radial mixing and spherically asymmetric instabilities during \oxygen[]-\carbon[] shell mergers \citep{meakinActiveCarbonOxygen2006,andrassy3DHydrodynamicSimulations2020, yadavLargescaleMixingViolent2020a}. 

This has consequences for \pnuc{} production. Nuclei with $A > 110$ are primarily synthesized during the merger, not during explosive burning, regardless of the peak CCSN energy \citep{robertiGprocessNucleosynthesisCorecollapse2023,robertiGprocessNucleosynthesisCorecollapse2024b}. 
To explore the impact of the macrophysical uncertainties in the \oxygen[] shell during a merger, we adopt a 3D hydrodynamic inspired set of modified radial mixing profiles and ingestion rates to determine the impact of mixing on the \gprn.
We will also explore how varying the nuclear reaction rates impact the nucleosynthesis of the \pnucn{} as the \oxygen[] shell is a convective-reactive environment.

Section \ref{sec:methods} describes the post-processing of the \cite{ritterNuGridStellarData2018} model using 3D hydrodynamic-inspired mixing profiles and assesses the impact of varying nuclear reaction rates.
Section \ref{sec:convreacflow} examines how the convective-reactive environment produces the \pnucn{} and the role of \carbon[]-shell ingestion.
Sections \ref{sec:convdownturnimpact}{--}\ref{sec:goshimpact} explore how 3D hydrodynamic insights affect \pnucn{} production: \ref{sec:convdownturnimpact} analyzes the impact of downturns and boosted velocities in the \oxygen[] shell, \ref{sec:ingestionimpact} investigates reduced \carbon[]-shell ingestion, and \ref{sec:goshimpact} evaluates dips in the mixing efficiency due to quenching.
Section \ref{sec:nuclearimpact} presents the sensitivity to nuclear rates, their correlation with \pnucn{}, and their relation to mixing profiles and velocities.
Finally, Section \ref{sec:discussion_conclusion} summarizes our findings and their implications.

\section{Methodology} \label{sec:methods}

\subsection{Initial Model and Post-Processing Setup}

\begin{figure}
\includegraphics[width=\columnwidth]{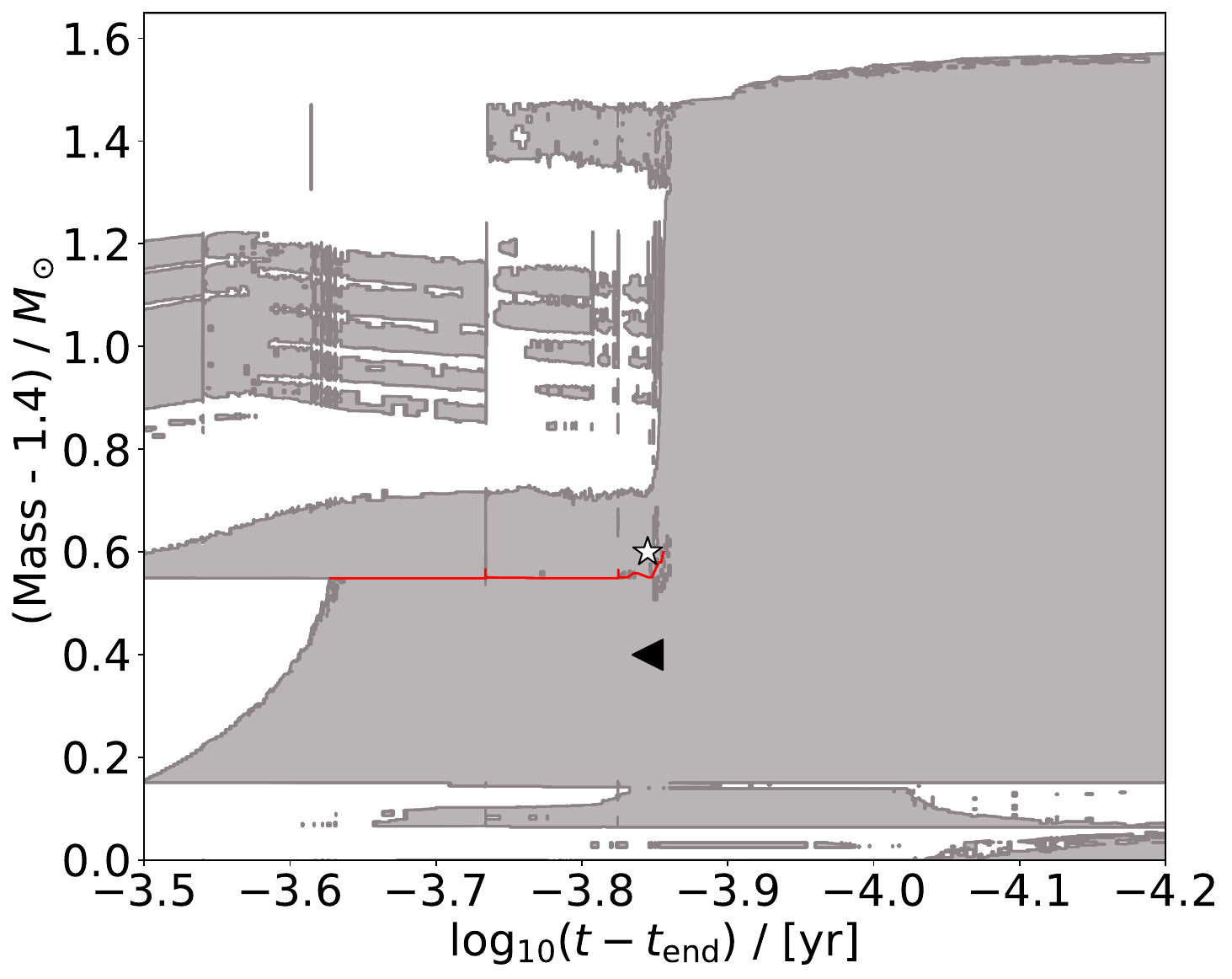}
\caption{Kippenhahn diagram showing the merger of the convective O and \carbon[]burning shells. The \oxygen[]burning shell extends from $\unit{1.55}{\Msun}$ to $\unit{1.95}{\Msun}$, and the first ingested \carbon[]-burning shell from $\unit{1.96}{\Msun}$ to $\unit{2.11}{\Msun}$. A red guideline has been provided to mark the thin radiative layer separating the \oxygen[] and \carbon[] shell. Other convective regions are also \carbon[]burning shells. The merger onsets at $\log_{10}(t-t_{\mathrm{end}}) /\mathrm{yr} \approx -3.85$ and reaches full extent at $\approx-4$. A black triangle marks where the initial composition is taken from and a white star marks the location where the ingested \carbon[]-shell material is taken from for this study.
\label{fig:kippenhahn}}
\end{figure}

We post-process the $\mzams=\unit{15}{\Msun}$, $Z=0.02$ massive stellar model from the NuGrid data set \citep{ritterNuGridStellarData2018}.
The 1D stellar model was computed with \MESA{} \citep{paxtonMODULESEXPERIMENTSLAR2010} without rotation and convective boundary mixing is treated using an exponential-diffusive prescription \citep{freytagHydrodynamicalModelsStellar1996,herwigEvolutionAGBStars2000} with an overshoot parameter of $f=0.022$ at all boundaries except at the base of convective shells where $f=0.005$ until the end of core \helium[] burning, after which $f=0$.
This model has a merger of its convective \oxygen[] and \carbon[]-burning shells late in its evolution as shown in Figure \ref{fig:kippenhahn}.
During this merger, the \carbon[]-burning ashes and stable isotopic material are ingested into the much hotter \oxygen[]-burning shell.

The detailed nucleosynthesis is calculated with the 1D multi-zone post-processing code \mppnp{} \citep{pignatariNuGridStellarData2016}.
\mppnp{} is a multi-zone post-processing code that uses stellar structure calculated by stellar evolution codes to calculate the full nucleosynthesis of a stellar model.
\mppnp{} treats mixing, nuclear burning, and ingestion separately rather than the coupled treatment by a code like \MESA{} \citep{paxtonMODULESEXPERIMENTSLAR2010}. 
A convergence test was performed by decreasing the timesteps and increasing the number of mass zones.
It found that a timestep of $\Delta t = \unit{0.01}{\second}$ and $400$ equidistant mass zones with $4$ additional zones at the bottom of the \oxygen[] shell were sufficient to resolve both the burning and mixing timescales with a decreasing mixing efficiency profile.
Calculations initially used a $5234$ isotope network, but many \nt-rich species were unnecessary.  
A network of only the necessary $1470$ isotopes was adopted, focusing on the  \nt-deficient isotopes, for faster calculations.  
Isomeric states were not included in this network so \tantalum[180m] is not calculated. 
Entrainment of \carbon[]-shell material is treated the same as described in \cite{denissenkovIprocessYieldsRapidly2019} and does not depend on the time step.
A single simulation costs approximately $8$ hours on $40$ cores, for a total of $274$ core years for this work.

The \mppnp{} code calculates both the undecayed mass fraction as a function of mass and the decayed mass-averaged mass fraction at a temperature of $T=100~\mathrm{MK}$ without explosive contributions.
The mass fraction $X_i$ of species $i$ is defined as the ratio of the mass of that species to the total mass of the stellar material, such that $\sum_i X_i = 1$ for all isotopes.

\begin{figure*}
\includegraphics[width=\textwidth]{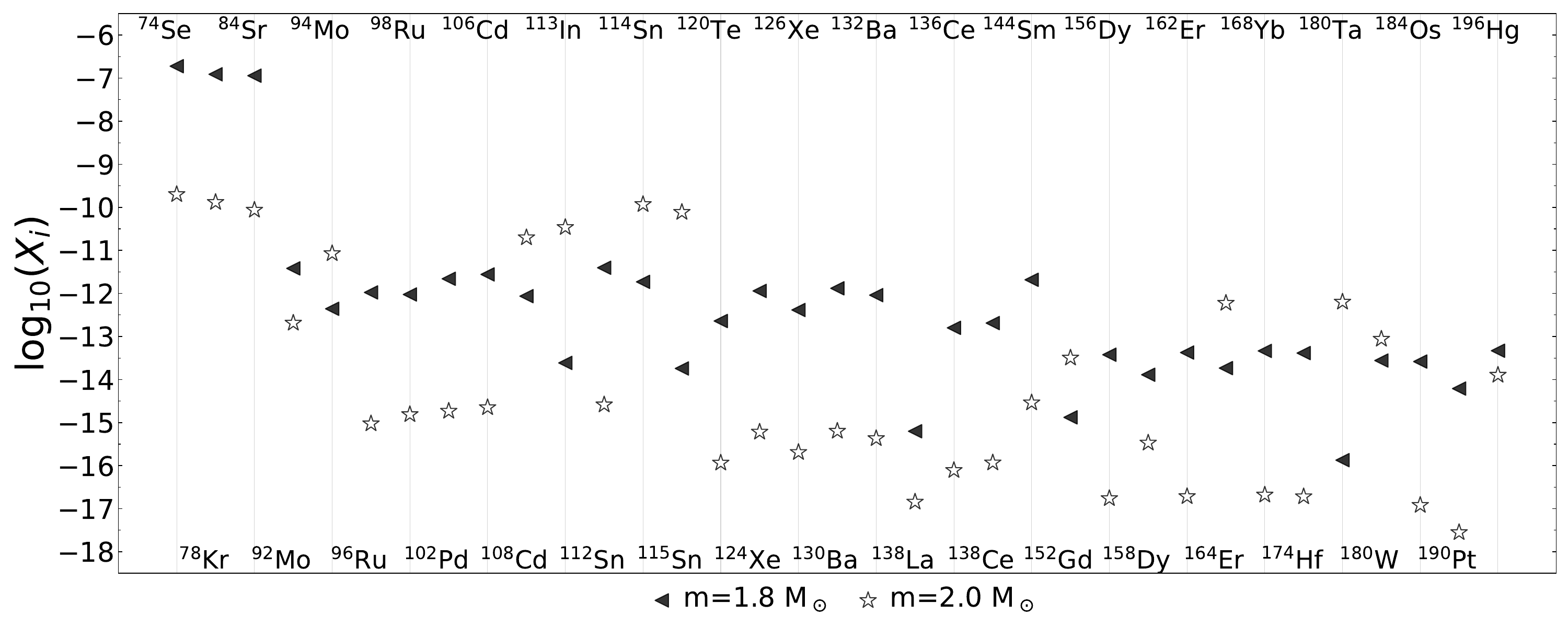}
\caption{Mass fractions of the \pnucn{} from $\log_{10}(t-t_{\mathrm{end}}) /\mathrm{yr}=-3.856$ used for initial \oxygen[]-shell composition and ingested \carbon[]-shell material. Markers are the same as Figure \ref{fig:kippenhahn}.
\label{fig:initial_comp}}
\end{figure*}

To analyze the impact of macrophysical uncertainties and varying the nuclear reaction rates during the merger, isotopic mass fractions are taken just before the onset of merger at $\log_{10}(t - t_{\mathrm{end}})/\mathrm{yr} = -3.845$ from $m = \unit{1.8}{\Msun}$, and the ingested \carbon[]-shell material is taken from $m = \unit{2.0}{\Msun}$ at the same time, as shown in s~\ref{fig:kippenhahn} and~\ref{fig:initial_comp}.  
Earlier in the model, there is some \pnuc{} production in the first convective \oxygen[] shell, which extends from $\unit{1.55}{\Msun}$ to $\unit{1.95}{\Msun}$ during $\log_{10}(t - t_{\mathrm{end}})/\mathrm{yr} = -1.76$ to $-2.16$.  
These nuclei are not processed by any further burning before the merger.  

The results are presented in terms of an overproduction \OP{} compared to the initial composition:
\begin{equation}\label{eq:OP}
    \OP = \log_{10}\Biggl(\frac{X_{f}}{X_{i}}\Biggr)
\end{equation}
where $X_{f}$ is the final mass-averaged decayed mass fraction of a species in the \oxygen[] shell and $X_{i}$ is the inital mass fraction.
The average overproduction factor is calculated as the arithmetic mean of \OP{}:
\begin{equation}
    \langle \OP \rangle = \frac{1}{N}\sum_i^{N} \OP_i
\end{equation}
where $N$ is the number of \pnucn{}.
The mean logarithmic value is defined similar to hs and ls for the \sprn{} as done by \cite{bussoNucleosynthesisAsymptoticGiant1999}.

\begin{figure} 
\includegraphics[width=\columnwidth]{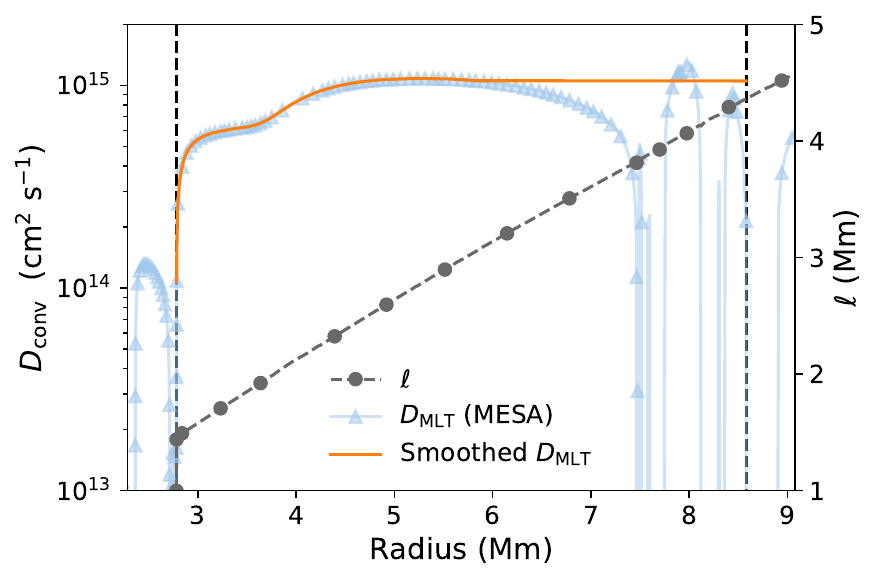}
\caption{The diffusion coefficient profile and mixing length at model number 9200 for the $M_{\mathrm{ZAMS}}=\unit{15}{\Msun}$, $Z=0.02$ model. The light blue line is $D$ from MESA, the orange line the smoothed $D$ used for the MLT mixing scenario in this paper, and the grey line is the mixing length. Black dashed lines mark the shell boundaries for this paper.
\label{fig:dmlt_forced}}
\end{figure}

The stellar structure used is from the onset of the merger at $\log_{10}(t-t_{\mathrm{end}}) /\mathrm{\yr}=-3.856$ and to clearly analyze the impact of mixing alone, the stellar structure is kept constant.
Although the structure is not static in the model, the change to the temperature, density, and entropy between the initial composition and where we take the structure from is less than 5\% during those $\unit{110}{\second}$.
The merger at $\log_{10}(t-t_{\mathrm{end}}) /\mathrm{yr}=-3.856$ is not fully developed, but MLT cannot accurately describe this region as the mixing length $\ell$ is too large \citep{renziniEmbarrassmentsCurrentTreatments1987, arnett3DSimulationsMLT2019}.
Because of this, the mixing efficiency profile is smoothed at the top as shown in Figure \ref{fig:dmlt_forced} to simulate a full merger.

\subsection{1D implementation of 3D macrophysics}\label{sec:mixing_methods}

MLT predictions of the radial mixing efficiency profile deviate from the more realistic predictions from 3D simulations.
3D convective \oxygen[] burning simulations show that the radial convective velocity profile gradually decreases near the shell boundary, in contrast to MLT predictions of a stiff boundary \citep{meakinTurbulentConvectionStellar2007,jonesIdealizedHydrodynamicSimulations2017}.  
This downturn is seen at both the bottom and top of convective shells \citep{herwigHydrodynamicSimulationsHe2006,meakinTurbulentConvectionStellar2007,jonesIdealizedHydrodynamicSimulations2017}.  
The decrease occurs because mixing is driven by convective plumes in these simulations, rather than the idealized convective blobs in MLT.  
Plumes exhibit strong radial velocities in the interior of the convective region but lose their radial component as they reach the boundary, while non-radial velocity components increase.  
This behavior contrasts with MLT, which predicts a sharp drop to zero velocity at the boundary.  
Using Equation 4 from \cite{jonesIdealizedHydrodynamicSimulations2017}, the downturn to the mixing efficiency profile can be implemented in 1D:
\begin{equation}
    D_{\mathrm{3D\text{-} insp.}} = \frac{1}{3} v_{\mathrm{MLT}}\times\min{(\ell,r-r_0)}
\end{equation}
where $\ell$ is the mixing length, $r_0$ is the Schwarzschild boundary at the bottom of the \oxygen[] shell, and an additional factor of $1/3$ is applied to match the MLT diffusion coefficient at the top of the shell.

\begin{figure} 
\includegraphics[width=\columnwidth]{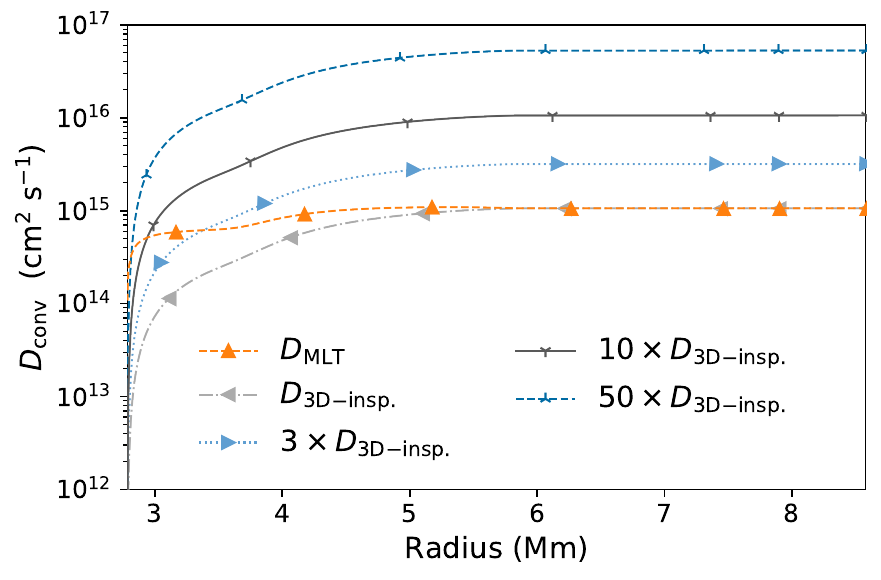}
\caption{The diffusion coefficient profiles for the MLT and 3D-inspired gradual downturn scenarios. The dashed orange line is $D_{\mathrm{MLT}}$ and the dashed light grey line, dotted light blue line, solid grey line, and dashed dark blue line are the downturn profiles with boost factors of $1$, $3$, $10$, and $50$ respectively. 
\label{fig:all_dconv}}
\end{figure}

MLT also underpredicts the strength of the convective velocities in the \oxygen[] shell.
\cite{jonesIdealizedHydrodynamicSimulations2017} found that convective velocities are stronger by a factor of $\sim 30$ compared to MLT. 
\cite{andrassy3DHydrodynamicSimulations2020} in their 3D \carbon[]-shell entrainment simulations show that their velocities could be up to $\sim 5$ times stronger than \cite{jonesIdealizedHydrodynamicSimulations2017} depending on the luminosity of \carbon[] and \oxygen[] burning.
It is possible the velocities could be even higher as these simulations do not include feedback from a nuclear network or treatment of radiation pressure \citep{jonesIdealizedHydrodynamicSimulations2017, andrassy3DHydrodynamicSimulations2020}.
\cite{rizzutiShellMergersLate2024a} finds that velocities are boosted by a factor of $\sim 10$ due to the feedback from new reactions with the ingested material in their 3D \oxygen[]-\carbon[] shell mergers.
We implement this by applying a boost factor of $3\times$, $10\times$, and $50\times$ to the convective velocities as shown in Figure \ref{fig:all_dconv}.

The entrainment rate of \carbon[]-shell material could be lower during the merger depending on the strength of burning \citep{jonesIdealizedHydrodynamicSimulations2017, andrassy3DHydrodynamicSimulations2020}.
To investigate the impact of this, a range of rates are considered: $\unit{\natlog{4}{-5}}{\Msun\second^{-1}}$, $\unit{\natlog{4}{-4}}{\Msun\second^{-1}}$, $\unit{\natlog{4}{-3}}{\Msun\second^{-1}}$, and a scenario with no entrainment.
The maximum mass of the convective \carbon[]-shell in our model is \unit{0.8}{\Msun}, and since our simulation goes from $\log_{10}(t-t_{\mathrm{end}})/\mathrm{yr}=-3.856$ to -3.845 (\unit{110}{\second}), the maximum entrainment rate would be $\unit{\natlog{7}{-3}}{\Msun\second^{-1}}$ similar to \cite{ritterConvectivereactiveNucleosynthesisSc2018}.

\begin{figure} 
\includegraphics[width=\columnwidth]{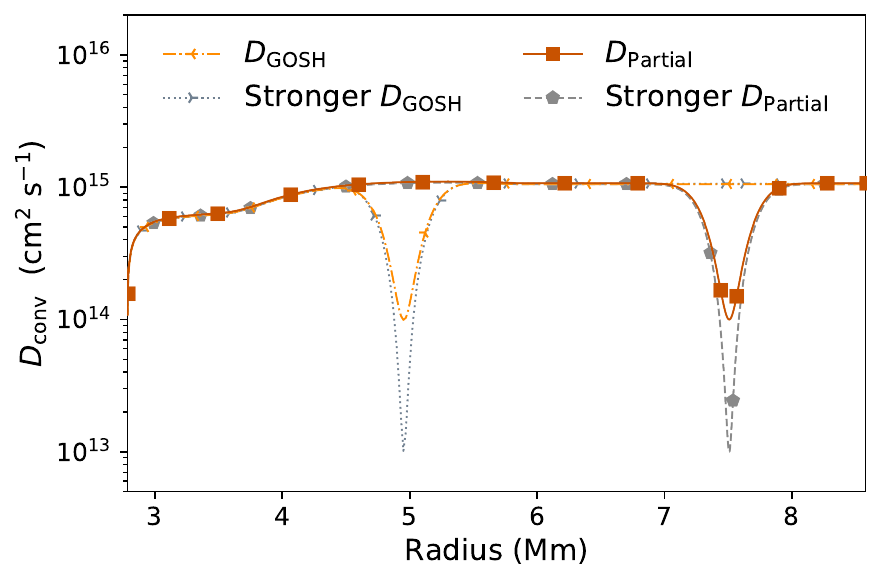}
\caption{The quenched mixing scenario convective profiles. The dashed light orange line and dotted light grey line are the GOSH-like profiles with a dip centred at $r=\unit{4.95}{\Mm}$. The solid red line and dashed dark grey line are the partial merger profiles with a dip centred $r=7.5~\mathrm{Mm}$.
\label{fig:dgosh_partial}}
\end{figure}

1D and 3D simulations show that the convective profile can be quenched as material is ingested as entropy changes due to entrainment and burning \citep{ibenThermalPulsesPcapture1975, sackmannCarbonEruptiveStars1974, herwigFormationHydrogendeficientPostAGB1999, millerbertolamiNewEvolutionaryCalculations2006, herwigCONVECTIVEREACTIVEPROTON2011,herwigGLOBALNONSPHERICALOSCILLATIONS2014}.
As an example, during \hydrogen[]-ingestion into a \helium[]-shell, the energy feedback from the ingested protons quickly burning can cause a split in the convective profile with a very small amount of entrainment \citep{herwigCONVECTIVEREACTIVEPROTON2011}.
\cite{herwigGLOBALNONSPHERICALOSCILLATIONS2014} found this effect could decrease the radial velocity profile and reduce the entrainment of species and labelled the event Global Oscillation of Shell Hydrogen-Ingestion (GOSH).
Similar effects during \carbon[]-shell entrainment could be possible, as \cite{andrassy3DHydrodynamicSimulations2020} found that strong oscillatory modes like GOSHs were present in their 3D simulations.
Energy feedback events like this could explain supernova observations \citep{smithPREPARINGEXPLOSIONHYDRODYNAMIC2014}.
There is no clear prescription for how to implement this effect into 1D models, but it is clear that there would be decreased mixing because of a split.
To investigate a possible convective quenching, we consider a GOSH-like event and a partial merger, where a Gaussian dip (but not full split) occurs in the MLT diffusion profile:
\begin{equation}
D_{\mathrm{quench}}=D_{\mathrm{MLT}}-(D_{\mathrm{MLT}}-c)\times\exp\Biggl[-\frac{(r-a)^2}{w^2}\Biggr]
\end{equation}
where $c$ is the maximum extent of the dip, $w$ is the width, and $a$ is the center of the dip.
The GOSH-like convective splitting could occur at the location where $D_\alpha=1$ for a significant burning event \citep{herwigCONVECTIVEREACTIVEPROTON2011}. 
We centre this event at $a=\unit{4.95}{\Mm}$, where $\unit{1}{\Mm}=\unit{10^6}{\meter}$, at a location of probable energetic feedback could occur.
The \oxygen[] shell could partially merge with the \carbon[] shell due to feedback effects, so we consider a partial merger at $a=\unit{7.5}{\Mm}$ where the unmerged MLT profile has a dip as seen in Figure \ref{fig:dmlt_forced}.
A width of $w=\unit{0.25}{\Mm}$ is used for both scenarios, which is approximately the distance between the top of the \oxygen[] shell and the convective bump above it in Figure \ref{fig:dmlt_forced}.
For both scenarios we consider a weaker dip to $c=\unit{10^{14}}{\cm^2\second^{-1}}$ and a stronger dip to $c=\unit{10^{13}}{\cm^2\second^{-1}}$ to investigate the impact on nucleosynthesis.
The profiles for the GOSH-like and partial merger scenarios are shown in Figure \ref{fig:dgosh_partial}.

\subsection{Determining impact of varying nuclear reactions}\label{sec:nuclear_methods}

Many of the reactions involving the unstable \nt-deficient isotopes from \selenium[] to \polonium[] have not been measured experimentally and are determined by theoretical models, and the uncertainties of unmeasured reactions for unstable isotopes are much greater than for stable isotopes.
To determine the impact of nuclear physics for the \gprn{}, we vary our $(\gamma, \alpha)$, $(\gamma, p)$, and $(\gamma, n)$ photo-disintegration rates used by the NuGrid code \citep{pignatariNuGridStellarData2016} for all unstable \nt-deficient isotopes from \selenium[] to \polonium[] by a random factor uniformly selected between 0.1 to 10 by a Monte Carlo method for 1000 cases.
This applies the same approach used for $(\nt,\gamma)$ rates during the \iprn{} developed by \cite{denissenkovImpactReactionRate2018} and \cite{denissenkovImpactNgReaction2021}.
This was done for the MLT and downturn mixing scenarios in Figure \ref{fig:all_dconv} with an ingestion rate of $\unit{\natlog{4}{-3}}{\Msun\second^{-1}}$.
This approach also allows for the identification of reaction rates that are relevant for the production of an isotope using correlations.
The Pearson coefficient describes correlations between $X/X_{\mathrm{no~variation}}$ and the variation factors where $X$ is the final mass-averaged and decayed mass fraction for a Monte Carlo case and $X_{\mathrm{no~variation}}$ is the same for the default case where all variation factors are $1$.
All correlations with $|r_\mathrm{P}| \geq 0.15$ are reported in this study.
In addition to the Pearson coefficient, a logarithmic slope $\zeta$ is also reported to determine the importance of a reaction on the final mass fraction of an isotope, which is discussed along with caveats about correlation rates in Appendix \ref{sec:appendixNuclearRates}.

\section{Results} \label{sec:results}

An overview of our results in terms of overproduction factors and average spreads of overproduction factors are presented in Table \ref{tab:mixing_scenarios}.

\begin{deluxetable*}{lccccc}
\tablecaption{$\langle\OP\rangle$ for each mixing scenario and the average spread $(\OP_{\max}-\OP_{\min})$ for the Monte Carlo simulations for the \pnucn. All Monte Carlo simulations are calculated with an ingestion rate of \unit{\natlog{4}{-3}}{ \Msun\second^{-1}}.
\label{tab:mixing_scenarios}}
\tablehead{Scenario & No Ingestion & \unit{\natlog{4}{-5}}{ \Msun\second^{-1}} & \unit{\natlog{4}{-4}}{ \Msun\second^{-1}} & \unit{\natlog{4}{-3}}{ \Msun\second^{-1}} & Monte Carlo Spread}
\startdata
    MLT & $-0.11$ & $1.06$ & $1.92$ & $2.24$ & $0.56$ \\
    $1\times$ Downturn & $0.05$ & $1.12$ & $1.98$ & $2.58$  & $0.59$  \\
    $3\times$ Downturn & $-0.23$ & $1.28$ & $2.18$ & $2.83$ & $0.69$ \\
    $10\times$ Downturn & $-1.23$ & $1.18$ & $2.10$ & $2.89$ & $0.76$ \\
    $50\times$ Downturn & $-5.47$ & $0.88$ & $1.81$  & $2.72$ & $0.79$  \\
    GOSH-like  & \textendash & \textendash & \textendash & $2.06$ & \textendash \\
    Stronger GOSH-like  & \textendash & \textendash & \textendash & $1.78$ &  \textendash \\
    Partial Merger  & \textendash & \textendash & \textendash & $2.13$ & \textendash \\
    Stronger Partial Merger  & \textendash & \textendash & \textendash & $1.91$ & \textendash  \\
\enddata
\end{deluxetable*}

\subsection{Convective-reactive production of the \texorpdfstring{\pnucn}{p nuclei}} \label{sec:convreacflow}

\begin{figure} 
\includegraphics[width=\columnwidth]{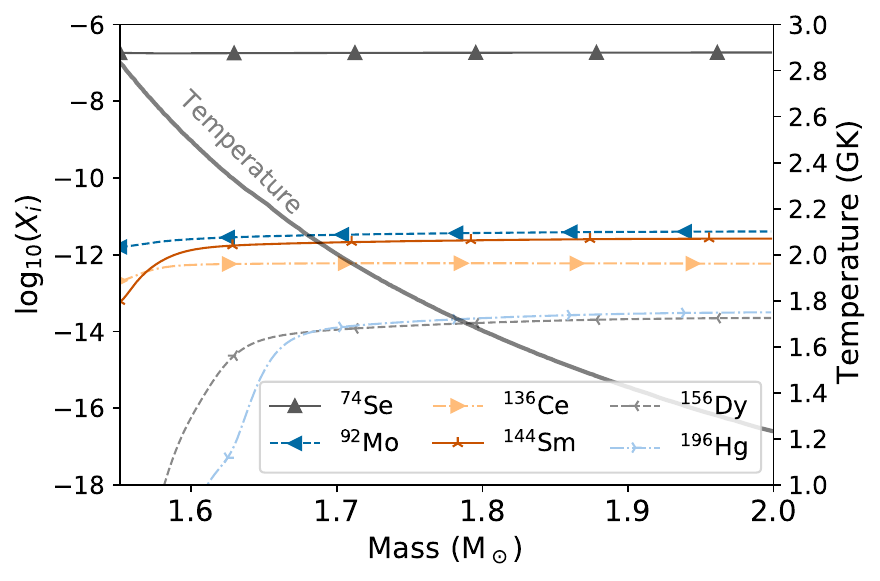}
\caption{Undecayed mass fractions at $t=110~\mathrm{sec}$ for the MLT mixing scenario without \carbon[]-shell ingestion.
\label{fig:temperatureprofile}}
\end{figure}

Convective-reactive nucleosynthesis is characterized by a region where the timescales for advection and nuclear reactions are similar.
In the \gprn{}, heavier species are produced at cooler temperatures of $\unit{2.3{-}2.5}{\Giga\Kelvin}$ and destroyed at higher temperatures, and lighter species are produced in temperatures up to $\unit{3.5}{\Giga\Kelvin}$ \citep{rauscherConstrainingAstrophysicalOrigin2013}.
Whether an isotope undergoes $(\gamma,\nt)$ and $(\nt,\gamma)$ reactions or contributes to the production of lighter \pnucn{} by $(\gamma,\pt)$ and $(\gamma,\alpha)$ reactions depends on the temperature at that position. 
The convective-reactive environment of the \oxygen[] shell allows for the production of both light and heavy \pnucn{} because the shell is not well-mixed, which allows the shell to produce most of the \pnucn{}, although temperatures in the \oxygen[] shell are too cool to sufficiently produce those with $A<110$.
Figure \ref{fig:temperatureprofile} shows how different mass ranges of \pnucn{} can be produced and peak at different positions in the \oxygen[] shell.

\begin{figure}
\includegraphics[width=\columnwidth]{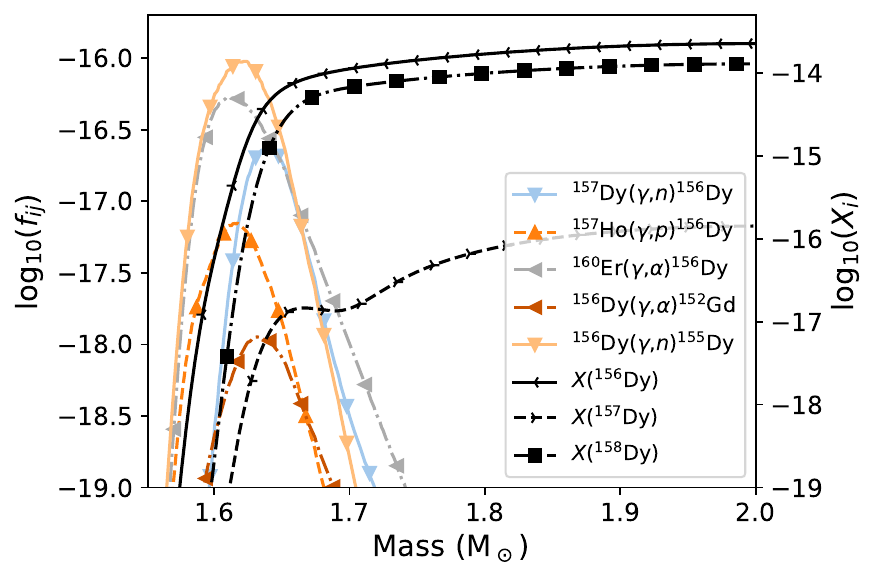}
\caption{Reaction fluxes $f_{ij}$ for \dysprosium[156] and undecayed mass fractions $X_i$ for \dysprosium[156-158] for the MLT scenario with no ingestion after $t=\unit{110}{\second}$. Reactions in the legend are written in the direction $i\rightarrow j$}.
\label{fig:dy156_fluxes_zeroingest}
\end{figure}

Locations of peak destruction can be identified by a sudden drop in mass fraction as seen for \dysprosium[156], \mercury[196], and to a lesser extent \samarium[144].
This is the location where $D_\alpha=1$, which is different for each isotope \citep{herwigCONVECTIVEREACTIVEPROTON2011}.
This is contrary to the normal assumption of a well-mixed convective environment where $D_\alpha \ll 1$ or radiative burning where little to no mixing occurs, and is a key reason why the \gprn{} in the \oxygen[] shell produces the whole range of \pnucn.
Figure \ref{fig:dy156_fluxes_zeroingest} shows the dominant reactions that \dysprosium[156] is produced and destroyed by in the simulation without \carbon[]-shell ingestion.
These reactions are shown as reaction fluxes $f_{ij}$ which are defined as the net flux of a reaction between species $i$ and $j$:
\begin{equation}
    f_{ij} = \frac{X_i X_j}{A_i A_j} \rho N_A \Biggl(\langle \sigma v \rangle_{ij} - \langle \sigma v \rangle_{ji}\Biggr)
\end{equation}
where $X$ is the mass fraction, $A$ is the atomic mass, $\rho$ is the density, $N_A$ is Avogadro's number, and $\langle \sigma v \rangle_{ij}$ is the reaction rate between species $i$ and $j$.

The reactions that \dysprosium[156] undergoes in the \oxygen[] shell depend on the location in the shell, however Figure \ref{fig:dy156_fluxes_zeroingest} shows that the dominant destruction channel $\dysprosium[156](\gamma,\nt)\dysprosium[155]$ net destroys \dysprosium[156], as Figure \ref{fig:ingest_MLT} will show.
It is also evident that the mass fraction is not well-mixed as the gradient of the mass fraction is steep at the location of peak destruction.

\begin{figure}
\includegraphics[width=\columnwidth]{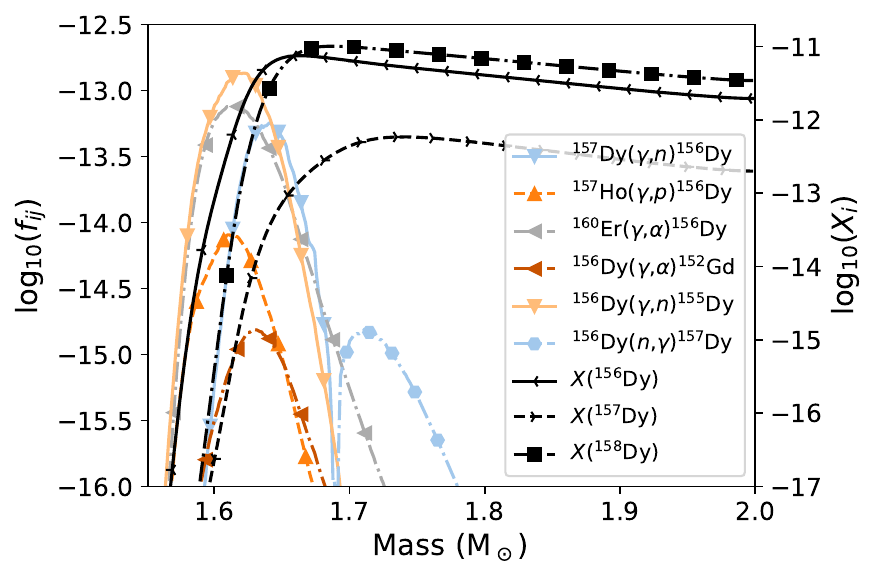}
\caption{Reaction fluxes $f_{ij}$ for \dysprosium[156] and undecayed mass fractions $X_i$ for \dysprosium[156-158] for the MLT scenario with an ingestion rate of $\unit{\natlog{4}{-3}}{\Msun\second^{-1}}$ after $t=\unit{110}{\second}$. Reactions in the legend are written in the direction $i\rightarrow j$.
\label{fig:dy156_fluxes_ingest}}
\end{figure}

Entraining \carbon[]-shell material is important for this process, as heavier species can be gradually depleted by $(\gamma,\alpha)$ and $(\gamma,\pt)$ reactions.
Figure \ref{fig:dy156_fluxes_ingest} shows the same as Figure \ref{fig:dy156_fluxes_zeroingest}, but with a \carbon[]-shell ingestion rate of $\unit{\natlog{4}{-3}}{\Msun\second^{-1}}$, the maximum considered in this study.
Since the initial amount of \dysprosium[156] in the ingested \carbon[] shell is negligible as shown in Figure \ref{fig:initial_comp}, the role of the merger is to provide species in the \carbon[] shell that are critical for the production of \dysprosium[156] such as the stable \dysprosium[] isotopes that undergo a $(\gamma, \nt)$ photodisintegration chain and the stable \erbium[] isotopes that do a sequence of $(\gamma, \nt)$ until the $\erbium[160](\gamma, \alpha)\dysprosium[156]$.

\begin{figure}
\includegraphics[width=\columnwidth]{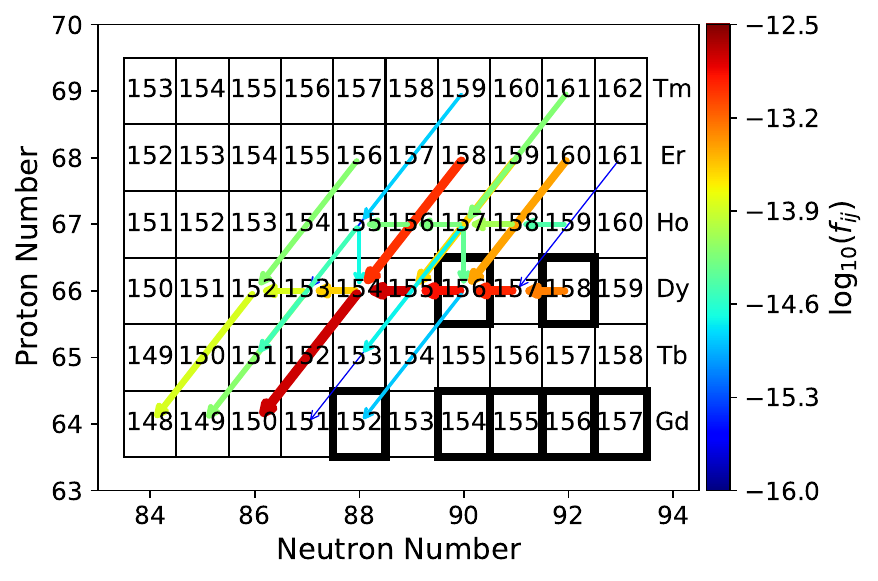}
\caption{Chart of reactions between isotopes at $m=\unit{1.64}{\Msun}$ [$T=\unit{2.28}{\Giga\Kelvin}$] for the same conditions as Figure \ref{fig:dy156_fluxes_ingest}. Both arrow colour and size indicate $\log_{10}(f_{ij})$, and arrows point in the direction of the reaction. The range of $\log_{10}(f_{ij})$ is the same as Figure \ref{fig:dy156_fluxes_ingest}.
\label{fig:dy156_network}}
\end{figure}

There are several differences between Figures \ref{fig:dy156_fluxes_zeroingest} and \ref{fig:dy156_fluxes_ingest}.
First, $f_{ij}$ and $X_i$ are larger by several orders of magnitude and \dysprosium[156] has a net production in the shell, as Figure \ref{fig:ingest_MLT} will show in Section \ref{sec:ingestionimpact}.
Second, the mass fraction of \dysprosium[156] has a tilt up where it is net produced and then drops off sharply at the location of peak destruction instead of the decline seen in Figure \ref{fig:dy156_fluxes_zeroingest}.

\dysprosium[156] is produced because the ingestion of \carbon[]-shell material allows for \dysprosium[158] to be replenished as it advects into the shell.
With a continual supply of \dysprosium[158], the chain $\dysprosium[158](\gamma,\nt)\dysprosium[157](\gamma,\nt)\dysprosium[156]$ can occur and significantly contribute to the production of \dysprosium[156] equal to $\erbium[160](\gamma,\alpha)\dysprosium[156]$.

Figure \ref{fig:dy156_fluxes_ingest} also shows another feature of this convective-reactive environment: \dysprosium[156] and \dysprosium[157] co-produce each other. 
\dysprosium[156] is advected from its location of peak production at $\sim$\unit{1.63}{\Msun} both deeper into the shell where it is fully destroyed, and toward the top where it undergoes $\dysprosium[156](\nt,\gamma)\dysprosium[157]$, which mildly contributes to the production of \dysprosium[157] and peaks at $\sim$\unit{1.73}{\Msun}.

Figures \ref{fig:dy156_fluxes_zeroingest} and \ref{fig:dy156_fluxes_ingest} demonstrate that in a convective-reactive environment isotopes are not well-mixed, and that the relevance of a reaction rate depends on their location in the shell.
Figure \ref{fig:dy156_fluxes_ingest} also show how isotopes can contribute to the production of each other at different locations in the shell.
Finally, comparing Figure \ref{fig:dy156_fluxes_ingest} to Figure \ref{fig:dy156_fluxes_zeroingest} shows the importance of ingesting \carbon[]-shell material for significant production of the \pnucn{} in the \oxygen[] shell.

\subsection{Impact from a downturn and boosting mixing speeds}\label{sec:convdownturnimpact}

\begin{figure*}
\includegraphics[width=\textwidth]{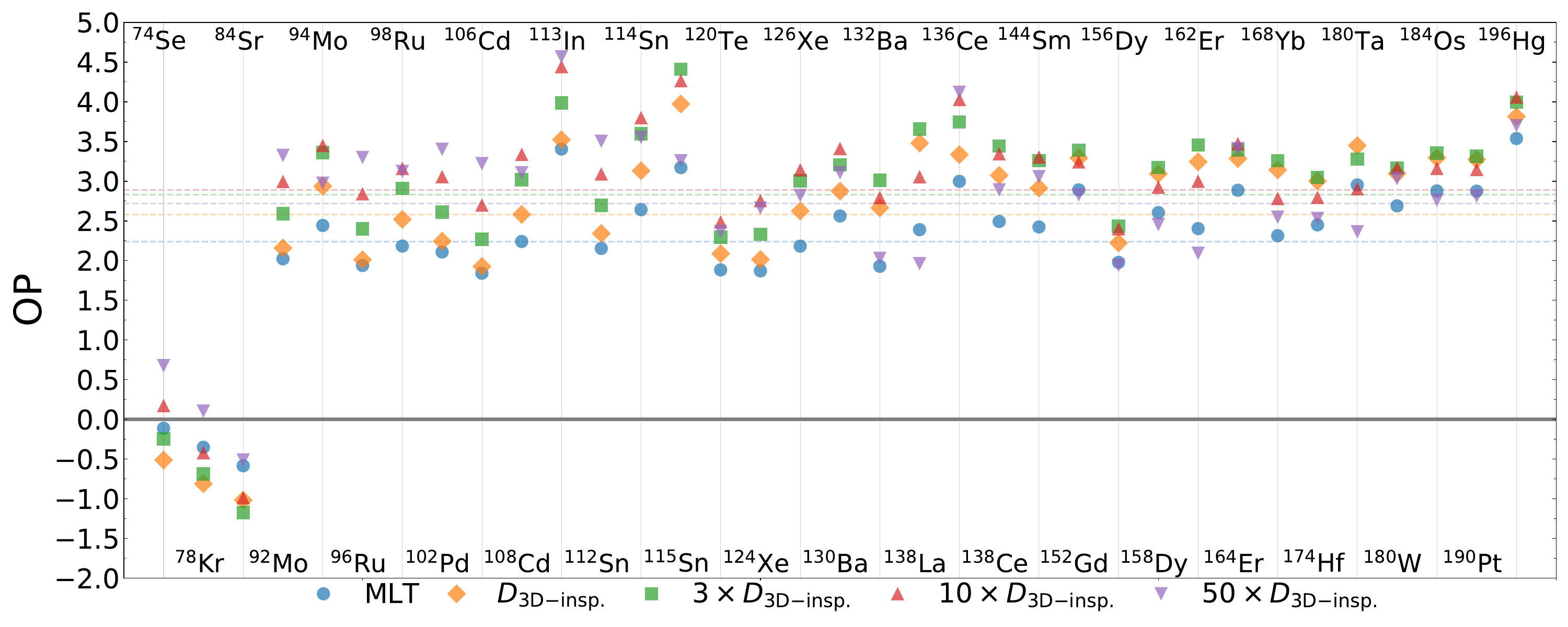}
\caption{The overproduction compared to initial of the \pnucn{} for the MLT and  3D-inspired mixing scenarios. The average spread in production $\OP_{\max} - \OP_{\min} = \unit{0.96}{\dex}$. $\OP=0\dex$ is the initial amount. Average \OP{} for each scenario are provided as dashed lines and corresponds to the values presented in Table \ref{tab:mixing_scenarios}.}
\label{fig:impactmixingcases}
\end{figure*}

Here we present the impact of a 3D-inspired gradual downturn at the lower boundary and boosting mixing speeds as explained in Section \ref{sec:mixing_methods} with the diffusion coefficients in Figure \ref{fig:all_dconv}. 
These cases are calculated with an ingestion rate of $\unit{\natlog{4}{-3}}{\Msun\second^{-1}}$ and the results are shown in Figure \ref{fig:impactmixingcases}.

The MLT simulation has an $\langle\OP\rangle$ of $\unit{2.24}{\dex}$, and the downturn scenarios have $\langle\OP\rangle$ of $\unit{2.58}{\dex}$, $\unit{2.83}{\dex}$, $\unit{2.89}{\dex}$, and $\unit{2.72}{\dex}$ for the $1\times$, $3\times$, $10\times$, and $50\times$3D-inspired mixing scenarios respectively.
The average spread in production for each isotope $\OP_{\max} - \OP_{\min} = \unit{0.96}{\dex}$, which shows that mixing speeds are important for the production of the \pnucn{}.
This \oxygen[] shell during the merger significantly produces all \pnucn{} except \selenium[74], \krypton[78], and \strontium[84].

The 3D-inspired $1\times$ scenario favours the production of the heavier \pnucn{} compared to the MLT scenario because $\tau_{\mathrm{mix}}$ decreases as the temperature increases.
Because of this, more reactions occur at the cooler temperatures where the $(\nt, \gamma)$ and $(\gamma, \nt)$ reactions are more favoured than the $(\gamma, \alpha)$ and $(\gamma, \pt)$ reactions.
All isotopes are comparably produced to MLT or more produced except \selenium[74], \krypton[78], and \strontium[84] who require the hottest temperatures for their production.

As mixing speeds increase, the production increases in a non-linear and non-monotonic way.
For the $50\times$ case, the average production of all \pnucn{} is lower than the $3\times$ and $10\times$ scenarios. 
This is because the mixing speeds are high enough that material is advected to the bottom of the \oxygen[] shell fast enough despite the downturn, and correspondingly the lighter \pnucn{} are generally more favoured including \selenium[74], \krypton[78], and \strontium[84].
Production for individual isotopes also can be non-linear and non-monotonic. 
For example, \tin[115], \barium[132], and \lanthanum[138] all increase for the $1\times$ and $3\times$ scenarios, but then their production is not as strong for the $10\times$ and $50\times$ scenarios.

Another result is that isotopic pairs of the same element are not affected the same way by the downturn compared to MLT, nor by the increase in mixing speed and we find that the ratio between these isotopes is dependent on the mixing scenario.
Since these isotopes are connected by $(\gamma,\nt)$ and $(\nt,\gamma)$ reactions, if the location of $D_\alpha = 1$ for a reaction changes because of a change to the mixing speed or the presence of the decreasing diffusion profile, then the production of these isotopes will change.
\molybdenum[94] is produced more for all mixing scenarios except for the $50\times$ scenario, where \molybdenum[92] has a larger \OP, and likewise \tin[115] exhibits this behaviour when compared to \tin[112] and \tin[114].
We find that it is possible that the ratio can tend to unity as mixing speeds increase as seen for \ruthenium[96,98], \cadmium[106,108], and \tin[112, 114].
For \cerium[136,138] the lighter isotope is always favoured as mixing speed increases and for \dysprosium[156,158] the opposite is true.
This demonstrates the importance of the decreasing diffusion profile and increasing mixing speed  is for the production of the \pnucn{}.

\subsection{Impact from varying the ingestion rate}\label{sec:ingestionimpact}

\begin{figure*} 
\includegraphics[width=\textwidth]{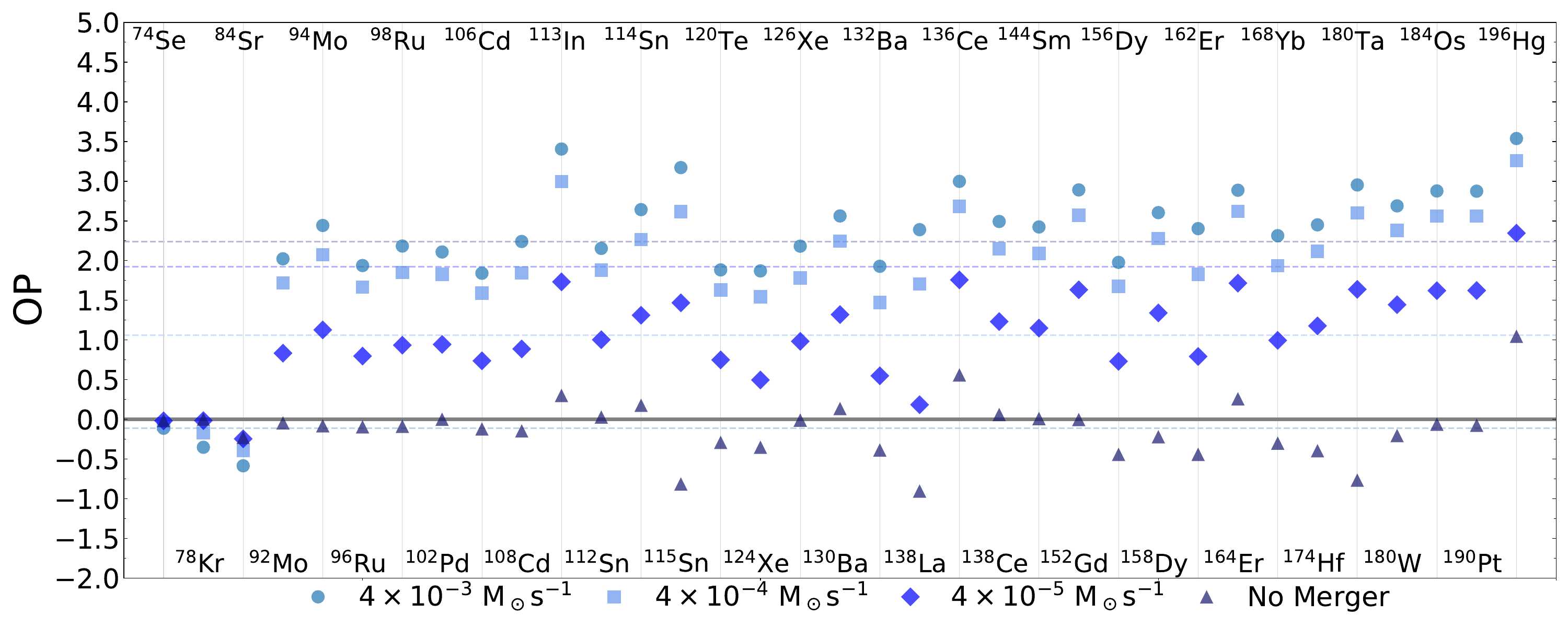}
\caption{The overproduction compared to initial of the \pnucn{} for the MLT mixing scenario for no ingestion, $\unit{\natlog{4}{-5}}{\Msun\second^{-1}}$, $\unit{\natlog{4}{-4}}{\Msun\second^{-1}}$, and $\unit{\natlog{4}{-3}}{\Msun\second^{-1}}$. The average spread in production $\OP_{\max} - \OP_{\min} = \unit{1.22}{\dex}$ excluding the no ingestion case. $\OP=0$ is the initial amount. Average \OP{} for each scenario are provided as dashed lines and corresponds to the values presented in Table \ref{tab:mixing_scenarios}.}
\label{fig:ingest_MLT}
\end{figure*}

Here we present the impact of entraining \carbon[]-shell material with $\unit{\natlog{4}{-5}}{\Msun\second^{-1}}$, $\unit{\natlog{4}{-4}}{\Msun\second^{-1}}$, $\unit{\natlog{4}{-3}}{\Msun\second^{-1}}$ as well as no entrainment as explained in Section \ref{sec:mixing_methods}.
Figure \ref{fig:ingest_MLT} shows the results for the MLT mixing scenario and Figures \ref{fig:ingest_3D}{--}\ref{fig:ingest_50x3D} in Appendix \ref{sec:appendixResultsIngestion} show the results for all downturn scenarios.

The results show that the production of the \pnucn{} is monotonically increased by the ingestion of \carbon[]-shell material for all isotopes except \selenium[74], \krypton[78], and \strontium[84] who exhibit the opposite behaviour for all mixing scenarios.
The only exception is that \selenium[74] production increases for the $10\times$ and $50\times$3D-inspired scenarios with ingestion rate.
This is because with higher ingestion rates, the more stable isotopes enter the shell as demonstrated in Section \ref{sec:convreacflow}.
The lightest three have the opposite behaviour because without ingestion they are destroyed less by $(\nt,\gamma)$ reactions.
The difference in \OP{} between two ingestion rates is largely uniformly for \molybdenum[92]{--}\mercury[196].
The average spread $\OP_{\max} - \OP_{\min}$ between the different ingestion rates for the MLT and 3D-inspired scenarios is $1.22$, $1.58$, $1.64$, $1.78$, and $1.84 \dex$.
This shows that the non-linear and non-monotonic behaviour in Section \ref{sec:convdownturnimpact} is because of the changing location of peak burning in the convective-reactive environment and not just more material being present.

Without ingestion of \carbon[]-shell material, the MLT, $1\times$, and $3\times$ scenarios see little to no production, but for the $10\times$ and $50\times$ scenarios there is a significant underproduction of the heavier \pnucn{}. 
This is because in the fastest cases the species are advected to the bottom of the shell where the temperatures are purely destructive and there is no replenishment from the \carbon[] shell.
This shows that entrainment of \carbon[]-shell material is necessary for significant contributions from pre-explosive \gprn.

Another important feature of ingestion is that \nt-heavier isotopes are more favoured as the rate increases for all mixing speeds in our study. 
For isotopic pairs, production of the heavier isotope with respect to the lighter increases with ingestion rate due to the increase in released neutrons from burning \carbon[] and \oxygen[].
Figure \ref{fig:ingest_MLT} shows, for example, how \cadmium[106,108] are similar for the no merger case but as ingestion rate increases \cadmium[108] becomes increasingly more abundant compared to \cadmium[106].
Other isotopic pairs exhibit an increase of the heavier isotope when ingestion is present but \OP{} between the two does not change with ingestion rate, such as \dysprosium[156,158].

The sign of the ratio of isotopic pairs is largely preserved across the ingestion rates for the MLT, $1\times$, $3\times$, and $50\times$ scenarios, but the magnitude of the ratio does change.
The few exceptions are \erbium[162,164] in the $1\times$ and $3\times$ scenarios who become roughly equal for the fastest ingestion rate, \tin[115] which becomes more abundant than \tin[112,114] in the $3\times$ scenario.
In the $10\times$ scenario production of isotopic pairs is largely equal for the slower ingestion rates, but at the fastest ingestion rate the ratio can become unequal.
This shows that the the ingestion rate matters for the magnitude of the ratio between isotopic pairs, but that the sign of the ratio depends on which diffusion profile is used.

\subsection{Impact from dips from convective quenching}\label{sec:goshimpact}

\begin{figure*} 
\includegraphics[width=\textwidth]{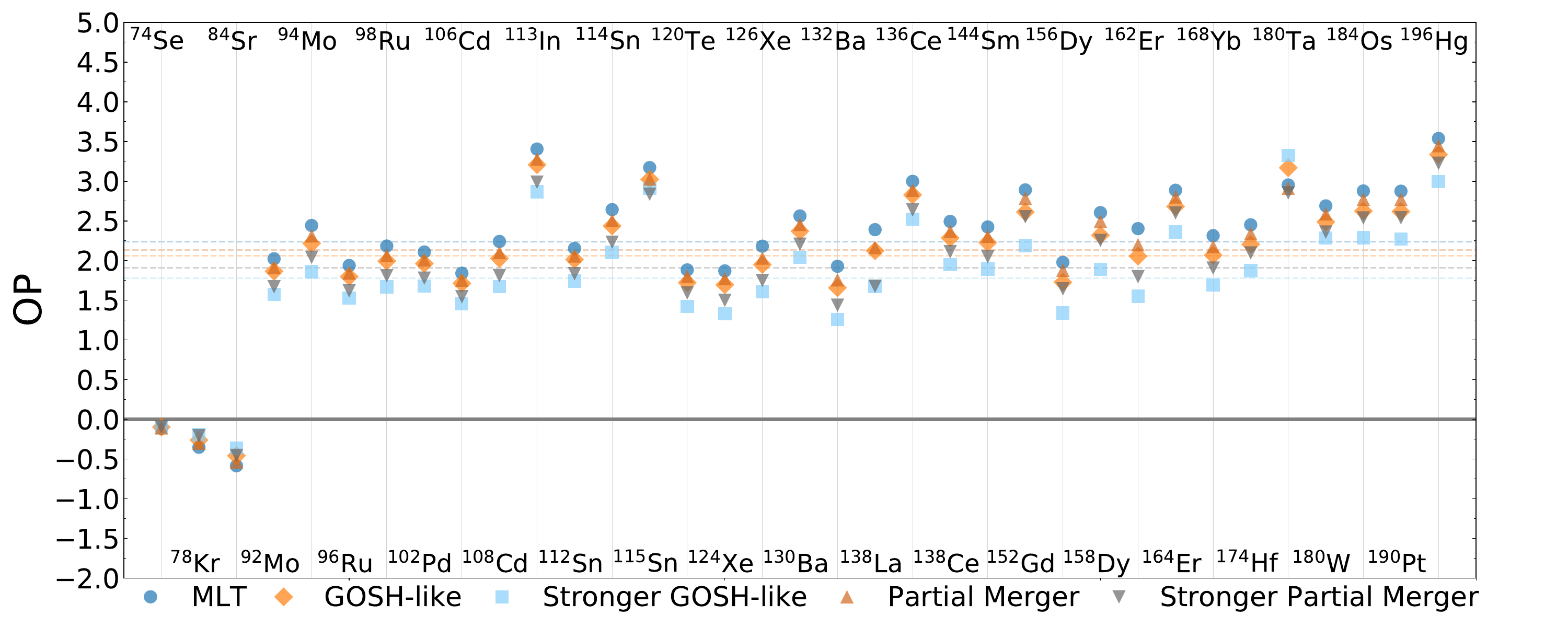}
\caption{The overproduction compared to initial of the \pnucn{} for the MLT, GOSH-like, and partial merger scenarios with ingestion rate $\unit{\natlog{4}{-3}}{\Msun\second^{-1}}$. The average spread in production $\OP_{\max} - \OP_{\min} = \unit{0.51}{\dex}$. $\OP=0$ is the initial amount. Average \OP{} for each scenario are provided as dashed lines and corresponds to the values presented in Table \ref{tab:mixing_scenarios}.}
\label{fig:impactmixingcases_GOSH_partial}
\end{figure*}

Here we present the impact of dips from convective quenching using the profiles shown in Figure \ref{fig:dgosh_partial} from Section \ref{sec:mixing_methods} which represent GOSH-like feedback and a partial merger.
The results are shown in Figure \ref{fig:impactmixingcases_GOSH_partial}.

The MLT simulation has $\langle\OP\rangle$ of $\unit{2.24}{\dex}$, the GOSH-like scenarios have $\OP=\unit{2.06}{\dex}$ and $\unit{1.78}{\dex}$, and the partial merger scenarios have $\OP=\unit{2.13}{\dex}$ and $\unit{1.91}{\dex}$.
We find that the GOSH-like scenarios suppresses the production more than the partial merger scenarios of equal dip depths, and that the deeper dips are suppress production more than the shallow with an average spread $\OP_{\max}-\OP_{\min}=0.51$.

The dip functions as a barrier for the convective-reactive flow, and can section off parts of the \oxygen[] shell.
The partial merger scenario both limits the ingested \carbon[]-shell material and slows down the material from deeper in the \oxygen[] shell from advecting up to the top of the shell where very few reactions occur.
The GOSH-like dip slows down the material from reaching their preferential temperatures, and also prevents material at the deepest part of the \oxygen[] shell from mixing up to the top of the shell which keeps it at hotter temperatures where it can be destroyed.
This is also why the deeper dips suppress production more, as the velocities are slower by an additional factor of $10$ at the deepest point of the dip.

The suppression of production is largely uniform for all isotopes except for \selenium[74], \krypton[78], \strontium[84], and \tantalum[180].
Additionally, it appears that isotopes $A\geq138$ are more strongly affected by the stronger GOSH-like dip. 
The isotopes \selenium[74], \krypton[78], and \strontium[84] have a minor increase from these dips because the location of their production is at the bottom of the \oxygen[] shell where the dips are not present.
The boost in production of \tantalum[180] is because the peak production and destruction locations happen to be centered at the exact same location as the GOSH dip, which lowers its destruction and slightly boosts its production.
Isotopic ratios are not significantly affected by the dips, although the magnitude of the ratios can increase slightly.
This demonstrates the importance of both the location and magnitude of the convective dips in the \oxygen[] shell for convective-reactive \gprn.

\subsection{Nuclear physics impact and mixing dependencies}\label{sec:nuclearimpact}

\begin{figure*}
\includegraphics[width=\textwidth]{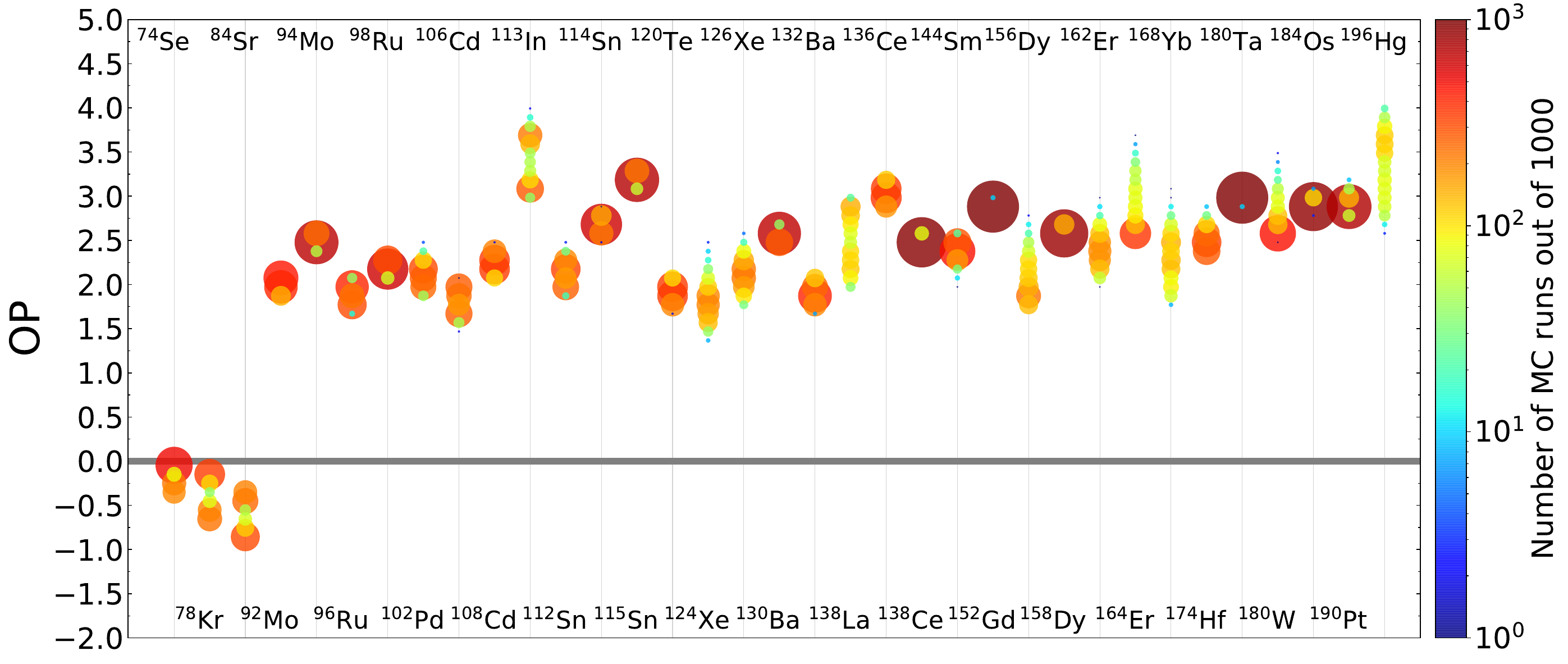}
\caption{Histogram showing the spread due to varying $(\gamma,\pt)$, $(\gamma,\nt)$, $(\gamma,\alpha)$ and corresponding capture rates for unstable \nt-deficient isotopes from \selenium[]{--}\polonium[] for the MLT mixing scenario. Colour and size both correspond to the logarthimic binning of Monte Carlo runs. The average spread $\OP_{\max}-\OP_{\min}=\unit{0.56}{\dex}$. $\OP=0$ is the initial amount.
\label{fig:nuclearimpact_MLT}}
\end{figure*}

Here we present the impact of varying our adopted nuclear physics rates for the MLT and 3D-inspired mixing scenarios for an ingestion rate of $\unit{\natlog{4}{-3}}{\Msun\second^{-1}}$.
The result for the MLT mixing scenario is shown in Figure \ref{fig:nuclearimpact_MLT} and Table \ref{tab:mlt_corr}, and the 3D-inspired mixing scenarios are shown in Figures \ref{fig:nuclearimpact_3D}{--}\ref{fig:nuclearimpact_50x3D} and Tables \ref{tab:mlt_corr}{--}\ref{tab:50x3d_corr} which can be found in Appendix \ref{sec:appendixResultsNuclear}.

The MLT mixing scenario has an average spread in production of $\OP_{\max}-\OP_{\min}=\unit{0.56}{\dex}$, and the 3D-inspired scenarios have an average spread of $\unit{0.59}{\dex}$, $\unit{0.69}{\dex}$, $\unit{0.76}{\dex}$, and $\unit{0.79}{\dex}$ for the $1\times$, $3\times$, $10\times$, and $50\times$ scenarios respectively.
We find that the spread in production increases with mixing speed because the material is able to reach hotter temperatures and the number of possible nucleosynthetic pathways changes.

In the MLT mixing scenario about a third of the isotopes are not affected in any significant way, but in the 3D-inspired scenarios only 5 are not affected although the specific isotopes vary.
The isotopes that appear the least affected across mixing scenario are \cerium[138], \gadolinium[152], \dysprosium[158], and \tantalum[180].

The spread of an individual isotope is dependent on the mixing scenario.
Species like \cadmium[106], \dysprosium[156], and \tungsten[180] have a different spread in production for each mixing scenario.
While the change can be monotonic for species like \tungsten[180], \osmium[184], and \platnium[190], for \cadmium[106] and \tin[112], and \barium[130] they decrease for the $50\times$3D-inspired scenario.

This mixing scenario dependence for the spread is also seen for the distribution of \OP{} as no isotope is double peaked across all mixing scenarios.
As examples, \selenium[74], \indium[113], and \lanthanum[138] in some mixing scenarios clearly are double peaked, indicating that there are distinctive branches in the nucleosynthetic pathways, but do not have it in others.
Additionally, the magnitude of which peak is favoured is also dependent on the mixing scenario as seen for \krypton[78] and \strontium[84].

Whether a particular reaction rate is correlated an isotope's final mass fraction along with the strength of the correlation is dependent on the mixing scenario.
Table \ref{tab:unique_rates} lists the rates unique to single mixing scenario.

\begin{deluxetable}{ll}
\tablecaption{Reactions correlated with the production/destruction of an isotope unique to an individual mixing scenario.
\label{tab:unique_rates}}
\tablehead{
\multicolumn{1}{l}{\textbf{Isotope}} & \multicolumn{1}{l}{\textbf{Unique Correlated Reaction Rates}}}
\startdata
\multicolumn{2}{l}{\textbf{MLT Mixing Case}} \\ 
\hline
$^{138}\mathrm{Ce}$ & $^{139}\mathrm{Pr}(\gamma,\pt)$ \\
$^{168}\mathrm{Yb}$ & $^{176}\mathrm{W}(\gamma,\alpha)$ \\
$^{174}\mathrm{Hf}$ & $^{178}\mathrm{W}(\gamma,\nt)$ \\
$^{184}\mathrm{Os}$ & $^{186}\mathrm{Pt}(\gamma,\nt)$~~$^{188}\mathrm{Pt}(\gamma,\nt)$ \\
\hline
\multicolumn{2}{l}{\textbf{3D-inspired Mixing Scenario}} \\ 
\hline
$^{113}\mathrm{In}$ & $^{114}\mathrm{In}(\gamma,\nt)$ \\
$^{152}\mathrm{Gd}$ & $^{150}\mathrm{Gd}(\gamma,\alpha)$~~$^{196}\mathrm{Pb}(\gamma,\nt)$ \\
$^{180}\mathrm{Ta}$ & $^{179}\mathrm{Ta}(\gamma,\alpha)$ \\
\hline
\multicolumn{2}{l}{$\mathbf{ 3\times}$\textbf{3D-inspired Mixing Scenario}} \\ 
\hline
$^{180}\mathrm{W}$ & $^{181}\mathrm{Os}(\gamma,\nt)$ \\
\hline
\multicolumn{2}{l}{$\mathbf{ 10\times}$\textbf{3D-inspired Mixing Scenario}} \\ 
\hline
$^{84}\mathrm{Sr}$ & $^{84}\mathrm{Rb}(\gamma,\nt)$ \\
$^{120}\mathrm{Te}$ & $^{119}\mathrm{Te}(\gamma,\nt)$ \\
$^{126}\mathrm{Xe}$ & $^{122}\mathrm{Xe}(\gamma,\nt)$ \\
$^{130}\mathrm{Ba}$ & $^{126}\mathrm{Ba}(\gamma,\pt)$~~$^{128}\mathrm{Ba}(\gamma,\alpha)$~~$^{128}\mathrm{Ba}(\gamma,\pt)$ \\
$^{132}\mathrm{Ba}$ & $^{128}\mathrm{Ba}(\gamma,\alpha)$ \\
$^{168}\mathrm{Yb}$ & $^{169}\mathrm{Hf}(\gamma,\nt)$ \\
$^{174}\mathrm{Hf}$ & $^{176}\mathrm{W}(\gamma,\alpha)$ \\
$^{184}\mathrm{Os}$ & $^{185}\mathrm{Pt}(\gamma,\alpha)$ \\
\hline
\multicolumn{2}{l}{$\mathbf{ 50\times}$\textbf{3D-inspired Mixing Scenario} } \\ 
\hline
$^{92}\mathrm{Mo}$ & $^{100}\mathrm{Pd}(\gamma,\alpha)$~~$^{100}\mathrm{Pd}(\gamma,\pt)$~~$^{110}\mathrm{Sn}(\gamma,\nt)$~~$^{110}\mathrm{Sn}(\gamma,\pt)$ \\
$^{96}\mathrm{Ru}$ & $^{97}\mathrm{Ru}(\gamma,\alpha)$~~$^{110}\mathrm{Sn}(\gamma,\alpha)$~~$^{110}\mathrm{Sn}(\gamma,\nt)$~~$^{110}\mathrm{Sn}(\gamma,\pt)$ \\
$^{102}\mathrm{Pd}$ & $^{104}\mathrm{Cd}(\gamma,\alpha)$~~$^{104}\mathrm{Cd}(\gamma,\pt)$ \\
$^{106}\mathrm{Cd}$ & $^{104}\mathrm{Cd}(\gamma,\pt)$~~$^{110}\mathrm{Sn}(\gamma,\pt)$ \\
$^{108}\mathrm{Cd}$ & $^{110}\mathrm{Sn}(\gamma,\alpha)$ \\
$^{112}\mathrm{Sn}$ & $^{110}\mathrm{Sn}(\gamma,\pt)$ \\
$^{115}\mathrm{Sn}$ & $^{122}\mathrm{Xe}(\gamma,\nt)$ \\
$^{120}\mathrm{Te}$ & $^{120}\mathrm{Xe}(\gamma,\alpha)$ \\
$^{126}\mathrm{Xe}$ & $^{127}\mathrm{Ba}(\gamma,\nt)$ \\
$^{130}\mathrm{Ba}$ & $^{132}\mathrm{Ce}(\gamma,\alpha)$~~$^{132}\mathrm{Ce}(\gamma,\nt)$~~$^{132}\mathrm{Ce}(\gamma,\pt)$\\
&$^{134}\mathrm{Ce}(\gamma,\alpha)$~~$^{134}\mathrm{Ce}(\gamma,\nt)$ \\
$^{132}\mathrm{Ba}$ & $^{132}\mathrm{Ce}(\gamma,\alpha)$~~$^{132}\mathrm{Ce}(\gamma,\nt)$~~$^{132}\mathrm{Ce}(\gamma,\pt)$\\
& $^{133}\mathrm{Ce}(\gamma,\nt)$~~$^{134}\mathrm{Ce}(\gamma,\alpha)$ \\
$^{138}\mathrm{Ce}$ & $^{139}\mathrm{Nd}(\gamma,\nt)$ \\
$^{156}\mathrm{Dy}$ & $^{156}\mathrm{Er}(\gamma,\alpha)$~~$^{158}\mathrm{Er}(\gamma,\alpha)$~~$^{158}\mathrm{Er}(\gamma,\nt)$ \\
$^{162}\mathrm{Er}$ & $^{168}\mathrm{Hf}(\gamma,\nt)$~~$^{162}\mathrm{Yb}(\gamma,\alpha)$~~$^{164}\mathrm{Yb}(\gamma,\alpha)$ \\
$^{184}\mathrm{Os}$ & $^{184}\mathrm{Pt}(\gamma,\nt)$ \\
\enddata
\end{deluxetable}

It is clearly important to consider the mixing conditions if an experiment is to be proposed to measure a reaction rate.
As Table \ref{tab:unique_rates} shows for the $50\times$3D-inspired scenario, there are many reactions even for a single isotope that can be correlated uniquely in that mixing scenario.
This clearly shows that a decreasing radial velocity profile and the exact magnitude of the mixing speeds are crucial for understanding the nuclear reactions in this convective-reactive environment.

There are also correlated reactions that all mixing scenarios share as shown in Table \ref{tab:shared_rates}.
Additionally, all downturn cases share correlations not found in the MLT scenario: $X(\tin[115])$ with $\tin[110](\gamma,\alpha)$ and $X(\cerium[138])$ with $\neodymium[138](\gamma,\pt)$. 
However, the shared correlations are not of equal strength across all mixing scenarios. 
As an example the final mass fraction of \dysprosium[156] is correlated with \erbium[160]$(\gamma,\alpha)$, but in the $50\times$3D-inspired the correlation is much weaker.
This underscores the possible difficulties in using 1D astrophysical sites to identify important reactions for nuclear physics experiments.

\begin{deluxetable}{ll}
\tablecaption{Reactions correlated with the production/destruction of an isotope shared across all mixing scenarios.
\label{tab:shared_rates}}
\tablehead{
\multicolumn{1}{l}{\textbf{Isotope}} & \multicolumn{1}{l}{\textbf{Shared Correlated Reaction Rates}}}
\startdata
$^{74}\mathrm{Se}$ & $^{75}\mathrm{Se}(\gamma,\nt)$ \\
$^{78}\mathrm{Kr}$ & $^{79}\mathrm{Kr}(\gamma,\nt)$ \\
$^{84}\mathrm{Sr}$ & $^{85}\mathrm{Sr}(\gamma,\nt)$ \\
$^{92}\mathrm{Mo}$ & $^{93}\mathrm{Mo}(\gamma,\nt)$ \\
$^{94}\mathrm{Mo}$ & $^{93}\mathrm{Mo}(\gamma,\nt)$ \\
$^{96}\mathrm{Ru}$ & $^{97}\mathrm{Ru}(\gamma,\nt)$ \\
$^{98}\mathrm{Ru}$ & $^{100}\mathrm{Pd}(\gamma,\alpha)$~~$^{100}\mathrm{Pd}(\gamma,\pt)$ \\
$^{102}\mathrm{Pd}$ & $^{100}\mathrm{Pd}(\gamma,\alpha)$~~$^{100}\mathrm{Pd}(\gamma,\pt)$~~$^{103}\mathrm{Pd}(\gamma,\nt)$ \\
$^{106}\mathrm{Cd}$ & $^{107}\mathrm{Cd}(\gamma,\nt)$~~$^{110}\mathrm{Sn}(\gamma,\alpha)$ \\
$^{108}\mathrm{Cd}$ & $^{107}\mathrm{Cd}(\gamma,\nt)$ \\
$^{113}\mathrm{In}$ & $^{113}\mathrm{Sn}(\gamma,\nt)$ \\
$^{112}\mathrm{Sn}$ & $^{113}\mathrm{Sn}(\gamma,\nt)$ \\
$^{114}\mathrm{Sn}$ & $^{110}\mathrm{Sn}(\gamma,\alpha)$~~$^{113}\mathrm{Sn}(\gamma,\nt)$~~$^{122}\mathrm{Xe}(\gamma,\alpha)$ \\
$^{115}\mathrm{Sn}$ & $^{113}\mathrm{Sn}(\gamma,\nt)$ \\
$^{120}\mathrm{Te}$ & $^{122}\mathrm{Xe}(\gamma,\alpha)$~~$^{122}\mathrm{Xe}(\gamma,\pt)$ \\
$^{124}\mathrm{Xe}$ & $^{122}\mathrm{Xe}(\gamma,\alpha)$~~$^{122}\mathrm{Xe}(\gamma,\pt)$ \\
$^{138}\mathrm{La}$ & $^{137}\mathrm{La}(\gamma,\nt)$ \\
$^{136}\mathrm{Ce}$ & $^{138}\mathrm{Nd}(\gamma,\nt)$~~$^{138}\mathrm{Nd}(\gamma,\pt)$~~$^{140}\mathrm{Nd}(\gamma,\alpha)$ \\
$^{144}\mathrm{Sm}$ & $^{196}\mathrm{Pb}(\gamma,\nt)$~~$^{142}\mathrm{Sm}(\gamma,\nt)$~~$^{142}\mathrm{Sm}(\gamma,\pt)$ \\ 
 & $^{143}\mathrm{Sm}(\gamma,\nt)$ \\
$^{152}\mathrm{Gd}$ & $^{152}\mathrm{Dy}(\gamma,\alpha)$ \\
$^{156}\mathrm{Dy}$ & $^{160}\mathrm{Er}(\gamma,\alpha)$ \\
$^{164}\mathrm{Er}$ & $^{164}\mathrm{Yb}(\gamma,\alpha)$~~$^{164}\mathrm{Yb}(\gamma,\nt)$ \\
$^{174}\mathrm{Hf}$ & $^{174}\mathrm{W}(\gamma,\alpha)$ \\
$^{180}\mathrm{Ta}$ & $^{179}\mathrm{Ta}(\gamma,\nt)$ \\
$^{180}\mathrm{W}$ & $^{180}\mathrm{Os}(\gamma,\alpha)$~~$^{180}\mathrm{Os}(\gamma,\nt)$~~$^{196}\mathrm{Pb}(\gamma,\nt)$ \\
$^{184}\mathrm{Os}$ & $^{196}\mathrm{Pb}(\gamma,\nt)$~~$^{184}\mathrm{Pt}(\gamma,\alpha)$ \\
$^{190}\mathrm{Pt}$ & $^{190}\mathrm{Hg}(\gamma,\alpha)$~~$^{190}\mathrm{Hg}(\gamma,\nt)$~~$^{196}\mathrm{Pb}(\gamma,\nt)$ \\
$^{196}\mathrm{Hg}$ & $^{196}\mathrm{Pb}(\gamma,\nt)$~~$^{202}\mathrm{Pb}(\gamma,\nt)$ \\
\enddata
\end{deluxetable}

The spread in production for varying our adopted nuclear physics rates for each of the mixing scenarios is comparable to the spread seen from varying the mixing conditions.
The results in Section \ref{sec:goshimpact} have a spread of $\unit{0.51}{\dex}$, Section \ref{sec:convdownturnimpact} $\unit{0.96}{\dex}$, and the maximum spread in Section \ref{sec:ingestionimpact} is $\unit{1.84}{\dex}$.
The average spread in production from varying the nuclear physics rates ranges from $0.56${--}$\unit{0.79}{\dex}$.

\cite{rauscherUncertaintiesProductionNuclei2016} studied the impact of nuclear uncertainties for the explosive \gpr{} for a $\unit{15}{\Msun}$ and two $\unit{25}{\Msun}$ models with solar metallicity using a Monte Carlo method.
The spread in their 90\% probability interval for the $\unit{15}{\Msun}$ model was $\unit{0.61}{\dex}$, $\unit{0.63}{\dex}$ for the $\unit{25}{\Msun}$ KEPLER model, and $\unit{0.99}{\dex}$ for the model from \cite{hashimotoExplosiveNucleosynthesisSupernova1989}.
This is comparable to the maximum spread found in this work, although we are considering pre-explosive \gprn{}.
\cite{rauscherUncertaintiesProductionNuclei2016} also provide tables of correlated rates, but only a handful of these rates appear in our Tables \ref{tab:mlt_corr}{--}\ref{tab:50x3d_corr} including those they find with $r_\mathrm{P} \geq |0.65|$.
This is because the convective-reactive environment we consider allows for different reaction pathways to be favoured depending on the mixing conditions.

\section{Discussions and Conclusion} \label{sec:discussion_conclusion}

In this paper, we have shown that understanding the mixing details in the \oxygen[]-burning shell during an \oxygen[]-\carbon[] shell merger is crucial for accurately modelling the nucleosynthesis of the \pnucn.
This work raises the importance of 3D hydrodynamic simulations for understanding the nucleosynthesis in \oxygen[]-\carbon[] shell mergers \citep{bazanConvectionNucleosynthesisCore1994, yadavLargescaleMixingViolent2020a, rizzutiShellMergersLate2024a}.
We have demonstrated the convective-reactive nature of the nucleosynthesis in the \oxygen[] shell, where the timescales for advection and reaction are comparable and our results show that:
\begin{itemize}
\item A gradual downturn motivated by 3D simulations increases production of the \pnucn{}, but increasing mixing speeds impacts the production in a non-linear and non-monotonic way.
\item The ratio of isotopic pairs is sensitive to the mixing scenario.
\item Increasing the entrainment rate of \carbon[]-shell material increases the production of the \pnucn{}.
\item Without entrainment, the \pnucn{} are negligibly produced or net destroyed.
\item A dip in the convective profile can suppress the production of the \pnucn{}.
\item The location and magnitude of the convective dip impacts the nucleosynthesis of the \pnucn{}.
\item Varying the adopted reactions rates have a comparable spread in production to changing mixing conditions.
\item The spread due to varying the reaction rates is dependent on the mixing scenario.
\item Whether a reaction rate is correlated with an isotope is dependent on the mixing scenario, and some are unique to a single mixing scenario.
\end{itemize}

\begin{figure*} 
\includegraphics[width=\textwidth]{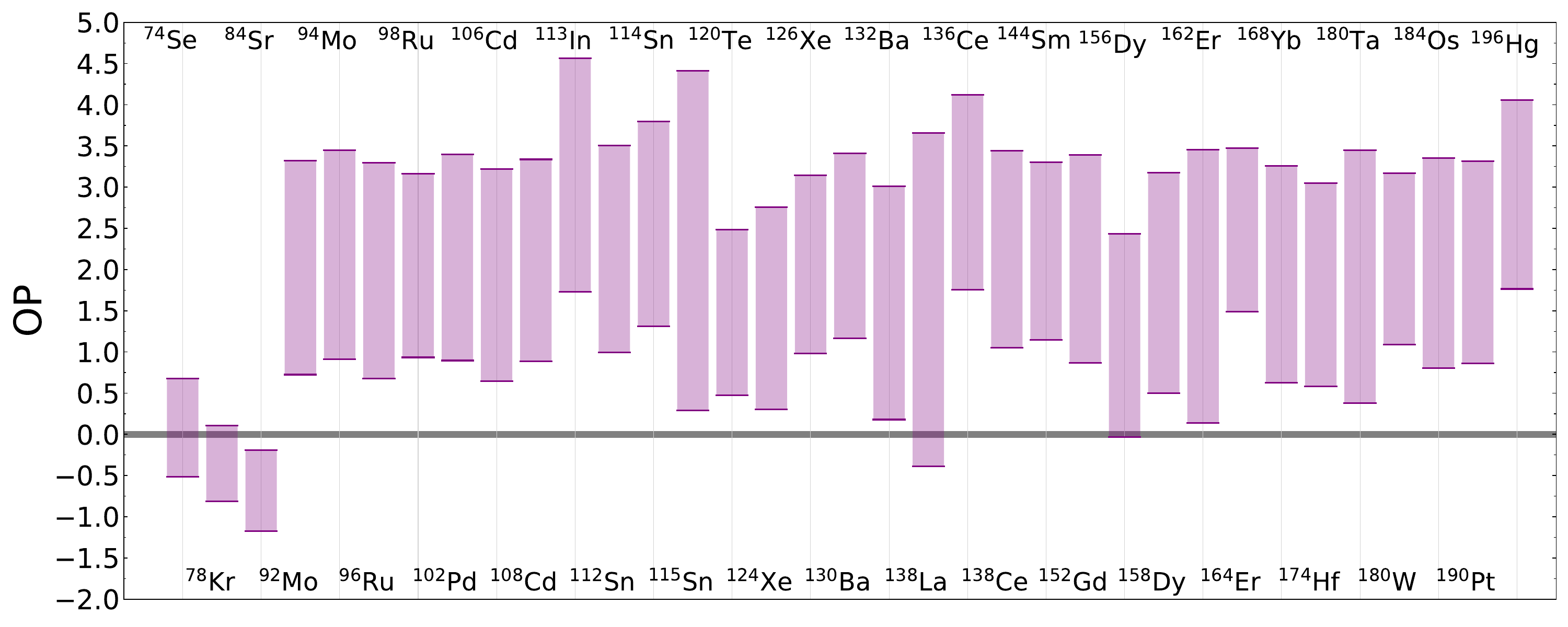}
\caption{Bars representing the maximum and minimum \OP{} across all mixing scenarios, excluding those without \carbon[]-shell ingestion.
The average spread $\OP_{\max} - \OP_{\min} = \unit{2.45}{\dex}$. $\OP=0$ is the initial amount.
\label{fig:impact_mixing_cases_bars_all}}
\end{figure*}

Figure \ref{fig:impact_mixing_cases_bars_all} shows the maximum spread for the \pnucn{} across all mixing scenarios considered in this paper, excluding the case of no merger, have an average spread of \unit{2.45}{\dex}.
This shows the significant impact of the macrophysical uncertainties in 1D stellar models on nucleosynthesis, and this highlights the importance of understanding hydrodynamic models better as the shell merger dominates production of the \pnucn{} \citep{robertiGprocessNucleosynthesisCorecollapse2023,robertiGprocessNucleosynthesisCorecollapse2024b}.

Although we find that the mixing conditions significantly impact the results of Section \ref{sec:nuclearimpact}, not all scenarios are equally likely to represent the conditions in a merger.
The MLT mixing scenario, $1\times$, and $50\times$ case may not be representative of the conditions for realistic \oxygen[]-\carbon[] shell mergers.
3D hydrodynamic simulations show the \oxygen[] shell has a downturn to radial convective velocities and mixing speeds roughly $3-10$ times larger than what MLT predicts \citep{jonesIdealizedHydrodynamicSimulations2017,andrassy3DHydrodynamicSimulations2020, rizzutiShellMergersLate2024a}.
This suggests that the $3\times$ and $10\times$ downturn mixing scenarios are likely more representative of the conditions in a merger.
The exploration done in this paper shows the importance of understanding the mixing conditions, but not all scenarios are equally likely to represent the conditions in a merger.

The results in this work are important for the interpretation of presolar grains.
As shown in this work, the ratio of isotopic pairs is sensitive to the mixing conditions in the \oxygen[] shell. 
This means that comparing the grains data to the results of this work can be used as a diagnostic tool to constrain the mixing details of \oxygen[]-\carbon[] shell mergers and connect measured isotopic ratios to 3D hydrodynamics.

There are limitations to the results provided in this work and further extensions that could be done.
We have focused on the impact of mixing conditions and varying reaction rates, but do not investigate how different stellar conditions are relevant to the nucleosynthesis or if changing the stable seeds from the \carbon[] shell impact these results.
The distribution of stable seeds and the metallicity of the star are of significant importance for the nucleosynthesis of the \pnucn{} \citep{travaglioTestingRoleSNe2015, battinoHeavyElementsNucleosynthesis2020}.
It is possible that earlier in stellar evolution that weak \sprn{} in the \carbon[] shell could modify the stable seeds \citep{pignatariWEAKSPROCESSMASSIVE2010}, which could be relevant for the lighter \pnucn.

Other stellar models could have different \oxygen[]-shell sizes and temperature profiles due to different mixing prescriptions, initial mass, and rotations, which would directly affect locations of peak burning and how the mixing conditions impact the nucleosynthesis.
However, if the shell is convective-reactive, a spread in production comparing mixing scenarios would still be expected.

Finally, \oxygen[]-\carbon[] shell mergers are crucial to the nucleosynthesis of a massive star prior to the CCSN regardless of explosive energy \citep{ robertiGprocessNucleosynthesisCorecollapse2024b}.
Even if the results in Figure \ref{fig:impact_mixing_cases_bars_all} do not represent the whole of the nucleosynthesis of the \pnucn{}, they are still crucial for understanding the nucleosynthesis in the \oxygen[] shell prior to the CCSN.

The results of this work have implications beyond the \pnucn.
The light odd-Z elements \phosphorus[], \chlorine[], \potassium[], and \scandium[] are also produced during \oxygen[]-\carbon[] shell mergers and based on our preliminary results are likewise impacted by varying the mixing conditions \citep{ritterConvectivereactiveNucleosynthesisSc2018, robertiOccurrenceImpactCarbonOxygen2025}.
The long-lived radioactive isotope \potassium[40] ($t_{1/2} = \unit{\natlog{1.25}{9}}{\yr}$), which is relevant to the heating of planets early in their formation \citep{frankRadiogenicHeatingEvolution2014, oneillEffectGalacticChemical2020}, is also affected along with the stable \potassium[] isotopes \potassium[39] and \potassium[41]. 
Finally, observations have found \phosphorus[]-enhanced stars \citep{masseronPhosphorusrichStarsUnusual2020, braunerUnveilingChemicalFingerprint2023, braunerUnveilingChemicalFingerprint2024} which could be explained by a \oxygen[]-\carbon[] shell merger from a previous massive star.

\oxygen[]-\carbon[] shell mergers have a potentially huge impact on galactic chemical evolution models \citep{ritterConvectivereactiveNucleosynthesisSc2018}, and massive star models show this feature regardless of metallicity and stellar evolution model \citep{robertiOccurrenceImpactCarbonOxygen2025}.
Further work is needed to understand the impact of the macrophysical uncertainties in 1D stellar models on the nucleosynthesis of these light odd-Z elements.

\begin{acknowledgements}
The authors would like to thank Stephen Mojzsis for the useful discussions.
MP acknowledges the support to NuGrid from the ``Lendulet-2023'' Program of the Hungarian Academy of Sciences (LP2023-10, Hungary), the ERC Synergy Grant Programme (Geoastronomy, grant agreement number 101166936, Germany), the ERC Consolidator Grant funding scheme (Project RADIOSTAR, G.A. n. 724560, Hungary), the ChETEC COST Action (CA16117), supported by the European Cooperation in Science and Technology, and the IReNA network supported by NSF AccelNet (Grant No. OISE-1927130). MP also thanks the support from NKFI via K-project 138031 (Hungary).
FH is supported by a Natural Sciences and Engineering Research Council of Canada (NSERC) Discovery Grant and acknowledges support from the NSERC award SAPPJ-797 2021-00032 $\emph{Nuclear physics of the dynamic origin of the elements}$.
We acknowledge support from the ChETEC-INFRA project funded by the European Union’s Horizon 2020 Research and Innovation programme (Grant Agreement No 101008324). This research has used the Astrohub online virtual research environment (\url{https://astrohub.uvic.ca}), developed and operated by the Computational Stellar Astrophysics group at the University of Victoria and hosted on the Digital Research Alliance of Canada Arbutus Cloud at the University of Victoria. This work benefited from interactions and workshops co-organized by The Center for Nuclear astrophysics Across Messengers (CeNAM) which is supported by the U.S. Department of Energy, Office of Science, Office of Nuclear Physics, under Award Number DE-SC0023128.
\end{acknowledgements}

\software{The data and analysis tools are available on Zenodo under an open-source Creative Commons Attribution license: \dataset[doi:10.5281/zenodo.17576026]{https://doi.org/10.5281/zenodo.17576026}. They are also available for download from the NuGrid collaboration website: \url{download.nugridstars.org}.}

\bibliography{paper}{}
\bibliographystyle{aasjournal}

\appendix\label{sec:appendix}

\section{Correlations of nuclear reaction rates}\label{sec:appendixNuclearRates}

\renewcommand{\thefigure}{A\arabic{figure}}
\renewcommand{\thetable}{A\arabic{table}}
\setcounter{figure}{0}
\setcounter{table}{0}

\makeatletter
\renewcommand{\theHfigure}{appendix.\arabic{figure}}
\renewcommand{\theHtable}{appendix.\arabic{table}}
\makeatother

\begin{figure}[!ht]
\includegraphics[width=\columnwidth]{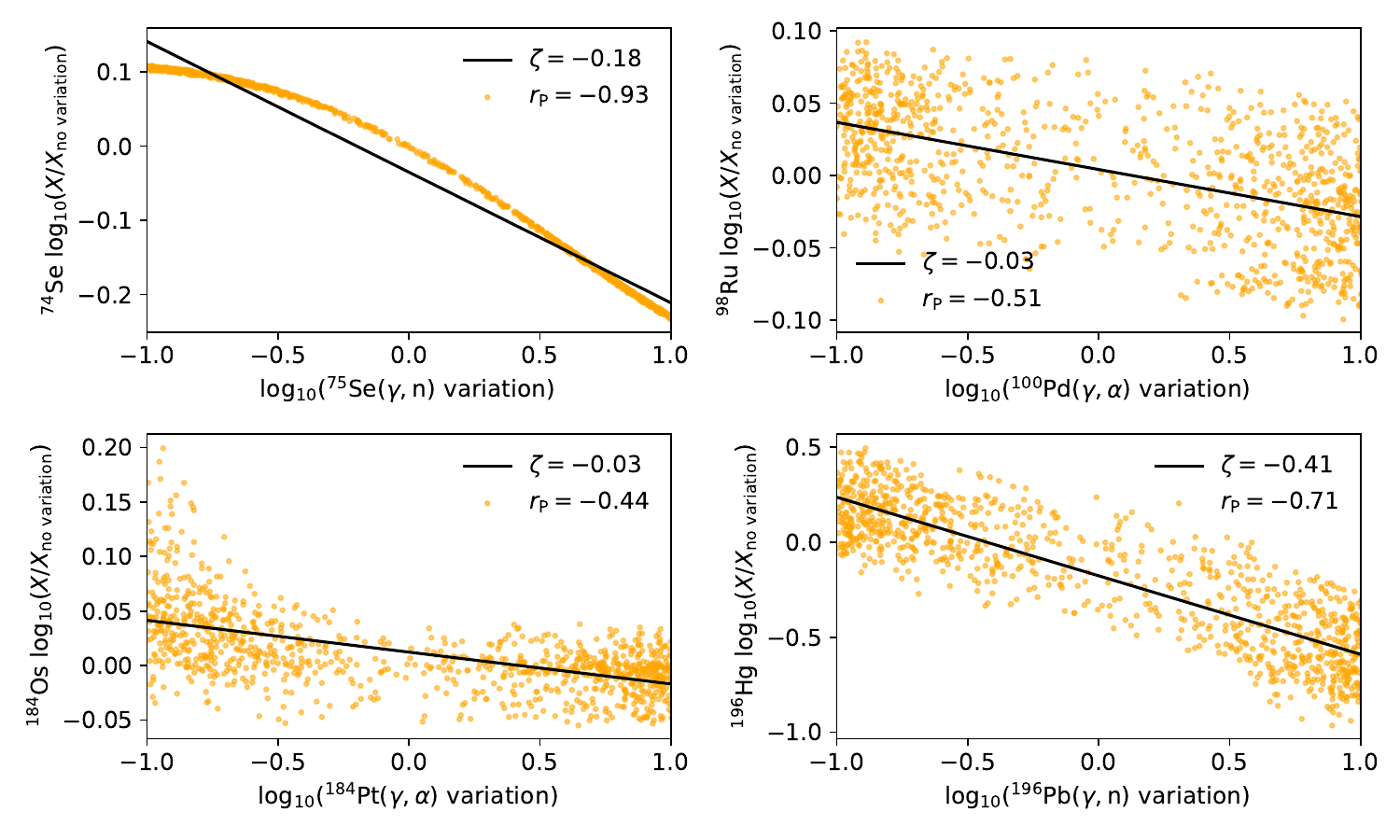}
\caption{Examples of strong correlations between mass fractions and reaction rates for four species under the MLT scenario. Orange dots indicate mass fractions for each variation factor, and the black line shows the linear fit to $\log_{10}(X/X_\mathrm{no~variation})$ versus $\log_{10}(\mathrm{variation~factor})$. Top left and bottom right: strong correlation and significant mass fraction changes for \selenium[74] and \mercury[196]. Top right: strong correlation for \ruthenium[98] with large scatter. Bottom left: correlation for \osmium[184] with a weak slope and asymmetric impact.
\label{fig:corr_ex}}
\end{figure}

The Pearson correlation coefficient, $r_\mathrm{P}$, is insufficient to assess the importance of a correlated rate.
As shown in Figure \ref{fig:corr_ex}, a strong correlation does not necessarily imply a significant impact on the final mass fraction of a species. To better quantify this impact, we use $\zeta$, which is defined as the slope of the linear regression between $\log_{10}(X/X_\mathrm{no~variation})$ and $\log_{10}(\mathrm{variation~factor})$, where $X$ is the final, mass-averaged, and decayed mass fraction for a variation, and $X_\mathrm{no~variation}$ is the corresponding default case with no rate variation.

Figure \ref{fig:corr_ex} demonstrates that strong correlations do not always imply significance, nor does a strong $\zeta$ guarantee it. 
For instance, the bottom right panel shows \mercury[196] with both a strong correlation and slope, while the bottom left shows a strong correlation for \osmium[184] but a weak slope. 
Only rates with both high $r_\mathrm{P}$ and $\zeta$ substantially affect final abundances.

A caveat of $r_\mathrm{P}$ for this method of varying the reaction rates is that it not distinguish between the photo-disintegration and corresponding capture rate because the same variation factor is applied to both.
All correlated rates are reported according to their photo-disintegration rates, but as shown by the upper left plot of Figure \ref{fig:corr_ex} for \selenium[74] and $\selenium[75](\gamma,\nt)\selenium[74]$ this results in a production term having a negative correlation because $\selenium[74](\nt,\gamma)\selenium[75]$ is also modified in the same way.
As explained in Section \ref{sec:convreacflow}, the reactions in this shell are not balanced and both a destruction and production term could be relevant at different locations, although as Figure \ref{fig:dy156_fluxes_ingest} shows for heavier species the $(\gamma,\nt)$ rate is typically much stronger.

\section{Results of varying the ingestion rate}\label{sec:appendixResultsIngestion}

\renewcommand{\thefigure}{B\arabic{figure}}
\renewcommand{\thetable}{B\arabic{table}}
\setcounter{figure}{0}
\setcounter{table}{0}

\makeatletter
\renewcommand{\theHfigure}{appendix.\arabic{figure}}
\renewcommand{\theHtable}{appendix.\arabic{table}}
\makeatother

Here we provide the results for varying the ingestion rates for the scenarios with a downturn in the mixing efficiency profile as shown in Figure \ref{fig:all_dconv}.

\begin{figure*}[!h] 
\includegraphics[width=\textwidth]{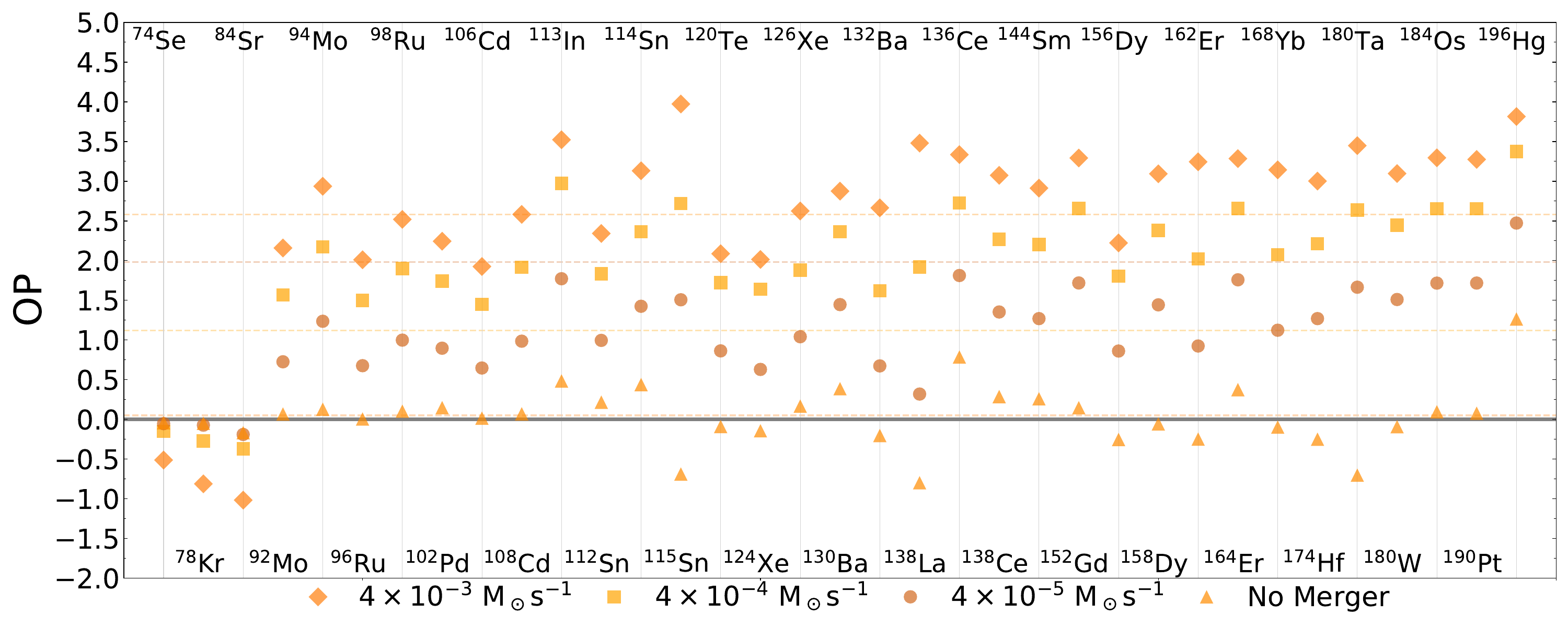}
\caption{The overproduction compared to initial of the \pnucn{} for the 3D-inspired mixing scenario for no ingestion, $\unit{\natlog{4}{-5}}{\Msun\second^{-1}}$, $\unit{\natlog{4}{-4}}{\Msun\second^{-1}}$, and $\unit{\natlog{4}{-3}}{\Msun\second^{-1}}$. The average spread in production $\OP_{\max} - \OP_{\min} = \unit{1.58}{\dex}$. $\OP=0$ is the initial amount. Average \OP{} for each scenario are provided as dashed lines and corresponds to the values presented in Table \ref{tab:mixing_scenarios}.}
\label{fig:ingest_3D}
\end{figure*} 

\begin{figure*}[!h]
\includegraphics[width=\textwidth]{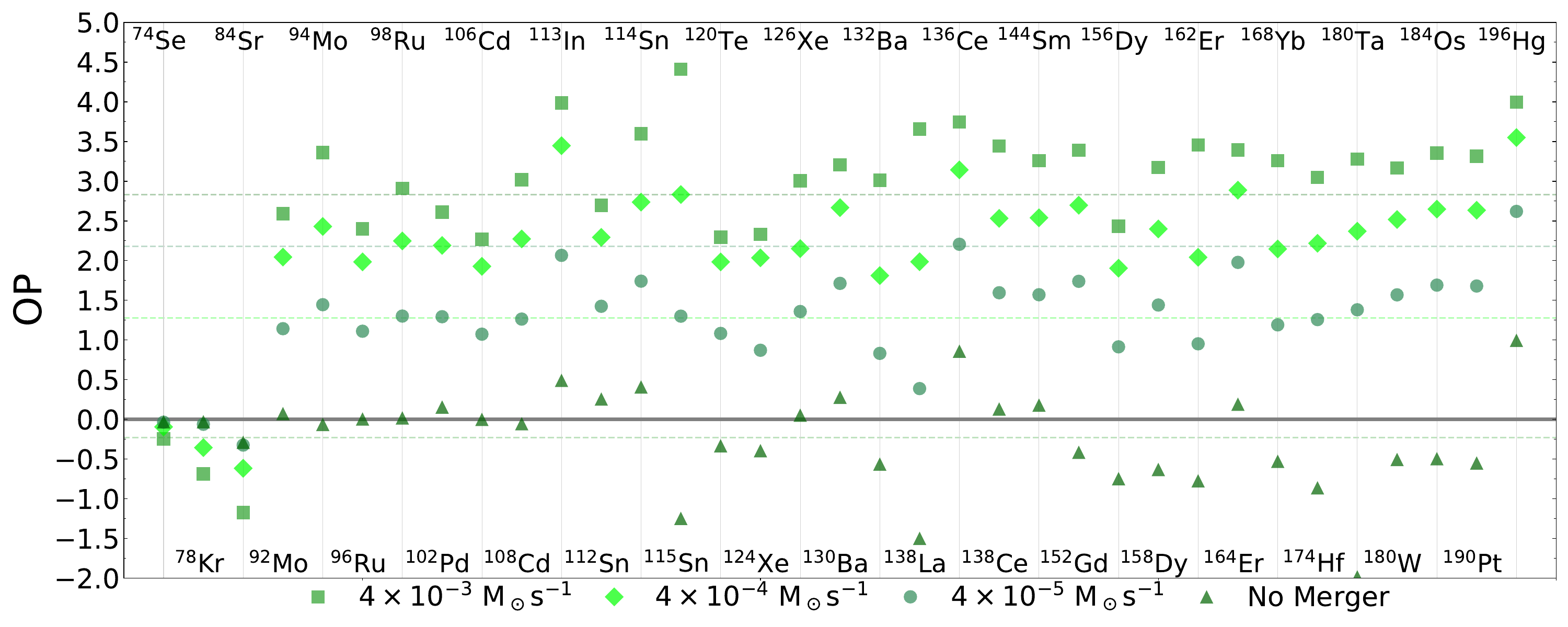}
\caption{The overproduction compared to initial of the \pnucn{} for the $3\times$3D-inspired mixing scenario for no ingestion, $\unit{\natlog{4}{-5}}{\Msun\second^{-1}}$, $\unit{\natlog{4}{-4}}{\Msun\second^{-1}}$, and $\unit{\natlog{4}{-3}}{\Msun\second^{-1}}$. The average spread in production $\OP_{\max} - \OP_{\min} = \unit{1.64}{\dex}$ excluding the no ingestion case. $\OP=0$ is the initial amount. Average \OP{} for each scenario are provided as dashed lines and corresponds to the values presented in Table \ref{tab:mixing_scenarios}.}
\label{fig:ingest_3x3D}
\end{figure*}

\begin{figure*}[!h]
\includegraphics[width=\textwidth]{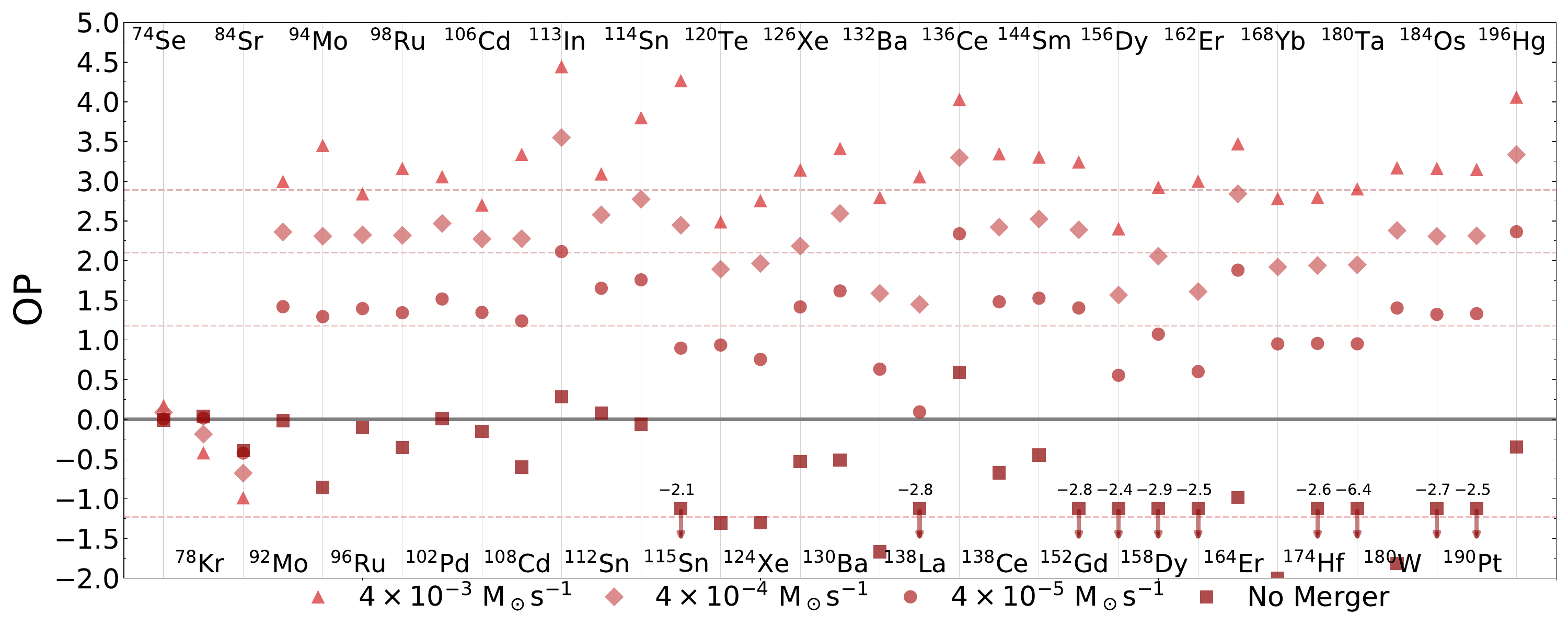}
\caption{The overproduction compared to initial of the \pnucn{} for the $10\times$3D-inspired mixing scenario for no ingestion, $\unit{\natlog{4}{-5}}{\Msun\second^{-1}}$, $\unit{\natlog{4}{-4}}{\Msun\second^{-1}}$, and $\unit{\natlog{4}{-3}}{\Msun\second^{-1}}$. Arrows denote \OP{} out of bounds and the true \OP{} is written above. The average spread in production $\OP_{\max} - \OP_{\min} = \unit{1.78}{\dex}$ excluding the no ingestion case. $\OP=0$ is the initial amount. Average \OP{} for each scenario are provided as dashed lines and corresponds to the values presented in Table \ref{tab:mixing_scenarios}.}
\label{fig:ingest_10x3D} 
\end{figure*}

\begin{figure*}[!h]
\includegraphics[width=\textwidth]{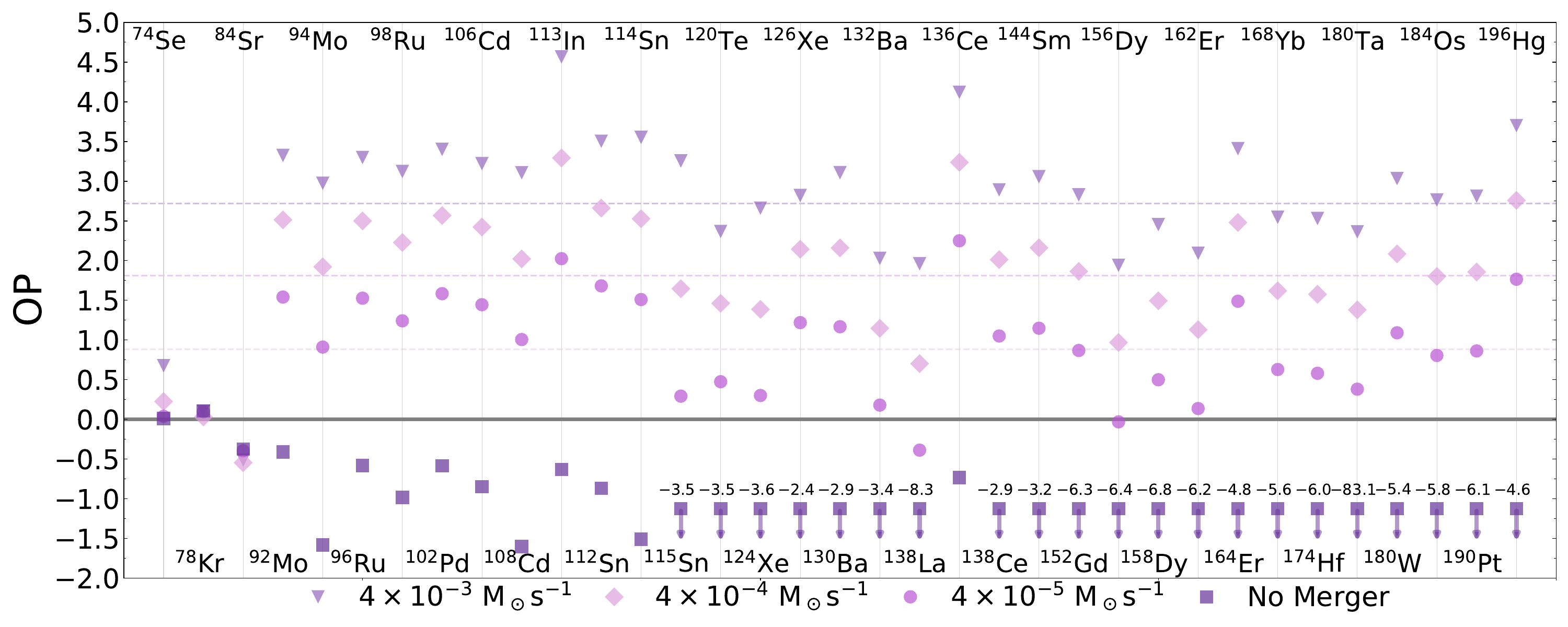}
\caption{The overproduction compared to initial of the \pnucn{} for the $50\times$3D-inspired mixing scenario for no ingestion, $\unit{\natlog{4}{-5}}{\Msun\second^{-1}}$, $\unit{\natlog{4}{-4}}{\Msun\second^{-1}}$, and $\unit{\natlog{4}{-3}}{\Msun\second^{-1}}$. Arrows denote \OP{} out of bounds and the true \OP{} is written above. The average spread in production $\OP_{\max} - \OP_{\min} = \unit{1.84}{\dex}$ excluding the no ingestion case. $\OP=0$ is the initial amount. Average \OP{} for each scenario excluding ``No Merger" are provided as dashed lines and corresponds to the values presented in Table \ref{tab:mixing_scenarios}.}
\label{fig:ingest_50x3D}
\end{figure*}

\clearpage

\section{Results of varying the input nuclear reactions}\label{sec:appendixResultsNuclear}

\renewcommand{\thefigure}{C\arabic{figure}}
\renewcommand{\thetable}{C\arabic{table}}
\setcounter{figure}{0}
\setcounter{table}{0}

\makeatletter
\renewcommand{\theHfigure}{appendix.\arabic{figure}}
\renewcommand{\theHtable}{appendix.\arabic{table}}
\makeatother

Here we provide the results for varying the input nuclear reactions for the scenarios with a downturn in the mixing efficiency profile as shown in Figure \ref{fig:all_dconv} and the reaction rate correlation tables for the MLT and downturn scenarios.

\begin{figure*}[!h] 
\includegraphics[width=\textwidth]{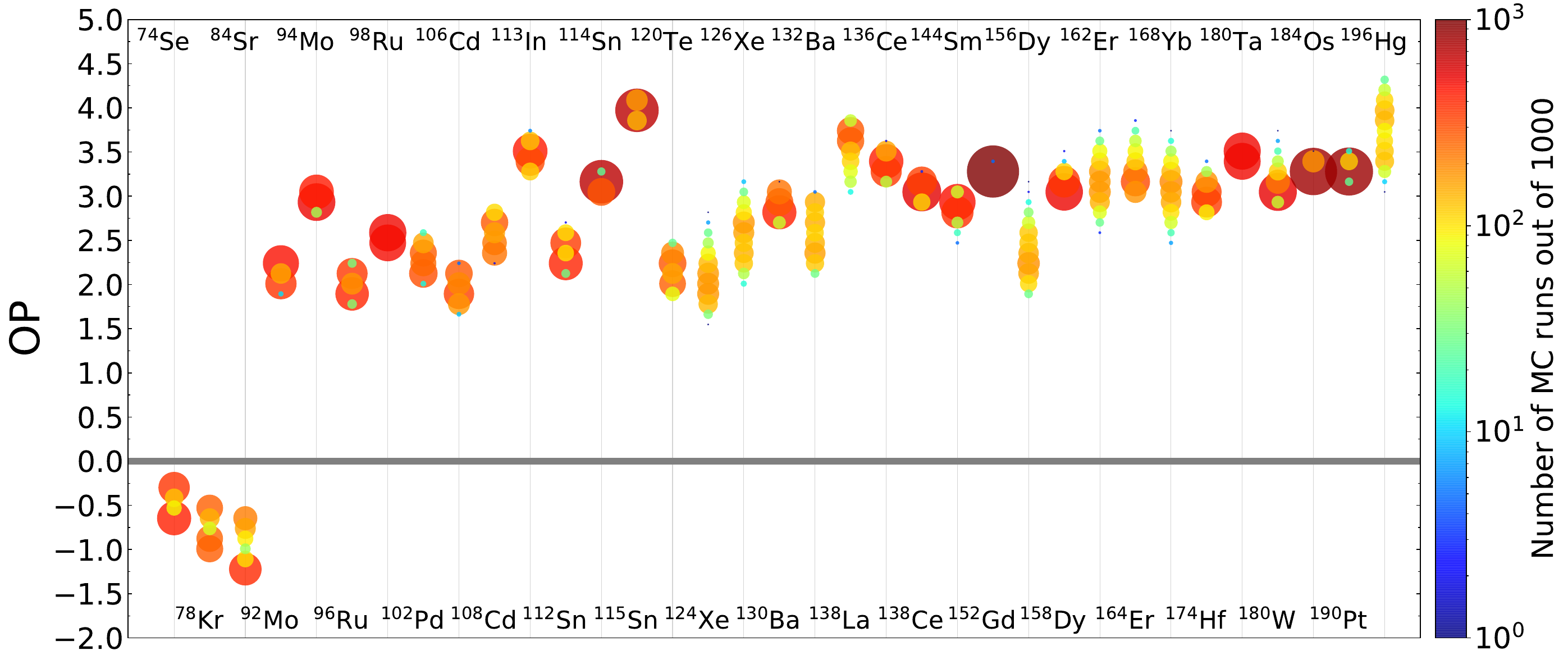}
\caption{Histogram showing the spread due to varying $(\gamma,\pt)$, $(\gamma,\nt)$, $(\gamma,\alpha)$ and corresponding capture rates for unstable \nt-deficient isotopes from \selenium[]{--}\polonium[] for the 3D-inspired mixing scenario. Colour and size both correspond to the logarithmic binning of Monte Carlo runs.  The average spread $\OP_{\max}-\OP_{\min}=\unit{0.59}{\dex}$. $\OP=0$ is the initial amount.
\label{fig:nuclearimpact_3D}}
\end{figure*}

\begin{figure*} [!h]
\includegraphics[width=\textwidth]{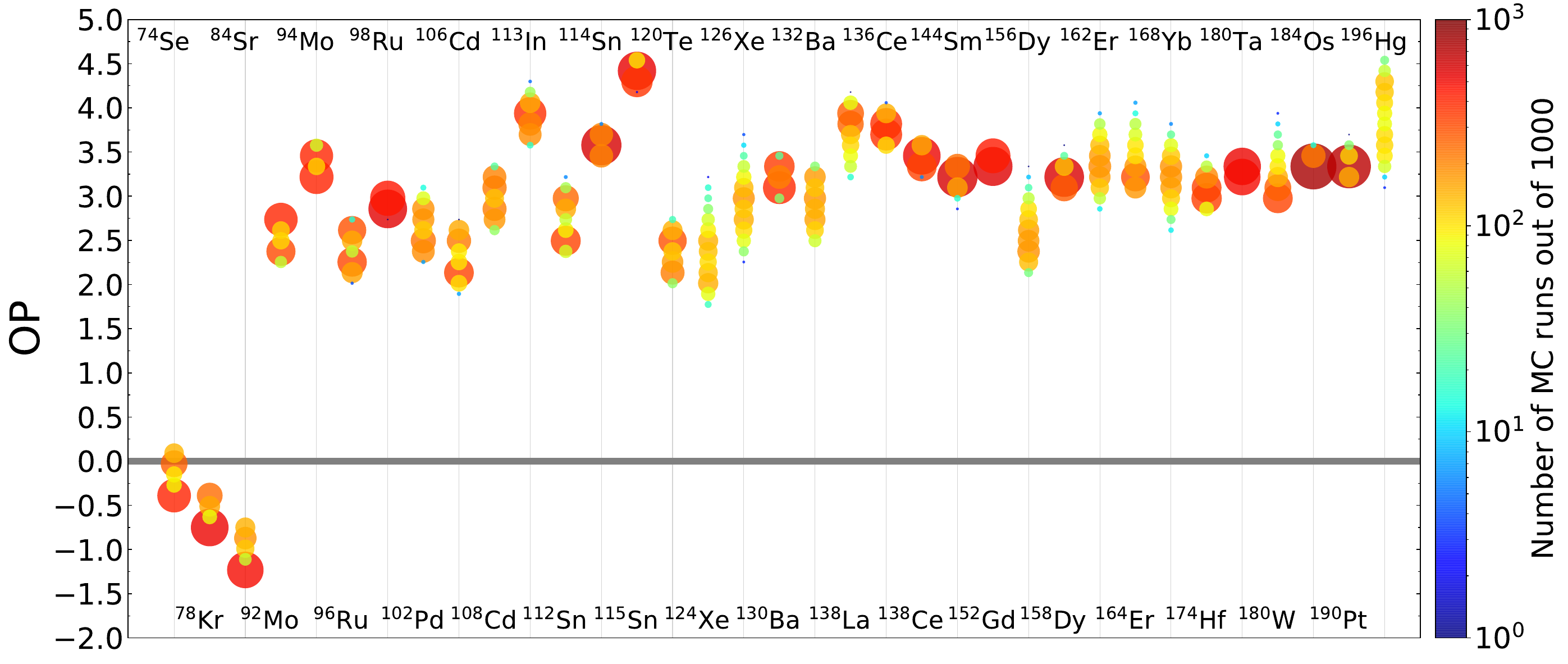}
\caption{Histogram showing the spread due to varying $(\gamma,\pt)$, $(\gamma,\nt)$, $(\gamma,\alpha)$ and corresponding capture rates for unstable \nt-deficient isotopes from \selenium[]{--}\polonium[] for the $3\times$3D-inspired mixing scenario. Colour and size both correspond to the logarithmic binning of Monte Carlo runs. The average spread $\OP_{\max}-\OP_{\min}=\unit{0.69}{\dex}$. $\OP=0$ is the initial amount.
\label{fig:nuclearimpact_3x3D}}
\end{figure*}

\begin{figure*} 
\includegraphics[width=\textwidth]{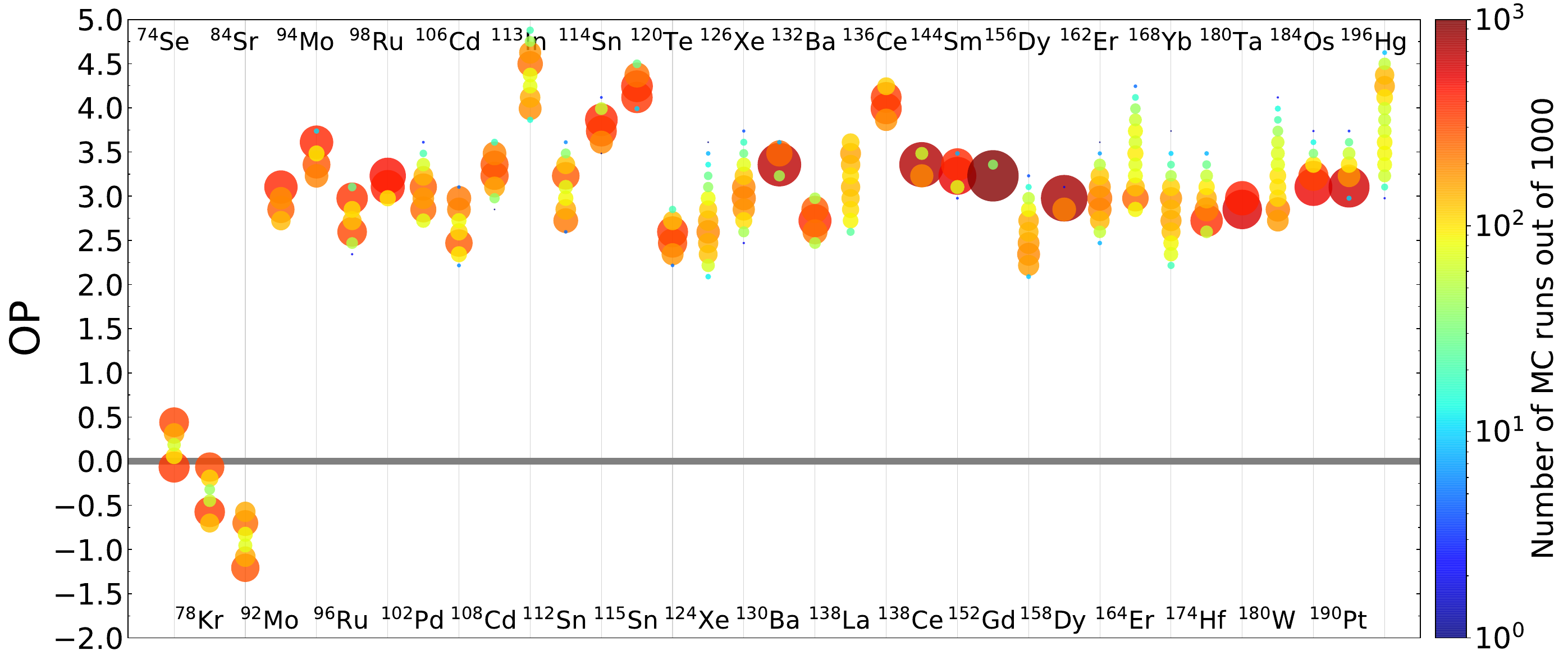}
\caption{Histogram showing the spread due to varying $(\gamma,\pt)$, $(\gamma,\nt)$, $(\gamma,\alpha)$ and corresponding capture rates for unstable \nt-deficient isotopes from \selenium[]{--}\polonium[] for the $10\times$3D-inspired mixing scenario. Colour and size both correspond to the logarithmic binning of Monte Carlo runs. The average spread $\OP_{\max}-\OP_{\min}=\unit{0.76}{\dex}$. $\OP=0$ is the initial amount.
\label{fig:nuclearimpact_10x3D}}
\end{figure*}

\begin{figure*} [!h]
\includegraphics[width=\textwidth]{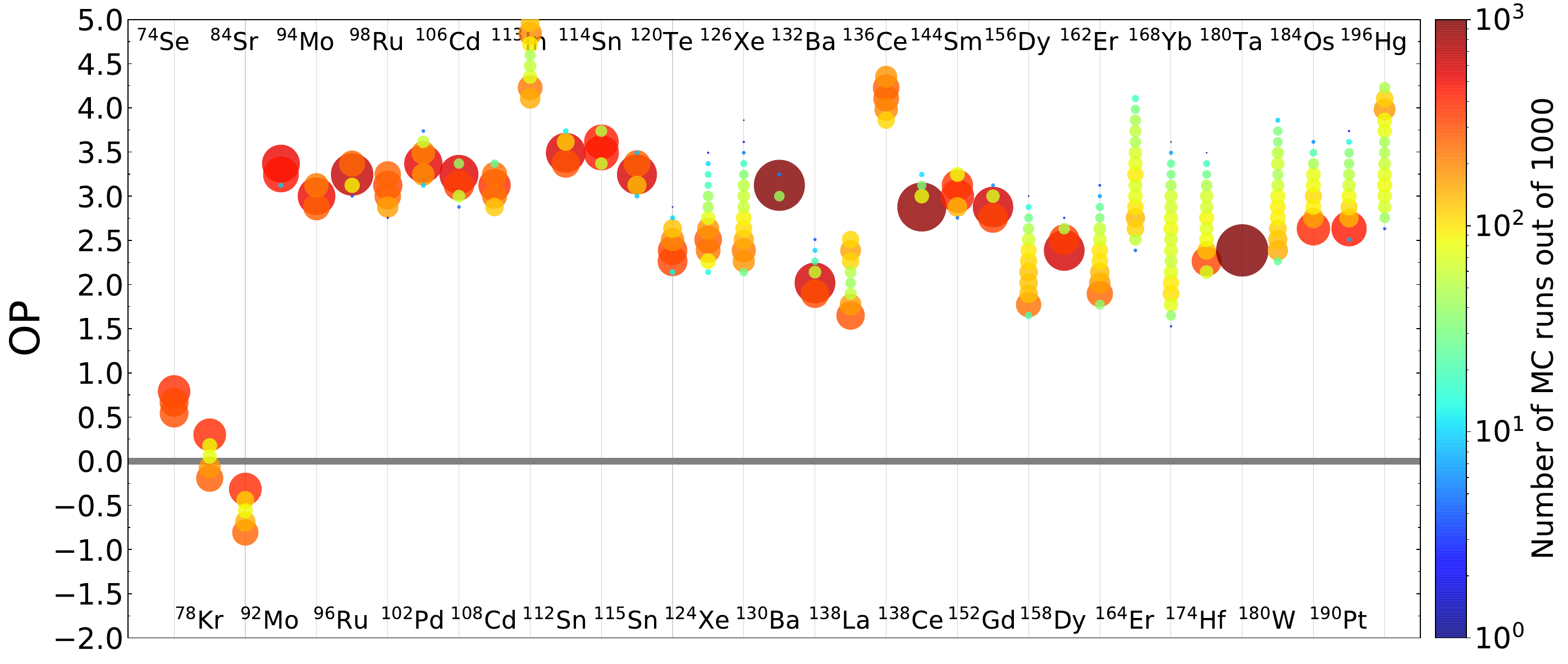}
\caption{Histogram showing the spread due to varying $(\gamma,\pt)$, $(\gamma,\nt)$, $(\gamma,\alpha)$ and corresponding capture rates for unstable \nt-deficient isotopes from \selenium[]{--}\polonium[] for the $50\times$3D-inspired mixing scenario. Colour and size both correspond to the logarithmic binning of Monte Carlo runs. The average spread $\OP_{\max}-\OP_{\min}=\unit{0.79}{\dex}$. $\OP=0$ is the initial amount.
\label{fig:nuclearimpact_50x3D}}
\end{figure*}

\startlongtable
\begin{deluxetable*}{llcc llcc}
\tablewidth{0pt}
\tablecaption{Correlations and $\zeta$ slopes between mass fraction and reaction rates for the MLT mixing scenario.
\label{tab:mlt_corr}}
\tablehead{
\colhead{\textbf{Isotope}} & \colhead{\textbf{Reaction}} & \colhead{$\mathbf{r_\mathrm{\mathbf{P}}}$} & \colhead{$\mathbf{\zeta}$} &\colhead{\textbf{Isotope}} & \colhead{\textbf{Reaction}} & \colhead{$\mathbf{r_\mathrm{\mathbf{P}}}$} & \colhead{$\mathbf{\zeta}$}
}
\startdata
$^{74}\mathrm{Se}$ & $^{75}\mathrm{Se}(\gamma,\nt)$ & $-0.93$ & $-0.18$ & $^{144}\mathrm{Sm}$ & $^{142}\mathrm{Sm}(\gamma,\nt)$ & $-0.19$ & $-0.02$ \\ 
$^{78}\mathrm{Kr}$ & $^{79}\mathrm{Kr}(\gamma,\nt)$ & $-0.88$ & $-0.28$ & $ $ & $^{142}\mathrm{Sm}(\gamma,\pt)$ & $-0.17$ & $-0.02$ \\ 
$^{84}\mathrm{Sr}$ & $^{85}\mathrm{Sr}(\gamma,\nt)$ & $-0.88$ & $-0.28$ & $ $ & $^{143}\mathrm{Sm}(\gamma,\nt)$ & $-0.25$ & $-0.03$ \\ 
$^{92}\mathrm{Mo}$ & $^{93}\mathrm{Mo}(\gamma,\nt)$ & $-0.94$ & $-0.07$ & $ $ & $^{146}\mathrm{Sm}(\gamma,\nt)$ & $0.20$ & $0.03$ \\ 
$ $ & $^{110}\mathrm{Sn}(\gamma,\alpha)$ & $0.16$ & $0.01$ & $ $ & $^{150}\mathrm{Gd}(\gamma,\nt)$ & $0.17$ & $0.02$ \\ 
$^{94}\mathrm{Mo}$ & $^{93}\mathrm{Mo}(\gamma,\nt)$ & $0.97$ & $0.06$ & $ $ & $^{150}\mathrm{Gd}(\gamma,\alpha)$ & $-0.15$ & $-0.02$ \\ 
$^{96}\mathrm{Ru}$ & $^{97}\mathrm{Ru}(\gamma,\nt)$ & $-0.88$ & $-0.12$ & $ $ & $^{196}\mathrm{Pb}(\gamma,\nt)$ & $0.47$ & $0.06$ \\ 
$ $ & $^{100}\mathrm{Pd}(\gamma,\alpha)$ & $0.20$ & $0.03$ & $ $ & $^{202}\mathrm{Pb}(\gamma,\nt)$ & $0.21$ & $0.03$ \\ 
$^{98}\mathrm{Ru}$ & $^{97}\mathrm{Ru}(\gamma,\nt)$ & $0.36$ & $0.02$ & $^{152}\mathrm{Gd}$ & $^{152}\mathrm{Dy}(\gamma,\alpha)$ & $-0.40$ & $-0.01$ \\ 
$ $ & $^{100}\mathrm{Pd}(\gamma,\pt)$ & $0.62$ & $0.04$ & $ $ & $^{154}\mathrm{Dy}(\gamma,\alpha)$ & $-0.15$ & $-0.00$ \\ 
$ $ & $^{100}\mathrm{Pd}(\gamma,\alpha)$ & $-0.51$ & $-0.03$ & $ $ & $^{160}\mathrm{Er}(\gamma,\alpha)$ & $0.39$ & $0.00$ \\ 
$^{102}\mathrm{Pd}$ & $^{100}\mathrm{Pd}(\gamma,\pt)$ & $-0.29$ & $-0.05$ & $^{156}\mathrm{Dy}$ & $^{159}\mathrm{Er}(\gamma,\nt)$ & $-0.18$ & $-0.06$ \\ 
$ $ & $^{100}\mathrm{Pd}(\gamma,\alpha)$ & $-0.30$ & $-0.05$ & $ $ & $^{160}\mathrm{Er}(\gamma,\alpha)$ & $0.74$ & $0.26$ \\ 
$ $ & $^{103}\mathrm{Pd}(\gamma,\nt)$ & $-0.66$ & $-0.12$ & $ $ & $^{202}\mathrm{Pb}(\gamma,\nt)$ & $0.18$ & $0.05$ \\ 
$^{106}\mathrm{Cd}$ & $^{107}\mathrm{Cd}(\gamma,\nt)$ & $-0.86$ & $-0.16$ & $^{158}\mathrm{Dy}$ & $^{158}\mathrm{Er}(\gamma,\alpha)$ & $-0.23$ & $-0.01$ \\ 
$ $ & $^{110}\mathrm{Sn}(\gamma,\alpha)$ & $0.23$ & $0.05$ & $ $ & $^{160}\mathrm{Er}(\gamma,\alpha)$ & $0.56$ & $0.01$ \\ 
$^{108}\mathrm{Cd}$ & $^{107}\mathrm{Cd}(\gamma,\nt)$ & $0.72$ & $0.10$ & $ $ & $^{196}\mathrm{Pb}(\gamma,\nt)$ & $0.16$ & $0.00$ \\ 
$ $ & $^{109}\mathrm{Cd}(\gamma,\nt)$ & $-0.46$ & $-0.06$ & $ $ & $^{202}\mathrm{Pb}(\gamma,\nt)$ & $0.18$ & $0.00$ \\ 
$ $ & $^{110}\mathrm{Sn}(\gamma,\pt)$ & $0.16$ & $0.02$ & $^{162}\mathrm{Er}$ & $^{159}\mathrm{Er}(\gamma,\nt)$ & $-0.18$ & $-0.06$ \\ 
$^{113}\mathrm{In}$ & $^{113}\mathrm{Sn}(\gamma,\nt)$ & $0.91$ & $0.37$ & $ $ & $^{160}\mathrm{Er}(\gamma,\nt)$ & $-0.18$ & $-0.06$ \\ 
$^{112}\mathrm{Sn}$ & $^{110}\mathrm{Sn}(\gamma,\alpha)$ & $-0.17$ & $-0.03$ & $ $ & $^{160}\mathrm{Er}(\gamma,\alpha)$ & $-0.26$ & $-0.07$ \\ 
$ $ & $^{113}\mathrm{Sn}(\gamma,\nt)$ & $-0.83$ & $-0.15$ & $ $ & $^{161}\mathrm{Er}(\gamma,\nt)$ & $0.21$ & $0.06$ \\ 
$^{114}\mathrm{Sn}$ & $^{110}\mathrm{Sn}(\gamma,\alpha)$ & $-0.15$ & $-0.01$ & $ $ & $^{166}\mathrm{Yb}(\gamma,\alpha)$ & $0.53$ & $0.14$ \\ 
$ $ & $^{113}\mathrm{Sn}(\gamma,\nt)$ & $0.74$ & $0.06$ & $ $ & $^{196}\mathrm{Pb}(\gamma,\nt)$ & $0.25$ & $0.07$ \\ 
$ $ & $^{122}\mathrm{Xe}(\gamma,\nt)$ & $-0.18$ & $-0.01$ & $ $ & $^{202}\mathrm{Pb}(\gamma,\nt)$ & $0.17$ & $0.05$ \\ 
$ $ & $^{122}\mathrm{Xe}(\gamma,\pt)$ & $0.20$ & $0.02$ & $^{164}\mathrm{Er}$ & $^{164}\mathrm{Yb}(\gamma,\nt)$ & $-0.24$ & $-0.09$ \\ 
$ $ & $^{122}\mathrm{Xe}(\gamma,\alpha)$ & $0.41$ & $0.04$ & $ $ & $^{164}\mathrm{Yb}(\gamma,\alpha)$ & $-0.58$ & $-0.32$ \\ 
$^{115}\mathrm{Sn}$ & $^{113}\mathrm{Sn}(\gamma,\nt)$ & $0.80$ & $0.05$ & $ $ & $^{196}\mathrm{Pb}(\gamma,\nt)$ & $0.17$ & $0.05$ \\ 
$ $ & $^{122}\mathrm{Xe}(\gamma,\pt)$ & $0.17$ & $0.01$ & $^{168}\mathrm{Yb}$ & $^{168}\mathrm{Hf}(\gamma,\alpha)$ & $-0.28$ & $-0.14$ \\ 
$ $ & $^{122}\mathrm{Xe}(\gamma,\alpha)$ & $0.35$ & $0.02$ & $ $ & $^{172}\mathrm{Hf}(\gamma,\alpha)$ & $0.60$ & $0.24$ \\ 
$^{120}\mathrm{Te}$ & $^{121}\mathrm{Te}(\gamma,\nt)$ & $-0.71$ & $-0.09$ & $ $ & $^{176}\mathrm{W}(\gamma,\alpha)$ & $0.20$ & $0.07$ \\ 
$ $ & $^{122}\mathrm{Xe}(\gamma,\pt)$ & $0.45$ & $0.05$ & $ $ & $^{196}\mathrm{Pb}(\gamma,\nt)$ & $0.21$ & $0.07$ \\ 
$ $ & $^{122}\mathrm{Xe}(\gamma,\alpha)$ & $-0.32$ & $-0.04$ & $^{174}\mathrm{Hf}$ & $^{174}\mathrm{W}(\gamma,\nt)$ & $-0.21$ & $-0.03$ \\ 
$^{124}\mathrm{Xe}$ & $^{122}\mathrm{Xe}(\gamma,\nt)$ & $-0.17$ & $-0.04$ & $ $ & $^{174}\mathrm{W}(\gamma,\alpha)$ & $-0.40$ & $-0.08$ \\ 
$ $ & $^{122}\mathrm{Xe}(\gamma,\pt)$ & $-0.24$ & $-0.06$ & $ $ & $^{178}\mathrm{W}(\gamma,\nt)$ & $-0.15$ & $-0.03$ \\ 
$ $ & $^{122}\mathrm{Xe}(\gamma,\alpha)$ & $-0.45$ & $-0.15$ & $ $ & $^{178}\mathrm{W}(\gamma,\alpha)$ & $0.42$ & $0.07$ \\ 
$ $ & $^{123}\mathrm{Xe}(\gamma,\nt)$ & $0.16$ & $0.03$ & $^{180}\mathrm{Ta}$ & $^{179}\mathrm{Ta}(\gamma,\nt)$ & $-0.91$ & $-0.02$ \\ 
$ $ & $^{125}\mathrm{Xe}(\gamma,\nt)$ & $-0.46$ & $-0.16$ & $^{180}\mathrm{W}$ & $^{180}\mathrm{Os}(\gamma,\nt)$ & $-0.28$ & $-0.07$ \\ 
$^{126}\mathrm{Xe}$ & $^{122}\mathrm{Xe}(\gamma,\alpha)$ & $-0.34$ & $-0.09$ & $ $ & $^{180}\mathrm{Os}(\gamma,\alpha)$ & $-0.52$ & $-0.21$ \\ 
$ $ & $^{125}\mathrm{Xe}(\gamma,\nt)$ & $0.49$ & $0.13$ & $ $ & $^{196}\mathrm{Pb}(\gamma,\nt)$ & $0.21$ & $0.05$ \\ 
$ $ & $^{127}\mathrm{Xe}(\gamma,\nt)$ & $-0.25$ & $-0.07$ & $^{184}\mathrm{Os}$ & $^{185}\mathrm{Os}(\gamma,\nt)$ & $0.33$ & $0.02$ \\ 
$ $ & $^{126}\mathrm{Ba}(\gamma,\pt)$ & $-0.18$ & $-0.04$ & $ $ & $^{184}\mathrm{Pt}(\gamma,\alpha)$ & $-0.44$ & $-0.03$ \\ 
$ $ & $^{126}\mathrm{Ba}(\gamma,\alpha)$ & $-0.32$ & $-0.09$ & $ $ & $^{186}\mathrm{Pt}(\gamma,\nt)$ & $0.17$ & $0.01$ \\ 
$^{130}\mathrm{Ba}$ & $^{126}\mathrm{Ba}(\gamma,\alpha)$ & $-0.17$ & $-0.01$ & $ $ & $^{186}\mathrm{Pt}(\gamma,\alpha)$ & $-0.16$ & $-0.01$ \\ 
$ $ & $^{128}\mathrm{Ba}(\gamma,\nt)$ & $-0.19$ & $-0.01$ & $ $ & $^{188}\mathrm{Pt}(\gamma,\nt)$ & $-0.17$ & $-0.01$ \\ 
$ $ & $^{129}\mathrm{Ba}(\gamma,\nt)$ & $0.23$ & $0.01$ & $ $ & $^{196}\mathrm{Pb}(\gamma,\nt)$ & $0.16$ & $0.01$ \\ 
$ $ & $^{131}\mathrm{Ba}(\gamma,\nt)$ & $-0.79$ & $-0.05$ & $^{190}\mathrm{Pt}$ & $^{190}\mathrm{Hg}(\gamma,\nt)$ & $-0.28$ & $-0.03$ \\ 
$^{132}\mathrm{Ba}$ & $^{131}\mathrm{Ba}(\gamma,\nt)$ & $0.66$ & $0.09$ & $ $ & $^{190}\mathrm{Hg}(\gamma,\alpha)$ & $-0.51$ & $-0.06$ \\ 
$ $ & $^{133}\mathrm{Ba}(\gamma,\nt)$ & $-0.61$ & $-0.08$ & $ $ & $^{196}\mathrm{Pb}(\gamma,\nt)$ & $0.20$ & $0.02$ \\ 
$^{138}\mathrm{La}$ & $^{137}\mathrm{La}(\gamma,\nt)$ & $-0.71$ & $-0.37$ & $^{196}\mathrm{Hg}$ & $^{196}\mathrm{Pb}(\gamma,\nt)$ & $-0.71$ & $-0.41$ \\ 
$^{136}\mathrm{Ce}$ & $^{138}\mathrm{Nd}(\gamma,\nt)$ & $-0.38$ & $-0.05$ & $ $ & $^{197}\mathrm{Pb}(\gamma,\nt)$ & $-0.16$ & $-0.06$ \\ 
$ $ & $^{138}\mathrm{Nd}(\gamma,\pt)$ & $0.65$ & $0.08$ & $ $ & $^{200}\mathrm{Pb}(\gamma,\nt)$ & $0.19$ & $0.10$ \\ 
$ $ & $^{138}\mathrm{Nd}(\gamma,\alpha)$ & $-0.16$ & $-0.02$ & $ $ & $^{202}\mathrm{Pb}(\gamma,\nt)$ & $0.30$ & $0.14$ \\ 
$ $ & $^{140}\mathrm{Nd}(\gamma,\alpha)$ & $0.34$ & $0.04$ & $ $ & $ $ &  &  \\ 
$^{138}\mathrm{Ce}$ & $^{137}\mathrm{Ce}(\gamma,\nt)$ & $0.51$ & $0.02$ & $ $ & $ $ &  &  \\ 
$ $ & $^{139}\mathrm{Ce}(\gamma,\nt)$ & $-0.43$ & $-0.01$ & $ $ & $ $ &  &  \\ 
$ $ & $^{139}\mathrm{Pr}(\gamma,\pt)$ & $0.29$ & $0.01$ & $ $ & $ $ &  &  \\ 
$ $ & $^{138}\mathrm{Nd}(\gamma,\nt)$ & $-0.30$ & $-0.01$ & $ $ & $ $ &  &  \\ 
$ $ & $^{138}\mathrm{Nd}(\gamma,\alpha)$ & $-0.22$ & $-0.01$ & $ $ & $ $ &  &  \\ 
\enddata
\tablecomments{The data is split into two sets of four columns.}
\end{deluxetable*}

\startlongtable
\begin{deluxetable*}{llcc llcc}
\tablewidth{0pt}
\tablecaption{Correlations and $\zeta$ slopes between mass fraction and reaction rates for the 3D-inspired mixing scenario.
\label{tab:3d_corr}}
\tablehead{
\colhead{\textbf{Isotope}} & \colhead{\textbf{Reaction}} & \colhead{$\mathbf{r_\mathrm{\mathbf{P}}}$} & \colhead{$\mathbf{\zeta}$} &\colhead{\textbf{Isotope}} & \colhead{\textbf{Reaction}} & \colhead{$\mathbf{r_\mathrm{\mathbf{P}}}$} & \colhead{$\mathbf{\zeta}$}
}
\startdata
$^{74}\mathrm{Se}$ & $^{75}\mathrm{Se}(\gamma,\nt)$ & $-0.83$ & $-0.22$ & $^{152}\mathrm{Gd}$ & $^{150}\mathrm{Gd}(\gamma,\alpha)$ & $-0.18$ & $-0.00$ \\ 
$^{78}\mathrm{Kr}$ & $^{79}\mathrm{Kr}(\gamma,\nt)$ & $-0.79$ & $-0.24$ & $ $ & $^{151}\mathrm{Gd}(\gamma,\nt)$ & $-0.20$ & $-0.00$ \\ 
$^{84}\mathrm{Sr}$ & $^{85}\mathrm{Sr}(\gamma,\nt)$ & $-0.80$ & $-0.34$ & $ $ & $^{152}\mathrm{Dy}(\gamma,\alpha)$ & $-0.31$ & $-0.01$ \\ 
$^{92}\mathrm{Mo}$ & $^{93}\mathrm{Mo}(\gamma,\nt)$ & $-0.91$ & $-0.14$ & $ $ & $^{154}\mathrm{Dy}(\gamma,\alpha)$ & $-0.17$ & $-0.00$ \\ 
$^{94}\mathrm{Mo}$ & $^{93}\mathrm{Mo}(\gamma,\nt)$ & $0.93$ & $0.10$ & $ $ & $^{160}\mathrm{Er}(\gamma,\alpha)$ & $0.39$ & $0.01$ \\ 
$^{96}\mathrm{Ru}$ & $^{97}\mathrm{Ru}(\gamma,\nt)$ & $-0.82$ & $-0.14$ & $ $ & $^{196}\mathrm{Pb}(\gamma,\nt)$ & $0.16$ & $0.00$ \\ 
$ $ & $^{100}\mathrm{Pd}(\gamma,\alpha)$ & $0.21$ & $0.04$ & $^{156}\mathrm{Dy}$ & $^{157}\mathrm{Dy}(\gamma,\nt)$ & $-0.28$ & $-0.12$ \\ 
$^{98}\mathrm{Ru}$ & $^{97}\mathrm{Ru}(\gamma,\nt)$ & $0.50$ & $0.02$ & $ $ & $^{159}\mathrm{Er}(\gamma,\nt)$ & $-0.17$ & $-0.05$ \\ 
$ $ & $^{100}\mathrm{Pd}(\gamma,\pt)$ & $0.58$ & $0.03$ & $ $ & $^{160}\mathrm{Er}(\gamma,\alpha)$ & $0.68$ & $0.26$ \\ 
$ $ & $^{100}\mathrm{Pd}(\gamma,\alpha)$ & $-0.31$ & $-0.01$ & $ $ & $^{196}\mathrm{Pb}(\gamma,\nt)$ & $0.16$ & $0.05$ \\ 
$ $ & $^{110}\mathrm{Sn}(\gamma,\alpha)$ & $0.22$ & $0.01$ & $^{158}\mathrm{Dy}$ & $^{157}\mathrm{Dy}(\gamma,\nt)$ & $0.24$ & $0.03$ \\ 
$^{102}\mathrm{Pd}$ & $^{100}\mathrm{Pd}(\gamma,\pt)$ & $-0.18$ & $-0.04$ & $ $ & $^{159}\mathrm{Dy}(\gamma,\nt)$ & $-0.21$ & $-0.03$ \\ 
$ $ & $^{100}\mathrm{Pd}(\gamma,\alpha)$ & $-0.21$ & $-0.04$ & $ $ & $^{159}\mathrm{Er}(\gamma,\nt)$ & $-0.17$ & $-0.02$ \\ 
$ $ & $^{103}\mathrm{Pd}(\gamma,\nt)$ & $-0.72$ & $-0.16$ & $ $ & $^{160}\mathrm{Er}(\gamma,\alpha)$ & $0.66$ & $0.07$ \\ 
$^{106}\mathrm{Cd}$ & $^{107}\mathrm{Cd}(\gamma,\nt)$ & $-0.75$ & $-0.16$ & $ $ & $^{196}\mathrm{Pb}(\gamma,\nt)$ & $0.16$ & $0.02$ \\ 
$ $ & $^{110}\mathrm{Sn}(\gamma,\alpha)$ & $0.31$ & $0.07$ & $ $ & $^{202}\mathrm{Pb}(\gamma,\nt)$ & $0.20$ & $0.02$ \\ 
$^{108}\mathrm{Cd}$ & $^{107}\mathrm{Cd}(\gamma,\nt)$ & $0.20$ & $0.05$ & $^{162}\mathrm{Er}$ & $^{159}\mathrm{Er}(\gamma,\nt)$ & $-0.25$ & $-0.10$ \\ 
$ $ & $^{109}\mathrm{Cd}(\gamma,\nt)$ & $-0.83$ & $-0.20$ & $ $ & $^{159}\mathrm{Er}(\gamma,\alpha)$ & $-0.16$ & $-0.05$ \\ 
$^{113}\mathrm{In}$ & $^{114}\mathrm{In}(\gamma,\nt)$ & $-0.32$ & $-0.05$ & $ $ & $^{160}\mathrm{Er}(\gamma,\nt)$ & $-0.23$ & $-0.09$ \\ 
$ $ & $^{110}\mathrm{Sn}(\gamma,\pt)$ & $-0.16$ & $-0.02$ & $ $ & $^{160}\mathrm{Er}(\gamma,\alpha)$ & $-0.44$ & $-0.15$ \\ 
$ $ & $^{110}\mathrm{Sn}(\gamma,\alpha)$ & $-0.27$ & $-0.04$ & $ $ & $^{161}\mathrm{Er}(\gamma,\nt)$ & $0.16$ & $0.06$ \\ 
$ $ & $^{113}\mathrm{Sn}(\gamma,\nt)$ & $0.60$ & $0.10$ & $ $ & $^{166}\mathrm{Yb}(\gamma,\alpha)$ & $0.33$ & $0.11$ \\ 
$ $ & $^{169}\mathrm{Lu}(\gamma,\nt)$ & $0.15$ & $0.02$ & $ $ & $^{196}\mathrm{Pb}(\gamma,\nt)$ & $0.27$ & $0.10$ \\ 
$^{112}\mathrm{Sn}$ & $^{113}\mathrm{Sn}(\gamma,\nt)$ & $-0.80$ & $-0.18$ & $ $ & $^{202}\mathrm{Pb}(\gamma,\nt)$ & $0.16$ & $0.06$ \\ 
$^{114}\mathrm{Sn}$ & $^{110}\mathrm{Sn}(\gamma,\alpha)$ & $-0.17$ & $-0.01$ & $^{164}\mathrm{Er}$ & $^{164}\mathrm{Yb}(\gamma,\nt)$ & $-0.26$ & $-0.07$ \\ 
$ $ & $^{113}\mathrm{Sn}(\gamma,\nt)$ & $0.65$ & $0.05$ & $ $ & $^{164}\mathrm{Yb}(\gamma,\alpha)$ & $-0.56$ & $-0.18$ \\ 
$ $ & $^{122}\mathrm{Xe}(\gamma,\pt)$ & $0.21$ & $0.02$ & $ $ & $^{196}\mathrm{Pb}(\gamma,\nt)$ & $0.21$ & $0.06$ \\ 
$ $ & $^{122}\mathrm{Xe}(\gamma,\alpha)$ & $0.44$ & $0.03$ & $ $ & $^{202}\mathrm{Pb}(\gamma,\nt)$ & $0.18$ & $0.05$ \\ 
$^{115}\mathrm{Sn}$ & $^{110}\mathrm{Sn}(\gamma,\alpha)$ & $-0.16$ & $-0.01$ & $^{168}\mathrm{Yb}$ & $^{164}\mathrm{Yb}(\gamma,\alpha)$ & $-0.19$ & $-0.09$ \\ 
$ $ & $^{113}\mathrm{Sn}(\gamma,\nt)$ & $0.74$ & $0.06$ & $ $ & $^{166}\mathrm{Yb}(\gamma,\nt)$ & $-0.22$ & $-0.08$ \\ 
$ $ & $^{122}\mathrm{Xe}(\gamma,\pt)$ & $0.16$ & $0.01$ & $ $ & $^{166}\mathrm{Yb}(\gamma,\alpha)$ & $-0.38$ & $-0.13$ \\ 
$ $ & $^{122}\mathrm{Xe}(\gamma,\alpha)$ & $0.33$ & $0.03$ & $ $ & $^{167}\mathrm{Yb}(\gamma,\nt)$ & $0.20$ & $0.07$ \\ 
$^{120}\mathrm{Te}$ & $^{121}\mathrm{Te}(\gamma,\nt)$ & $-0.76$ & $-0.19$ & $ $ & $^{172}\mathrm{Hf}(\gamma,\alpha)$ & $0.43$ & $0.16$ \\ 
$ $ & $^{122}\mathrm{Xe}(\gamma,\pt)$ & $0.26$ & $0.05$ & $ $ & $^{196}\mathrm{Pb}(\gamma,\nt)$ & $0.28$ & $0.09$ \\ 
$ $ & $^{122}\mathrm{Xe}(\gamma,\alpha)$ & $-0.23$ & $-0.05$ & $ $ & $^{202}\mathrm{Pb}(\gamma,\nt)$ & $0.19$ & $0.06$ \\ 
$^{124}\mathrm{Xe}$ & $^{122}\mathrm{Xe}(\gamma,\pt)$ & $-0.19$ & $-0.05$ & $^{174}\mathrm{Hf}$ & $^{172}\mathrm{Hf}(\gamma,\alpha)$ & $-0.40$ & $-0.07$ \\ 
$ $ & $^{122}\mathrm{Xe}(\gamma,\alpha)$ & $-0.39$ & $-0.15$ & $ $ & $^{174}\mathrm{W}(\gamma,\alpha)$ & $-0.19$ & $-0.04$ \\ 
$ $ & $^{125}\mathrm{Xe}(\gamma,\nt)$ & $-0.54$ & $-0.23$ & $ $ & $^{178}\mathrm{W}(\gamma,\alpha)$ & $0.51$ & $0.09$ \\ 
$^{126}\mathrm{Xe}$ & $^{122}\mathrm{Xe}(\gamma,\alpha)$ & $-0.35$ & $-0.14$ & $ $ & $^{182}\mathrm{Os}(\gamma,\alpha)$ & $0.18$ & $0.03$ \\ 
$ $ & $^{125}\mathrm{Xe}(\gamma,\nt)$ & $0.22$ & $0.09$ & $ $ & $^{196}\mathrm{Pb}(\gamma,\nt)$ & $0.19$ & $0.03$ \\ 
$ $ & $^{127}\mathrm{Xe}(\gamma,\nt)$ & $-0.63$ & $-0.27$ & $^{180}\mathrm{Ta}$ & $^{179}\mathrm{Ta}(\gamma,\nt)$ & $-0.91$ & $-0.07$ \\ 
$^{130}\mathrm{Ba}$ & $^{131}\mathrm{Ba}(\gamma,\nt)$ & $-0.82$ & $-0.13$ & $ $ & $^{179}\mathrm{Ta}(\gamma,\alpha)$ & $-0.15$ & $-0.01$ \\ 
$^{132}\mathrm{Ba}$ & $^{131}\mathrm{Ba}(\gamma,\nt)$ & $0.36$ & $0.14$ & $^{180}\mathrm{W}$ & $^{180}\mathrm{Os}(\gamma,\nt)$ & $-0.27$ & $-0.05$ \\ 
$ $ & $^{133}\mathrm{Ba}(\gamma,\nt)$ & $-0.76$ & $-0.30$ & $ $ & $^{180}\mathrm{Os}(\gamma,\alpha)$ & $-0.51$ & $-0.13$ \\ 
$^{138}\mathrm{La}$ & $^{133}\mathrm{La}(\gamma,\pt)$ & $-0.20$ & $-0.07$ & $ $ & $^{196}\mathrm{Pb}(\gamma,\nt)$ & $0.23$ & $0.04$ \\ 
$ $ & $^{135}\mathrm{La}(\gamma,\nt)$ & $-0.35$ & $-0.12$ & $^{184}\mathrm{Os}$ & $^{182}\mathrm{Os}(\gamma,\alpha)$ & $-0.38$ & $-0.02$ \\ 
$ $ & $^{136}\mathrm{La}(\gamma,\nt)$ & $0.26$ & $0.08$ & $ $ & $^{184}\mathrm{Pt}(\gamma,\alpha)$ & $-0.33$ & $-0.02$ \\ 
$ $ & $^{137}\mathrm{La}(\gamma,\nt)$ & $-0.34$ & $-0.14$ & $ $ & $^{186}\mathrm{Pt}(\gamma,\alpha)$ & $-0.15$ & $-0.01$ \\ 
$^{136}\mathrm{Ce}$ & $^{137}\mathrm{Ce}(\gamma,\nt)$ & $-0.49$ & $-0.07$ & $ $ & $^{188}\mathrm{Pt}(\gamma,\alpha)$ & $0.23$ & $0.01$ \\ 
$ $ & $^{138}\mathrm{Nd}(\gamma,\nt)$ & $-0.27$ & $-0.03$ & $ $ & $^{196}\mathrm{Pb}(\gamma,\nt)$ & $0.19$ & $0.01$ \\ 
$ $ & $^{138}\mathrm{Nd}(\gamma,\pt)$ & $0.48$ & $0.06$ & $^{190}\mathrm{Pt}$ & $^{190}\mathrm{Hg}(\gamma,\nt)$ & $-0.30$ & $-0.02$ \\ 
$ $ & $^{140}\mathrm{Nd}(\gamma,\alpha)$ & $0.38$ & $0.04$ & $ $ & $^{190}\mathrm{Hg}(\gamma,\alpha)$ & $-0.50$ & $-0.05$ \\ 
$^{138}\mathrm{Ce}$ & $^{137}\mathrm{Ce}(\gamma,\nt)$ & $0.65$ & $0.07$ & $ $ & $^{196}\mathrm{Pb}(\gamma,\nt)$ & $0.20$ & $0.02$ \\ 
$ $ & $^{139}\mathrm{Ce}(\gamma,\nt)$ & $-0.52$ & $-0.05$ & $^{196}\mathrm{Hg}$ & $^{196}\mathrm{Pb}(\gamma,\nt)$ & $-0.68$ & $-0.35$ \\ 
$ $ & $^{138}\mathrm{Nd}(\gamma,\pt)$ & $0.16$ & $0.01$ & $ $ & $^{197}\mathrm{Pb}(\gamma,\nt)$ & $-0.26$ & $-0.10$ \\ 
$^{144}\mathrm{Sm}$ & $^{142}\mathrm{Sm}(\gamma,\nt)$ & $-0.15$ & $-0.02$ & $ $ & $^{200}\mathrm{Pb}(\gamma,\nt)$ & $0.17$ & $0.07$ \\ 
$ $ & $^{142}\mathrm{Sm}(\gamma,\pt)$ & $-0.16$ & $-0.02$ & $ $ & $^{202}\mathrm{Pb}(\gamma,\nt)$ & $0.24$ & $0.10$ \\ 
$ $ & $^{143}\mathrm{Sm}(\gamma,\nt)$ & $-0.19$ & $-0.03$ & $ $ & $ $ &  &  \\ 
$ $ & $^{146}\mathrm{Sm}(\gamma,\nt)$ & $0.19$ & $0.03$ & $ $ & $ $ &  &  \\ 
$ $ & $^{146}\mathrm{Sm}(\gamma,\alpha)$ & $-0.22$ & $-0.03$ & $ $ & $ $ &  &  \\ 
$ $ & $^{150}\mathrm{Gd}(\gamma,\nt)$ & $0.21$ & $0.03$ & $ $ & $ $ &  &  \\ 
$ $ & $^{150}\mathrm{Gd}(\gamma,\alpha)$ & $-0.20$ & $-0.03$ & $ $ & $ $ &  &  \\ 
$ $ & $^{196}\mathrm{Pb}(\gamma,\nt)$ & $0.46$ & $0.06$ & $ $ & $ $ &  &  \\ 
$ $ & $^{202}\mathrm{Pb}(\gamma,\nt)$ & $0.20$ & $0.03$ & $ $ & $ $ &  &  \\ 
\enddata
\tablecomments{The data is split into two sets of four columns.}
\end{deluxetable*}

\startlongtable
\begin{deluxetable*}{llcc llcc}
\tablewidth{0pt}
\tablecaption{Correlations and $\zeta$ slopes between mass fraction and reaction rates for the $3\times$3D-inspired mixing scenario.
\label{tab:3x3d_corr}}
\tablehead{
\colhead{\textbf{Isotope}} & \colhead{\textbf{Reaction}} & \colhead{$\mathbf{r_\mathrm{\mathbf{P}}}$} & \colhead{$\mathbf{\zeta}$} &\colhead{\textbf{Isotope}} & \colhead{\textbf{Reaction}} & \colhead{$\mathbf{r_\mathrm{\mathbf{P}}}$} & \colhead{$\mathbf{\zeta}$}
}
\startdata
$^{74}\mathrm{Se}$ & $^{75}\mathrm{Se}(\gamma,\nt)$ & $-0.81$ & $-0.24$ & $^{144}\mathrm{Sm}$ & $^{142}\mathrm{Sm}(\gamma,\nt)$ & $-0.23$ & $-0.02$ \\ 
$^{78}\mathrm{Kr}$ & $^{79}\mathrm{Kr}(\gamma,\nt)$ & $-0.76$ & $-0.20$ & $ $ & $^{142}\mathrm{Sm}(\gamma,\pt)$ & $-0.23$ & $-0.03$ \\ 
$^{84}\mathrm{Sr}$ & $^{85}\mathrm{Sr}(\gamma,\nt)$ & $-0.75$ & $-0.26$ & $ $ & $^{143}\mathrm{Sm}(\gamma,\nt)$ & $-0.30$ & $-0.04$ \\ 
$^{92}\mathrm{Mo}$ & $^{93}\mathrm{Mo}(\gamma,\nt)$ & $-0.91$ & $-0.23$ & $ $ & $^{146}\mathrm{Sm}(\gamma,\alpha)$ & $-0.17$ & $-0.02$ \\ 
$^{94}\mathrm{Mo}$ & $^{93}\mathrm{Mo}(\gamma,\nt)$ & $0.92$ & $0.17$ & $ $ & $^{150}\mathrm{Gd}(\gamma,\nt)$ & $0.17$ & $0.02$ \\ 
$^{96}\mathrm{Ru}$ & $^{97}\mathrm{Ru}(\gamma,\nt)$ & $-0.84$ & $-0.25$ & $ $ & $^{150}\mathrm{Gd}(\gamma,\alpha)$ & $-0.16$ & $-0.02$ \\ 
$^{98}\mathrm{Ru}$ & $^{97}\mathrm{Ru}(\gamma,\nt)$ & $0.65$ & $0.05$ & $ $ & $^{196}\mathrm{Pb}(\gamma,\nt)$ & $0.44$ & $0.05$ \\ 
$ $ & $^{100}\mathrm{Pd}(\gamma,\pt)$ & $0.47$ & $0.03$ & $ $ & $^{202}\mathrm{Pb}(\gamma,\nt)$ & $0.17$ & $0.02$ \\ 
$ $ & $^{100}\mathrm{Pd}(\gamma,\alpha)$ & $-0.23$ & $-0.02$ & $^{152}\mathrm{Gd}$ & $^{151}\mathrm{Gd}(\gamma,\nt)$ & $-0.17$ & $-0.00$ \\ 
$ $ & $^{110}\mathrm{Sn}(\gamma,\alpha)$ & $0.21$ & $0.01$ & $ $ & $^{152}\mathrm{Dy}(\gamma,\alpha)$ & $-0.39$ & $-0.01$ \\ 
$^{102}\mathrm{Pd}$ & $^{100}\mathrm{Pd}(\gamma,\pt)$ & $-0.20$ & $-0.06$ & $ $ & $^{154}\mathrm{Dy}(\gamma,\alpha)$ & $-0.19$ & $-0.00$ \\ 
$ $ & $^{100}\mathrm{Pd}(\gamma,\alpha)$ & $-0.22$ & $-0.06$ & $ $ & $^{160}\mathrm{Er}(\gamma,\alpha)$ & $0.33$ & $0.01$ \\ 
$ $ & $^{103}\mathrm{Pd}(\gamma,\nt)$ & $-0.71$ & $-0.26$ & $^{156}\mathrm{Dy}$ & $^{157}\mathrm{Dy}(\gamma,\nt)$ & $-0.21$ & $-0.08$ \\ 
$^{106}\mathrm{Cd}$ & $^{107}\mathrm{Cd}(\gamma,\nt)$ & $-0.78$ & $-0.27$ & $ $ & $^{159}\mathrm{Er}(\gamma,\nt)$ & $-0.18$ & $-0.06$ \\ 
$ $ & $^{110}\mathrm{Sn}(\gamma,\alpha)$ & $0.21$ & $0.08$ & $ $ & $^{160}\mathrm{Er}(\gamma,\alpha)$ & $0.69$ & $0.25$ \\ 
$^{108}\mathrm{Cd}$ & $^{107}\mathrm{Cd}(\gamma,\nt)$ & $0.24$ & $0.08$ & $ $ & $^{196}\mathrm{Pb}(\gamma,\nt)$ & $0.15$ & $0.05$ \\ 
$ $ & $^{109}\mathrm{Cd}(\gamma,\nt)$ & $-0.80$ & $-0.24$ & $ $ & $^{202}\mathrm{Pb}(\gamma,\nt)$ & $0.15$ & $0.05$ \\ 
$^{113}\mathrm{In}$ & $^{110}\mathrm{Sn}(\gamma,\pt)$ & $-0.22$ & $-0.04$ & $^{158}\mathrm{Dy}$ & $^{157}\mathrm{Dy}(\gamma,\nt)$ & $0.20$ & $0.02$ \\ 
$ $ & $^{110}\mathrm{Sn}(\gamma,\alpha)$ & $-0.33$ & $-0.06$ & $ $ & $^{159}\mathrm{Er}(\gamma,\nt)$ & $-0.19$ & $-0.02$ \\ 
$ $ & $^{113}\mathrm{Sn}(\gamma,\nt)$ & $0.62$ & $0.15$ & $ $ & $^{160}\mathrm{Er}(\gamma,\alpha)$ & $0.67$ & $0.07$ \\ 
$ $ & $^{169}\mathrm{Lu}(\gamma,\nt)$ & $0.16$ & $0.03$ & $ $ & $^{196}\mathrm{Pb}(\gamma,\nt)$ & $0.16$ & $0.02$ \\ 
$^{112}\mathrm{Sn}$ & $^{113}\mathrm{Sn}(\gamma,\nt)$ & $-0.78$ & $-0.31$ & $ $ & $^{202}\mathrm{Pb}(\gamma,\nt)$ & $0.20$ & $0.02$ \\ 
$^{114}\mathrm{Sn}$ & $^{110}\mathrm{Sn}(\gamma,\pt)$ & $-0.17$ & $-0.02$ & $^{162}\mathrm{Er}$ & $^{159}\mathrm{Er}(\gamma,\nt)$ & $-0.29$ & $-0.12$ \\ 
$ $ & $^{110}\mathrm{Sn}(\gamma,\alpha)$ & $-0.26$ & $-0.03$ & $ $ & $^{159}\mathrm{Er}(\gamma,\alpha)$ & $-0.17$ & $-0.05$ \\ 
$ $ & $^{113}\mathrm{Sn}(\gamma,\nt)$ & $0.70$ & $0.08$ & $ $ & $^{160}\mathrm{Er}(\gamma,\nt)$ & $-0.22$ & $-0.09$ \\ 
$ $ & $^{122}\mathrm{Xe}(\gamma,\pt)$ & $0.16$ & $0.02$ & $ $ & $^{160}\mathrm{Er}(\gamma,\alpha)$ & $-0.40$ & $-0.13$ \\ 
$ $ & $^{122}\mathrm{Xe}(\gamma,\alpha)$ & $0.27$ & $0.03$ & $ $ & $^{161}\mathrm{Er}(\gamma,\nt)$ & $0.18$ & $0.08$ \\ 
$ $ & $^{169}\mathrm{Lu}(\gamma,\nt)$ & $0.15$ & $0.01$ & $ $ & $^{166}\mathrm{Yb}(\gamma,\alpha)$ & $0.32$ & $0.11$ \\ 
$^{115}\mathrm{Sn}$ & $^{110}\mathrm{Sn}(\gamma,\pt)$ & $-0.17$ & $-0.02$ & $ $ & $^{196}\mathrm{Pb}(\gamma,\nt)$ & $0.27$ & $0.10$ \\ 
$ $ & $^{110}\mathrm{Sn}(\gamma,\alpha)$ & $-0.25$ & $-0.03$ & $ $ & $^{202}\mathrm{Pb}(\gamma,\nt)$ & $0.15$ & $0.06$ \\ 
$ $ & $^{113}\mathrm{Sn}(\gamma,\nt)$ & $0.73$ & $0.08$ & $^{164}\mathrm{Er}$ & $^{164}\mathrm{Yb}(\gamma,\nt)$ & $-0.24$ & $-0.08$ \\ 
$ $ & $^{122}\mathrm{Xe}(\gamma,\alpha)$ & $0.24$ & $0.03$ & $ $ & $^{164}\mathrm{Yb}(\gamma,\alpha)$ & $-0.56$ & $-0.24$ \\ 
$ $ & $^{169}\mathrm{Lu}(\gamma,\nt)$ & $0.15$ & $0.01$ & $ $ & $^{165}\mathrm{Yb}(\gamma,\nt)$ & $-0.17$ & $-0.06$ \\ 
$^{120}\mathrm{Te}$ & $^{121}\mathrm{Te}(\gamma,\nt)$ & $-0.78$ & $-0.24$ & $ $ & $^{196}\mathrm{Pb}(\gamma,\nt)$ & $0.19$ & $0.06$ \\ 
$ $ & $^{122}\mathrm{Xe}(\gamma,\pt)$ & $0.24$ & $0.06$ & $ $ & $^{202}\mathrm{Pb}(\gamma,\nt)$ & $0.15$ & $0.06$ \\ 
$ $ & $^{122}\mathrm{Xe}(\gamma,\alpha)$ & $-0.22$ & $-0.06$ & $^{168}\mathrm{Yb}$ & $^{164}\mathrm{Yb}(\gamma,\alpha)$ & $-0.22$ & $-0.10$ \\ 
$^{124}\mathrm{Xe}$ & $^{122}\mathrm{Xe}(\gamma,\pt)$ & $-0.20$ & $-0.07$ & $ $ & $^{166}\mathrm{Yb}(\gamma,\nt)$ & $-0.18$ & $-0.07$ \\ 
$ $ & $^{122}\mathrm{Xe}(\gamma,\alpha)$ & $-0.36$ & $-0.18$ & $ $ & $^{166}\mathrm{Yb}(\gamma,\alpha)$ & $-0.34$ & $-0.12$ \\ 
$ $ & $^{125}\mathrm{Xe}(\gamma,\nt)$ & $-0.54$ & $-0.33$ & $ $ & $^{167}\mathrm{Yb}(\gamma,\nt)$ & $0.19$ & $0.07$ \\ 
$^{126}\mathrm{Xe}$ & $^{122}\mathrm{Xe}(\gamma,\pt)$ & $-0.19$ & $-0.07$ & $ $ & $^{172}\mathrm{Hf}(\gamma,\alpha)$ & $0.43$ & $0.16$ \\ 
$ $ & $^{122}\mathrm{Xe}(\gamma,\alpha)$ & $-0.39$ & $-0.18$ & $ $ & $^{196}\mathrm{Pb}(\gamma,\nt)$ & $0.29$ & $0.10$ \\ 
$ $ & $^{125}\mathrm{Xe}(\gamma,\nt)$ & $0.22$ & $0.11$ & $ $ & $^{202}\mathrm{Pb}(\gamma,\nt)$ & $0.19$ & $0.06$ \\ 
$ $ & $^{127}\mathrm{Xe}(\gamma,\nt)$ & $-0.57$ & $-0.29$ & $^{174}\mathrm{Hf}$ & $^{172}\mathrm{Hf}(\gamma,\alpha)$ & $-0.31$ & $-0.06$ \\ 
$^{130}\mathrm{Ba}$ & $^{126}\mathrm{Ba}(\gamma,\alpha)$ & $-0.16$ & $-0.02$ & $ $ & $^{174}\mathrm{W}(\gamma,\nt)$ & $-0.19$ & $-0.03$ \\ 
$ $ & $^{131}\mathrm{Ba}(\gamma,\nt)$ & $-0.81$ & $-0.15$ & $ $ & $^{174}\mathrm{W}(\gamma,\alpha)$ & $-0.30$ & $-0.07$ \\ 
$^{132}\mathrm{Ba}$ & $^{131}\mathrm{Ba}(\gamma,\nt)$ & $0.41$ & $0.15$ & $ $ & $^{178}\mathrm{W}(\gamma,\alpha)$ & $0.46$ & $0.08$ \\ 
$ $ & $^{133}\mathrm{Ba}(\gamma,\nt)$ & $-0.74$ & $-0.27$ & $ $ & $^{182}\mathrm{Os}(\gamma,\alpha)$ & $0.17$ & $0.03$ \\ 
$^{138}\mathrm{La}$ & $^{133}\mathrm{La}(\gamma,\pt)$ & $-0.19$ & $-0.08$ & $ $ & $^{196}\mathrm{Pb}(\gamma,\nt)$ & $0.20$ & $0.04$ \\ 
$ $ & $^{135}\mathrm{La}(\gamma,\nt)$ & $-0.31$ & $-0.12$ & $^{180}\mathrm{Ta}$ & $^{179}\mathrm{Ta}(\gamma,\nt)$ & $-0.89$ & $-0.07$ \\ 
$ $ & $^{136}\mathrm{La}(\gamma,\nt)$ & $0.25$ & $0.08$ & $^{180}\mathrm{W}$ & $^{180}\mathrm{Os}(\gamma,\nt)$ & $-0.26$ & $-0.08$ \\ 
$ $ & $^{137}\mathrm{La}(\gamma,\nt)$ & $-0.35$ & $-0.16$ & $ $ & $^{180}\mathrm{Os}(\gamma,\alpha)$ & $-0.50$ & $-0.21$ \\ 
$^{136}\mathrm{Ce}$ & $^{137}\mathrm{Ce}(\gamma,\nt)$ & $-0.47$ & $-0.07$ & $ $ & $^{181}\mathrm{Os}(\gamma,\nt)$ & $-0.16$ & $-0.04$ \\ 
$ $ & $^{138}\mathrm{Nd}(\gamma,\nt)$ & $-0.31$ & $-0.04$ & $ $ & $^{196}\mathrm{Pb}(\gamma,\nt)$ & $0.22$ & $0.06$ \\ 
$ $ & $^{138}\mathrm{Nd}(\gamma,\pt)$ & $0.52$ & $0.07$ & $^{184}\mathrm{Os}$ & $^{182}\mathrm{Os}(\gamma,\alpha)$ & $-0.19$ & $-0.02$ \\ 
$ $ & $^{140}\mathrm{Nd}(\gamma,\alpha)$ & $0.32$ & $0.04$ & $ $ & $^{185}\mathrm{Os}(\gamma,\nt)$ & $0.17$ & $0.01$ \\ 
$^{138}\mathrm{Ce}$ & $^{137}\mathrm{Ce}(\gamma,\nt)$ & $0.66$ & $0.08$ & $ $ & $^{184}\mathrm{Pt}(\gamma,\alpha)$ & $-0.45$ & $-0.04$ \\ 
$ $ & $^{139}\mathrm{Ce}(\gamma,\nt)$ & $-0.37$ & $-0.04$ & $ $ & $^{186}\mathrm{Pt}(\gamma,\alpha)$ & $-0.16$ & $-0.01$ \\ 
$ $ & $^{138}\mathrm{Nd}(\gamma,\nt)$ & $-0.16$ & $-0.02$ & $ $ & $^{188}\mathrm{Pt}(\gamma,\alpha)$ & $0.16$ & $0.01$ \\ 
$ $ & $^{138}\mathrm{Nd}(\gamma,\pt)$ & $0.28$ & $0.03$ & $ $ & $^{196}\mathrm{Pb}(\gamma,\nt)$ & $0.20$ & $0.01$ \\ 
$ $ & $^{140}\mathrm{Nd}(\gamma,\alpha)$ & $0.18$ & $0.02$ & $^{190}\mathrm{Pt}$ & $^{190}\mathrm{Hg}(\gamma,\nt)$ & $-0.30$ & $-0.04$ \\ 
$ $ & $ $ &  &  & $ $ & $^{190}\mathrm{Hg}(\gamma,\alpha)$ & $-0.51$ & $-0.07$ \\ 
$ $ & $ $ &  &  & $ $ & $^{196}\mathrm{Pb}(\gamma,\nt)$ & $0.19$ & $0.02$ \\ 
$ $ & $ $ &  &  & $^{196}\mathrm{Hg}$ & $^{196}\mathrm{Pb}(\gamma,\nt)$ & $-0.68$ & $-0.42$ \\ 
$ $ & $ $ &  &  & $ $ & $^{197}\mathrm{Pb}(\gamma,\nt)$ & $-0.28$ & $-0.13$ \\ 
$ $ & $ $ &  &  & $ $ & $^{202}\mathrm{Pb}(\gamma,\nt)$ & $0.20$ & $0.10$ \\ 
\enddata
\tablecomments{The data is split into two sets of four columns.}
\end{deluxetable*}

\startlongtable
\begin{deluxetable*}{llcc llcc}
\tablewidth{0pt}
\tablecaption{Correlations and $\zeta$ slopes between mass fraction and reaction rates for the $10\times$3D-inspired mixing scenario.
\label{tab:10x3d_corr}}
\tablehead{
\colhead{\textbf{Isotope}} & \colhead{\textbf{Reaction}} & \colhead{$\mathbf{r_\mathrm{\mathbf{P}}}$} & \colhead{$\mathbf{\zeta}$} &\colhead{\textbf{Isotope}} & \colhead{\textbf{Reaction}} & \colhead{$\mathbf{r_\mathrm{\mathbf{P}}}$} & \colhead{$\mathbf{\zeta}$}
}
\startdata
$^{74}\mathrm{Se}$ & $^{75}\mathrm{Se}(\gamma,\nt)$ & $-0.87$ & $-0.30$ & $^{138}\mathrm{Ce}$ & $^{137}\mathrm{Ce}(\gamma,\nt)$ & $0.58$ & $0.05$ \\ 
$^{78}\mathrm{Kr}$ & $^{79}\mathrm{Kr}(\gamma,\nt)$ & $-0.81$ & $-0.35$ & $ $ & $^{139}\mathrm{Ce}(\gamma,\nt)$ & $-0.19$ & $-0.01$ \\ 
$^{84}\mathrm{Sr}$ & $^{84}\mathrm{Rb}(\gamma,\nt)$ & $0.16$ & $0.06$ & $ $ & $^{138}\mathrm{Nd}(\gamma,\nt)$ & $-0.31$ & $-0.02$ \\ 
$ $ & $^{85}\mathrm{Sr}(\gamma,\nt)$ & $-0.81$ & $-0.33$ & $ $ & $^{138}\mathrm{Nd}(\gamma,\pt)$ & $0.37$ & $0.03$ \\ 
$^{92}\mathrm{Mo}$ & $^{93}\mathrm{Mo}(\gamma,\nt)$ & $-0.93$ & $-0.17$ & $ $ & $^{138}\mathrm{Nd}(\gamma,\alpha)$ & $-0.17$ & $-0.01$ \\ 
$^{94}\mathrm{Mo}$ & $^{93}\mathrm{Mo}(\gamma,\nt)$ & $0.94$ & $0.21$ & $ $ & $^{140}\mathrm{Nd}(\gamma,\alpha)$ & $0.20$ & $0.01$ \\ 
$^{96}\mathrm{Ru}$ & $^{97}\mathrm{Ru}(\gamma,\nt)$ & $-0.88$ & $-0.25$ & $^{144}\mathrm{Sm}$ & $^{142}\mathrm{Sm}(\gamma,\nt)$ & $-0.31$ & $-0.04$ \\ 
$^{98}\mathrm{Ru}$ & $^{97}\mathrm{Ru}(\gamma,\nt)$ & $0.68$ & $0.07$ & $ $ & $^{142}\mathrm{Sm}(\gamma,\pt)$ & $-0.29$ & $-0.03$ \\ 
$ $ & $^{100}\mathrm{Pd}(\gamma,\pt)$ & $0.45$ & $0.05$ & $ $ & $^{143}\mathrm{Sm}(\gamma,\nt)$ & $-0.41$ & $-0.05$ \\ 
$ $ & $^{100}\mathrm{Pd}(\gamma,\alpha)$ & $-0.29$ & $-0.03$ & $ $ & $^{196}\mathrm{Pb}(\gamma,\nt)$ & $0.35$ & $0.04$ \\ 
$ $ & $^{110}\mathrm{Sn}(\gamma,\alpha)$ & $0.18$ & $0.02$ & $^{152}\mathrm{Gd}$ & $^{152}\mathrm{Dy}(\gamma,\alpha)$ & $-0.50$ & $-0.02$ \\ 
$^{102}\mathrm{Pd}$ & $^{100}\mathrm{Pd}(\gamma,\pt)$ & $-0.29$ & $-0.08$ & $ $ & $^{154}\mathrm{Dy}(\gamma,\alpha)$ & $-0.16$ & $-0.00$ \\ 
$ $ & $^{100}\mathrm{Pd}(\gamma,\alpha)$ & $-0.30$ & $-0.08$ & $ $ & $^{158}\mathrm{Er}(\gamma,\nt)$ & $0.15$ & $0.00$ \\ 
$ $ & $^{103}\mathrm{Pd}(\gamma,\nt)$ & $-0.65$ & $-0.21$ & $ $ & $^{160}\mathrm{Er}(\gamma,\alpha)$ & $0.19$ & $0.00$ \\ 
$^{106}\mathrm{Cd}$ & $^{107}\mathrm{Cd}(\gamma,\nt)$ & $-0.83$ & $-0.30$ & $^{156}\mathrm{Dy}$ & $^{159}\mathrm{Er}(\gamma,\nt)$ & $-0.19$ & $-0.06$ \\ 
$ $ & $^{110}\mathrm{Sn}(\gamma,\alpha)$ & $0.19$ & $0.07$ & $ $ & $^{160}\mathrm{Er}(\gamma,\alpha)$ & $0.70$ & $0.26$ \\ 
$^{108}\mathrm{Cd}$ & $^{107}\mathrm{Cd}(\gamma,\nt)$ & $0.63$ & $0.15$ & $ $ & $^{202}\mathrm{Pb}(\gamma,\nt)$ & $0.16$ & $0.05$ \\ 
$ $ & $^{109}\mathrm{Cd}(\gamma,\nt)$ & $-0.50$ & $-0.10$ & $^{158}\mathrm{Dy}$ & $^{159}\mathrm{Dy}(\gamma,\nt)$ & $0.19$ & $0.01$ \\ 
$^{113}\mathrm{In}$ & $^{110}\mathrm{Sn}(\gamma,\pt)$ & $-0.17$ & $-0.05$ & $ $ & $^{158}\mathrm{Er}(\gamma,\nt)$ & $-0.16$ & $-0.01$ \\ 
$ $ & $^{110}\mathrm{Sn}(\gamma,\alpha)$ & $-0.24$ & $-0.08$ & $ $ & $^{158}\mathrm{Er}(\gamma,\alpha)$ & $-0.28$ & $-0.01$ \\ 
$ $ & $^{113}\mathrm{Sn}(\gamma,\nt)$ & $0.82$ & $0.34$ & $ $ & $^{159}\mathrm{Er}(\gamma,\nt)$ & $-0.18$ & $-0.01$ \\ 
$^{112}\mathrm{Sn}$ & $^{110}\mathrm{Sn}(\gamma,\alpha)$ & $-0.17$ & $-0.05$ & $ $ & $^{160}\mathrm{Er}(\gamma,\alpha)$ & $0.56$ & $0.02$ \\ 
$ $ & $^{113}\mathrm{Sn}(\gamma,\nt)$ & $-0.78$ & $-0.33$ & $ $ & $^{196}\mathrm{Pb}(\gamma,\nt)$ & $0.17$ & $0.01$ \\ 
$^{114}\mathrm{Sn}$ & $^{110}\mathrm{Sn}(\gamma,\pt)$ & $-0.18$ & $-0.02$ & $ $ & $^{202}\mathrm{Pb}(\gamma,\nt)$ & $0.17$ & $0.01$ \\ 
$ $ & $^{110}\mathrm{Sn}(\gamma,\alpha)$ & $-0.27$ & $-0.04$ & $^{162}\mathrm{Er}$ & $^{159}\mathrm{Er}(\gamma,\nt)$ & $-0.28$ & $-0.10$ \\ 
$ $ & $^{113}\mathrm{Sn}(\gamma,\nt)$ & $0.76$ & $0.12$ & $ $ & $^{159}\mathrm{Er}(\gamma,\alpha)$ & $-0.16$ & $-0.05$ \\ 
$ $ & $^{122}\mathrm{Xe}(\gamma,\alpha)$ & $0.15$ & $0.03$ & $ $ & $^{160}\mathrm{Er}(\gamma,\nt)$ & $-0.19$ & $-0.07$ \\ 
$ $ & $^{169}\mathrm{Lu}(\gamma,\nt)$ & $0.16$ & $0.02$ & $ $ & $^{160}\mathrm{Er}(\gamma,\alpha)$ & $-0.29$ & $-0.09$ \\ 
$^{115}\mathrm{Sn}$ & $^{110}\mathrm{Sn}(\gamma,\pt)$ & $-0.18$ & $-0.02$ & $ $ & $^{161}\mathrm{Er}(\gamma,\nt)$ & $0.18$ & $0.06$ \\ 
$ $ & $^{110}\mathrm{Sn}(\gamma,\alpha)$ & $-0.27$ & $-0.04$ & $ $ & $^{166}\mathrm{Yb}(\gamma,\alpha)$ & $0.40$ & $0.12$ \\ 
$ $ & $^{113}\mathrm{Sn}(\gamma,\nt)$ & $0.77$ & $0.12$ & $ $ & $^{196}\mathrm{Pb}(\gamma,\nt)$ & $0.26$ & $0.09$ \\ 
$ $ & $^{169}\mathrm{Lu}(\gamma,\nt)$ & $0.16$ & $0.02$ & $ $ & $^{202}\mathrm{Pb}(\gamma,\nt)$ & $0.15$ & $0.05$ \\ 
$^{120}\mathrm{Te}$ & $^{119}\mathrm{Te}(\gamma,\nt)$ & $-0.20$ & $-0.04$ & $^{164}\mathrm{Er}$ & $^{164}\mathrm{Yb}(\gamma,\nt)$ & $-0.23$ & $-0.10$ \\ 
$ $ & $^{121}\mathrm{Te}(\gamma,\nt)$ & $-0.66$ & $-0.13$ & $ $ & $^{164}\mathrm{Yb}(\gamma,\alpha)$ & $-0.59$ & $-0.43$ \\ 
$ $ & $^{122}\mathrm{Xe}(\gamma,\pt)$ & $0.42$ & $0.07$ & $ $ & $^{165}\mathrm{Yb}(\gamma,\nt)$ & $-0.17$ & $-0.09$ \\ 
$ $ & $^{122}\mathrm{Xe}(\gamma,\alpha)$ & $-0.35$ & $-0.06$ & $ $ & $^{196}\mathrm{Pb}(\gamma,\nt)$ & $0.16$ & $0.07$ \\ 
$^{124}\mathrm{Xe}$ & $^{122}\mathrm{Xe}(\gamma,\nt)$ & $-0.22$ & $-0.08$ & $^{168}\mathrm{Yb}$ & $^{168}\mathrm{Hf}(\gamma,\nt)$ & $-0.16$ & $-0.05$ \\ 
$ $ & $^{122}\mathrm{Xe}(\gamma,\pt)$ & $-0.27$ & $-0.10$ & $ $ & $^{168}\mathrm{Hf}(\gamma,\alpha)$ & $-0.40$ & $-0.21$ \\ 
$ $ & $^{122}\mathrm{Xe}(\gamma,\alpha)$ & $-0.44$ & $-0.23$ & $ $ & $^{169}\mathrm{Hf}(\gamma,\nt)$ & $-0.17$ & $-0.05$ \\ 
$ $ & $^{125}\mathrm{Xe}(\gamma,\nt)$ & $-0.40$ & $-0.22$ & $ $ & $^{172}\mathrm{Hf}(\gamma,\alpha)$ & $0.41$ & $0.19$ \\ 
$^{126}\mathrm{Xe}$ & $^{122}\mathrm{Xe}(\gamma,\nt)$ & $-0.21$ & $-0.06$ & $ $ & $^{196}\mathrm{Pb}(\gamma,\nt)$ & $0.22$ & $0.08$ \\ 
$ $ & $^{122}\mathrm{Xe}(\gamma,\pt)$ & $-0.26$ & $-0.08$ & $^{174}\mathrm{Hf}$ & $^{174}\mathrm{W}(\gamma,\nt)$ & $-0.23$ & $-0.06$ \\ 
$ $ & $^{122}\mathrm{Xe}(\gamma,\alpha)$ & $-0.46$ & $-0.19$ & $ $ & $^{174}\mathrm{W}(\gamma,\alpha)$ & $-0.52$ & $-0.19$ \\ 
$ $ & $^{125}\mathrm{Xe}(\gamma,\nt)$ & $0.38$ & $0.16$ & $ $ & $^{176}\mathrm{W}(\gamma,\alpha)$ & $-0.15$ & $-0.05$ \\ 
$ $ & $^{127}\mathrm{Xe}(\gamma,\nt)$ & $-0.26$ & $-0.11$ & $^{180}\mathrm{Ta}$ & $^{179}\mathrm{Ta}(\gamma,\nt)$ & $-0.88$ & $-0.03$ \\ 
$^{130}\mathrm{Ba}$ & $^{126}\mathrm{Ba}(\gamma,\pt)$ & $-0.18$ & $-0.02$ & $^{180}\mathrm{W}$ & $^{180}\mathrm{Os}(\gamma,\nt)$ & $-0.30$ & $-0.12$ \\ 
$ $ & $^{126}\mathrm{Ba}(\gamma,\alpha)$ & $-0.31$ & $-0.03$ & $ $ & $^{180}\mathrm{Os}(\gamma,\alpha)$ & $-0.54$ & $-0.39$ \\ 
$ $ & $^{128}\mathrm{Ba}(\gamma,\nt)$ & $-0.26$ & $-0.03$ & $ $ & $^{196}\mathrm{Pb}(\gamma,\nt)$ & $0.21$ & $0.08$ \\ 
$ $ & $^{128}\mathrm{Ba}(\gamma,\pt)$ & $-0.16$ & $-0.01$ & $^{184}\mathrm{Os}$ & $^{184}\mathrm{Pt}(\gamma,\alpha)$ & $-0.49$ & $-0.10$ \\ 
$ $ & $^{128}\mathrm{Ba}(\gamma,\alpha)$ & $-0.27$ & $-0.02$ & $ $ & $^{185}\mathrm{Pt}(\gamma,\alpha)$ & $-0.16$ & $-0.02$ \\ 
$ $ & $^{129}\mathrm{Ba}(\gamma,\nt)$ & $0.21$ & $0.02$ & $ $ & $^{196}\mathrm{Pb}(\gamma,\nt)$ & $0.16$ & $0.02$ \\ 
$ $ & $^{131}\mathrm{Ba}(\gamma,\nt)$ & $-0.51$ & $-0.05$ & $^{190}\mathrm{Pt}$ & $^{190}\mathrm{Hg}(\gamma,\nt)$ & $-0.27$ & $-0.06$ \\ 
$^{132}\mathrm{Ba}$ & $^{128}\mathrm{Ba}(\gamma,\alpha)$ & $-0.18$ & $-0.02$ & $ $ & $^{190}\mathrm{Hg}(\gamma,\alpha)$ & $-0.52$ & $-0.15$ \\ 
$ $ & $^{131}\mathrm{Ba}(\gamma,\nt)$ & $0.63$ & $0.12$ & $ $ & $^{196}\mathrm{Pb}(\gamma,\nt)$ & $0.18$ & $0.04$ \\ 
$ $ & $^{133}\mathrm{Ba}(\gamma,\nt)$ & $-0.56$ & $-0.10$ & $^{196}\mathrm{Hg}$ & $^{196}\mathrm{Pb}(\gamma,\nt)$ & $-0.75$ & $-0.53$ \\ 
$^{138}\mathrm{La}$ & $^{137}\mathrm{La}(\gamma,\nt)$ & $-0.65$ & $-0.35$ & $ $ & $^{202}\mathrm{Pb}(\gamma,\nt)$ & $0.20$ & $0.12$ \\ 
$^{136}\mathrm{Ce}$ & $^{138}\mathrm{Nd}(\gamma,\nt)$ & $-0.39$ & $-0.06$ & $ $ & $ $ &  &  \\ 
$ $ & $^{138}\mathrm{Nd}(\gamma,\pt)$ & $0.65$ & $0.10$ & $ $ & $ $ &  &  \\ 
$ $ & $^{138}\mathrm{Nd}(\gamma,\alpha)$ & $-0.16$ & $-0.03$ & $ $ & $ $ &  &  \\ 
$ $ & $^{140}\mathrm{Nd}(\gamma,\alpha)$ & $0.31$ & $0.05$ & $ $ & $ $ &  &  \\ 
\enddata
\tablecomments{The data is split into two sets of four columns.}
\end{deluxetable*}

\startlongtable
\begin{deluxetable*}{llcc llcc}
\tablewidth{0pt}
\tablecaption{Correlations and $\zeta$ slopes between mass fraction and reaction rates for the $50\times$3D-inspired mixing scenario.
\label{tab:50x3d_corr}}
\tablehead{
\colhead{\textbf{Isotope}} & \colhead{\textbf{Reaction}} & \colhead{$\mathbf{r_\mathrm{\mathbf{P}}}$} & \colhead{$\mathbf{\zeta}$} &\colhead{\textbf{Isotope}} & \colhead{\textbf{Reaction}} & \colhead{$\mathbf{r_\mathrm{\mathbf{P}}}$} & \colhead{$\mathbf{\zeta}$}
}
\startdata
$^{74}\mathrm{Se}$ & $^{75}\mathrm{Se}(\gamma,\nt)$ & $-0.95$ & $-0.10$ & $^{130}\mathrm{Ba}$ & $^{126}\mathrm{Ba}(\gamma,\alpha)$ & $-0.19$ & $-0.01$ \\ 
$^{78}\mathrm{Kr}$ & $^{79}\mathrm{Kr}(\gamma,\nt)$ & $-0.91$ & $-0.28$ & $ $ & $^{132}\mathrm{Ce}(\gamma,\nt)$ & $-0.23$ & $-0.01$ \\ 
$^{84}\mathrm{Sr}$ & $^{85}\mathrm{Sr}(\gamma,\nt)$ & $-0.92$ & $-0.26$ & $ $ & $^{132}\mathrm{Ce}(\gamma,\pt)$ & $0.32$ & $0.01$ \\ 
$^{92}\mathrm{Mo}$ & $^{93}\mathrm{Mo}(\gamma,\nt)$ & $-0.57$ & $-0.03$ & $ $ & $^{132}\mathrm{Ce}(\gamma,\alpha)$ & $-0.20$ & $-0.01$ \\ 
$ $ & $^{100}\mathrm{Pd}(\gamma,\pt)$ & $0.19$ & $0.01$ & $ $ & $^{134}\mathrm{Ce}(\gamma,\nt)$ & $-0.20$ & $-0.01$ \\ 
$ $ & $^{100}\mathrm{Pd}(\gamma,\alpha)$ & $0.20$ & $0.01$ & $ $ & $^{134}\mathrm{Ce}(\gamma,\alpha)$ & $0.41$ & $0.02$ \\ 
$ $ & $^{110}\mathrm{Sn}(\gamma,\nt)$ & $0.16$ & $0.01$ & $^{132}\mathrm{Ba}$ & $^{132}\mathrm{Ce}(\gamma,\nt)$ & $-0.28$ & $-0.04$ \\ 
$ $ & $^{110}\mathrm{Sn}(\gamma,\pt)$ & $0.23$ & $0.01$ & $ $ & $^{132}\mathrm{Ce}(\gamma,\pt)$ & $-0.25$ & $-0.03$ \\ 
$ $ & $^{110}\mathrm{Sn}(\gamma,\alpha)$ & $0.37$ & $0.02$ & $ $ & $^{132}\mathrm{Ce}(\gamma,\alpha)$ & $-0.31$ & $-0.04$ \\ 
$^{94}\mathrm{Mo}$ & $^{93}\mathrm{Mo}(\gamma,\nt)$ & $0.89$ & $0.08$ & $ $ & $^{133}\mathrm{Ce}(\gamma,\nt)$ & $-0.19$ & $-0.03$ \\ 
$^{96}\mathrm{Ru}$ & $^{97}\mathrm{Ru}(\gamma,\nt)$ & $-0.62$ & $-0.05$ & $ $ & $^{134}\mathrm{Ce}(\gamma,\alpha)$ & $-0.18$ & $-0.02$ \\ 
$ $ & $^{97}\mathrm{Ru}(\gamma,\alpha)$ & $-0.18$ & $-0.01$ & $^{138}\mathrm{La}$ & $^{137}\mathrm{La}(\gamma,\nt)$ & $-0.75$ & $-0.45$ \\ 
$ $ & $^{100}\mathrm{Pd}(\gamma,\alpha)$ & $0.33$ & $0.03$ & $^{136}\mathrm{Ce}$ & $^{138}\mathrm{Nd}(\gamma,\nt)$ & $-0.42$ & $-0.09$ \\ 
$ $ & $^{110}\mathrm{Sn}(\gamma,\nt)$ & $0.16$ & $0.01$ & $ $ & $^{138}\mathrm{Nd}(\gamma,\pt)$ & $0.66$ & $0.14$ \\ 
$ $ & $^{110}\mathrm{Sn}(\gamma,\pt)$ & $0.16$ & $0.01$ & $ $ & $^{140}\mathrm{Nd}(\gamma,\alpha)$ & $0.25$ & $0.06$ \\ 
$ $ & $^{110}\mathrm{Sn}(\gamma,\alpha)$ & $0.23$ & $0.02$ & $^{138}\mathrm{Ce}$ & $^{138}\mathrm{Nd}(\gamma,\nt)$ & $-0.33$ & $-0.04$ \\ 
$^{98}\mathrm{Ru}$ & $^{100}\mathrm{Pd}(\gamma,\pt)$ & $0.70$ & $0.13$ & $ $ & $^{138}\mathrm{Nd}(\gamma,\pt)$ & $-0.31$ & $-0.04$ \\ 
$ $ & $^{100}\mathrm{Pd}(\gamma,\alpha)$ & $-0.53$ & $-0.10$ & $ $ & $^{138}\mathrm{Nd}(\gamma,\alpha)$ & $-0.24$ & $-0.02$ \\ 
$^{102}\mathrm{Pd}$ & $^{100}\mathrm{Pd}(\gamma,\pt)$ & $-0.39$ & $-0.06$ & $ $ & $^{139}\mathrm{Nd}(\gamma,\nt)$ & $-0.31$ & $-0.04$ \\ 
$ $ & $^{100}\mathrm{Pd}(\gamma,\alpha)$ & $-0.38$ & $-0.05$ & $^{144}\mathrm{Sm}$ & $^{142}\mathrm{Sm}(\gamma,\nt)$ & $-0.36$ & $-0.06$ \\ 
$ $ & $^{103}\mathrm{Pd}(\gamma,\nt)$ & $-0.23$ & $-0.04$ & $ $ & $^{142}\mathrm{Sm}(\gamma,\pt)$ & $-0.29$ & $-0.04$ \\ 
$ $ & $^{104}\mathrm{Cd}(\gamma,\pt)$ & $0.17$ & $0.03$ & $ $ & $^{143}\mathrm{Sm}(\gamma,\nt)$ & $-0.50$ & $-0.09$ \\ 
$ $ & $^{104}\mathrm{Cd}(\gamma,\alpha)$ & $-0.27$ & $-0.04$ & $ $ & $^{196}\mathrm{Pb}(\gamma,\nt)$ & $0.19$ & $0.03$ \\ 
$^{106}\mathrm{Cd}$ & $^{104}\mathrm{Cd}(\gamma,\pt)$ & $-0.23$ & $-0.03$ & $^{152}\mathrm{Gd}$ & $^{152}\mathrm{Dy}(\gamma,\alpha)$ & $-0.50$ & $-0.05$ \\ 
$ $ & $^{107}\mathrm{Cd}(\gamma,\nt)$ & $-0.62$ & $-0.07$ & $ $ & $^{158}\mathrm{Er}(\gamma,\nt)$ & $0.18$ & $0.01$ \\ 
$ $ & $^{110}\mathrm{Sn}(\gamma,\pt)$ & $0.20$ & $0.02$ & $^{156}\mathrm{Dy}$ & $^{156}\mathrm{Er}(\gamma,\alpha)$ & $-0.44$ & $-0.27$ \\ 
$ $ & $^{110}\mathrm{Sn}(\gamma,\alpha)$ & $0.40$ & $0.05$ & $ $ & $^{158}\mathrm{Er}(\gamma,\nt)$ & $0.19$ & $0.09$ \\ 
$^{108}\mathrm{Cd}$ & $^{107}\mathrm{Cd}(\gamma,\nt)$ & $0.61$ & $0.11$ & $ $ & $^{158}\mathrm{Er}(\gamma,\alpha)$ & $-0.18$ & $-0.07$ \\ 
$ $ & $^{110}\mathrm{Sn}(\gamma,\pt)$ & $0.49$ & $0.08$ & $ $ & $^{160}\mathrm{Er}(\gamma,\alpha)$ & $0.25$ & $0.15$ \\ 
$ $ & $^{110}\mathrm{Sn}(\gamma,\alpha)$ & $-0.33$ & $-0.06$ & $^{158}\mathrm{Dy}$ & $^{158}\mathrm{Er}(\gamma,\nt)$ & $-0.28$ & $-0.02$ \\ 
$^{113}\mathrm{In}$ & $^{113}\mathrm{Sn}(\gamma,\nt)$ & $0.94$ & $0.44$ & $ $ & $^{158}\mathrm{Er}(\gamma,\alpha)$ & $-0.53$ & $-0.05$ \\ 
$^{112}\mathrm{Sn}$ & $^{110}\mathrm{Sn}(\gamma,\pt)$ & $-0.25$ & $-0.03$ & $^{162}\mathrm{Er}$ & $^{162}\mathrm{Yb}(\gamma,\alpha)$ & $-0.49$ & $-0.30$ \\ 
$ $ & $^{110}\mathrm{Sn}(\gamma,\alpha)$ & $-0.34$ & $-0.04$ & $ $ & $^{164}\mathrm{Yb}(\gamma,\alpha)$ & $-0.16$ & $-0.07$ \\ 
$ $ & $^{113}\mathrm{Sn}(\gamma,\nt)$ & $-0.65$ & $-0.08$ & $ $ & $^{168}\mathrm{Hf}(\gamma,\nt)$ & $0.17$ & $0.07$ \\ 
$^{114}\mathrm{Sn}$ & $^{110}\mathrm{Sn}(\gamma,\alpha)$ & $-0.21$ & $-0.02$ & $^{164}\mathrm{Er}$ & $^{164}\mathrm{Yb}(\gamma,\nt)$ & $-0.31$ & $-0.17$ \\ 
$ $ & $^{113}\mathrm{Sn}(\gamma,\nt)$ & $0.78$ & $0.08$ & $ $ & $^{164}\mathrm{Yb}(\gamma,\alpha)$ & $-0.56$ & $-0.51$ \\ 
$ $ & $^{122}\mathrm{Xe}(\gamma,\nt)$ & $-0.28$ & $-0.03$ & $ $ & $^{202}\mathrm{Pb}(\gamma,\nt)$ & $0.16$ & $0.09$ \\ 
$ $ & $^{122}\mathrm{Xe}(\gamma,\pt)$ & $0.16$ & $0.02$ & $^{168}\mathrm{Yb}$ & $^{168}\mathrm{Hf}(\gamma,\nt)$ & $-0.28$ & $-0.12$ \\ 
$ $ & $^{122}\mathrm{Xe}(\gamma,\alpha)$ & $0.23$ & $0.03$ & $ $ & $^{168}\mathrm{Hf}(\gamma,\alpha)$ & $-0.55$ & $-0.57$ \\ 
$ $ & $^{169}\mathrm{Lu}(\gamma,\nt)$ & $0.15$ & $0.01$ & $^{174}\mathrm{Hf}$ & $^{174}\mathrm{W}(\gamma,\nt)$ & $-0.30$ & $-0.13$ \\ 
$^{115}\mathrm{Sn}$ & $^{110}\mathrm{Sn}(\gamma,\alpha)$ & $-0.21$ & $-0.02$ & $ $ & $^{174}\mathrm{W}(\gamma,\alpha)$ & $-0.53$ & $-0.37$ \\ 
$ $ & $^{113}\mathrm{Sn}(\gamma,\nt)$ & $0.80$ & $0.08$ & $^{180}\mathrm{Ta}$ & $^{179}\mathrm{Ta}(\gamma,\nt)$ & $-0.87$ & $-0.00$ \\ 
$ $ & $^{122}\mathrm{Xe}(\gamma,\nt)$ & $-0.27$ & $-0.03$ & $^{180}\mathrm{W}$ & $^{180}\mathrm{Os}(\gamma,\nt)$ & $-0.38$ & $-0.21$ \\ 
$ $ & $^{122}\mathrm{Xe}(\gamma,\pt)$ & $0.15$ & $0.02$ & $ $ & $^{180}\mathrm{Os}(\gamma,\alpha)$ & $-0.52$ & $-0.46$ \\ 
$ $ & $^{122}\mathrm{Xe}(\gamma,\alpha)$ & $0.22$ & $0.03$ & $ $ & $^{196}\mathrm{Pb}(\gamma,\nt)$ & $0.19$ & $0.11$ \\ 
$ $ & $^{169}\mathrm{Lu}(\gamma,\nt)$ & $0.15$ & $0.01$ & $^{184}\mathrm{Os}$ & $^{184}\mathrm{Pt}(\gamma,\nt)$ & $-0.16$ & $-0.05$ \\ 
$^{120}\mathrm{Te}$ & $^{120}\mathrm{Xe}(\gamma,\alpha)$ & $-0.22$ & $-0.05$ & $ $ & $^{184}\mathrm{Pt}(\gamma,\alpha)$ & $-0.55$ & $-0.29$ \\ 
$ $ & $^{122}\mathrm{Xe}(\gamma,\pt)$ & $0.58$ & $0.11$ & $ $ & $^{196}\mathrm{Pb}(\gamma,\nt)$ & $0.16$ & $0.04$ \\ 
$ $ & $^{122}\mathrm{Xe}(\gamma,\alpha)$ & $-0.49$ & $-0.10$ & $^{190}\mathrm{Pt}$ & $^{190}\mathrm{Hg}(\gamma,\nt)$ & $-0.30$ & $-0.13$ \\ 
$^{124}\mathrm{Xe}$ & $^{122}\mathrm{Xe}(\gamma,\nt)$ & $-0.28$ & $-0.10$ & $ $ & $^{190}\mathrm{Hg}(\gamma,\alpha)$ & $-0.49$ & $-0.30$ \\ 
$ $ & $^{122}\mathrm{Xe}(\gamma,\pt)$ & $-0.31$ & $-0.11$ & $ $ & $^{196}\mathrm{Pb}(\gamma,\nt)$ & $0.20$ & $0.08$ \\ 
$ $ & $^{122}\mathrm{Xe}(\gamma,\alpha)$ & $-0.43$ & $-0.21$ & $^{196}\mathrm{Hg}$ & $^{196}\mathrm{Pb}(\gamma,\nt)$ & $-0.76$ & $-0.60$ \\ 
$ $ & $^{123}\mathrm{Xe}(\gamma,\nt)$ & $0.18$ & $0.06$ & $ $ & $^{202}\mathrm{Pb}(\gamma,\nt)$ & $0.18$ & $0.11$ \\ 
$^{126}\mathrm{Xe}$ & $^{126}\mathrm{Ba}(\gamma,\pt)$ & $-0.35$ & $-0.17$ & $ $ & $ $ &  &  \\ 
$ $ & $^{126}\mathrm{Ba}(\gamma,\alpha)$ & $-0.48$ & $-0.32$ & $ $ & $ $ &  &  \\ 
$ $ & $^{127}\mathrm{Ba}(\gamma,\nt)$ & $-0.22$ & $-0.12$ & $ $ & $ $ &  &  \\ 
\enddata
\tablecomments{The data is split into two sets of four columns.}
\end{deluxetable*}

\end{document}

%% file: units.tex
\newcommand{\numberspace}{\ensuremath{\;}}

\newcommand{\unitstyle}[1]{\ensuremath{\mathrm{#1}}}

\newcommand{\natlog}[2]{\ensuremath{#1\times 10^{#2}}} 

\newcommand{\centi}{\unitstyle{c}}

\newcommand{\Mega}{\unitstyle{M}}
\newcommand{\Giga}{\unitstyle{G}}

\newcommand{\meter}{\unitstyle{m}}

\newcommand{\second}{\unitstyle{s}}
\newcommand{\Kelvin}{\unitstyle{K}}
\newcommand{\K}{\Kelvin}  

\newcommand{\cm}{\centi\meter}

\newcommand{\Msun}{\ensuremath{\unitstyle{M}_\odot}}

\newcommand{\yr}{\unitstyle{yr}}        
\newcommand{\Mm}{\Mega\meter}   

\newcommand{\OP}{\ensuremath{\unitstyle{OP}}} 

\newcommand{\unit}[2]{\ensuremath{#1\numberspace\mathrm{#2}}}

%% file: nuclides.tex
\newcommand{\nuclei}[2]{\ensuremath{\mathrm{^{#1}#2}}}

\newcommand{\neutron}{\ensuremath{n}}
\newcommand{\nt}{\neutron}
\newcommand{\proton}{\ensuremath{p}}
\newcommand{\pt}{\proton}
\newcommand{\hydrogen}[1][1]{\nuclei{#1}{H}}
\newcommand{\helium}[1][4]{\nuclei{#1}{He}}

\newcommand{\carbon}[1][12]{\nuclei{#1}{C}}

\newcommand{\oxygen}[1][16]{\nuclei{#1}{O}}

\newcommand{\neon}[1][20]{\nuclei{#1}{Ne}}

\newcommand{\phosphorus}[1][31]{\nuclei{#1}{P}}

\newcommand{\chlorine}[1][35]{\nuclei{#1}{Cl}}

\newcommand{\potassium}[1][39]{\nuclei{#1}{K}}

\newcommand{\scandium}[1][45]{\nuclei{#1}{Sc}}

\newcommand{\selenium}[1][80]{\nuclei{#1}{Se}}

\newcommand{\krypton}[1][84]{\nuclei{#1}{Kr}}

\newcommand{\strontium}[1][88]{\nuclei{#1}{Sr}}

\newcommand{\molybdenum}[1][98]{\nuclei{#1}{Mo}}

\newcommand{\ruthenium}[1][102]{\nuclei{#1}{Ru}}

\newcommand{\cadmium}[1][114]{\nuclei{#1}{Cd}}
\newcommand{\indium}[1][115]{\nuclei{#1}{In}}
\newcommand{\tin}[1][120]{\nuclei{#1}{Sn}}

\newcommand{\barium}[1][138]{\nuclei{#1}{Ba}}
\newcommand{\lanthanum}[1][139]{\nuclei{#1}{La}}
\newcommand{\cerium}[1][140]{\nuclei{#1}{Ce}}

\newcommand{\neodymium}[1][142]{\nuclei{#1}{Nd}}

\newcommand{\samarium}[1][152]{\nuclei{#1}{Sm}}

\newcommand{\gadolinium}[1][158]{\nuclei{#1}{Gd}}

\newcommand{\dysprosium}[1][164]{\nuclei{#1}{Dy}}

\newcommand{\erbium}[1][168]{\nuclei{#1}{Er}}

\newcommand{\tantalum}[1][180]{\nuclei{#1}{Ta}}
\newcommand{\tungsten}[1][184]{\nuclei{#1}{W}}

\newcommand{\osmium}[1][192]{\nuclei{#1}{Os}}

\newcommand{\platnium}[1][195]{\nuclei{#1}{Pt}}

\newcommand{\mercury}[1][202]{\nuclei{#1}{Hg}}

\newcommand{\polonium}[1][210]{\nuclei{#1}{Po}}

%% file: formatting.tex
\newcommand{\isofont}[1]{{\mathrm{#1}}}
\newcommand{\isomass}[1]{{\ensuremath{\isofont{^{#1}}}}}
\newcommand{\isocharge}[1]{{\ensuremath{\isofont{_{#1}}}}}
\newcommand{\isotope}[3]{{\ensuremath{\isocharge{#1}\isomass{#2}\isofont{#3}}}}

\newcommand{\I}[2]{{\isotope{}{#1}{#2}}}

\newcommand{\Ep}[1]{{\ensuremath{10^{#1}}}}
\newcommand{\E}[1]{{\ensuremath{\powersep\Ep{#1}}}}

\newcommand{\powersep}{\times}

\newcommand{\mem}[1]{\ensuremath{\mathrm{ #1}}}


%% file: derivatives.tex
\newcommand{\D}{{\mathrm d}}



%% file: concepts.tex
\newcommand{\spr}{\mbox{$s$-process}}
\newcommand{\sprn}{\mbox{$s$ process}}

\newcommand{\iprn}{\mbox{$i$ process}}

\newcommand{\rpr}{\mbox{$r$-process}}
\newcommand{\rprn}{\mbox{$r$ process}}
\newcommand{\ppr}{\mbox{$p$-process}}
\newcommand{\pprn}{\mbox{$p$ process}}
\newcommand{\pnuc}{\mbox{$p$-nuclei}}
\newcommand{\pnucn}{\mbox{$p$ nuclei}}
\newcommand{\gpr}{\mbox{$\gamma$-process}}
\newcommand{\gprn}{\mbox{$\gamma$ process}}



%% file: code.tex
\newcommand{\code}[1]{\texttt{#1}}

\newcommand{\mesa}{\code{MESA}}
\newcommand{\MESA}{\mesa}

\newcommand{\mppnp}{\code{mppnp}} 


%% file: abbrev.tex
\newcommand{\n}{\ensuremath{\mem{n}}}

\newcommand{\p}{\ensuremath{\mem{p}}}
\newcommand{\e}{\ensuremath{\mem{e}}}

\newcommand{\h}{\ensuremath{\mem{H}}}

\newcommand{\nine}{\ensuremath{^{59}\mem{Ni}}}

\newcommand{\dex}{\ensuremath{\, \mathrm{dex}}} 
\newcommand{\mzams}{\ensuremath{M_{\rm ZAMS}}}

\newcommand{\beq}{\begin{equation}}
\newcommand{\beqa}{\begin{eqnarray}}
\newcommand{\eeq}{\end{equation}}
\newcommand{\eeqa}{\end{eqnarray}}
\newcommand{\bedis}{\begin{displaymath}}
\newcommand{\edis}{\end{displaymath}}

